\documentclass[iop]{emulateapj}
\usepackage{amssymb}
\usepackage{natbib}
\usepackage{graphicx}
\usepackage{epstopdf}
\newcommand{\msun}{\ensuremath{\rm M_\odot}}
\newcommand{\zsun}{\ensuremath{\rm Z_\odot}}
\newcommand{\msunyr}{\ensuremath{\rm M_{\odot}\;{\rm yr}^{-1}}}
\newcommand{\Ha}{\ensuremath{\rm H\alpha}}
\newcommand{\Hb}{\ensuremath{\rm H\beta}}
\newcommand{\lya}{\ensuremath{\rm Ly\alpha}}
\newcommand{\Ntwo}{[\ion{N}{2}]}
\newcommand{\kms}{km~s\ensuremath{^{-1}\,}}

\newcommand{\ztwo}{\ensuremath{z\sim2}}
\newcommand{\zthree}{\ensuremath{z\sim3}}
\newcommand{\Othree}{[\ion{O}{3}]}
\newcommand{\Otwo}{[\ion{O}{2}]}
\newcommand{\Cthree}{\ion{C}{3}]}
\newcommand{\Othreeuv}{\ion{O}{3}]}

\newcommand{\hetwo}{\ion{He}{2}}

\begin{document}

\title{PHYSICAL CONDITIONS IN A YOUNG, UNREDDENED,\\ LOW METALLICITY GALAXY AT HIGH REDSHIFT\altaffilmark{1}}
\author{{\sc Dawn K. Erb}\altaffilmark{2,3}, {\sc Max Pettini}\altaffilmark{4,5}, {\sc Alice E. Shapley}\altaffilmark{6},\\ {\sc  Charles C. Steidel}\altaffilmark{7}, {\sc David R. Law}\altaffilmark{6,8}, and {\sc Naveen A. Reddy}\altaffilmark{8,9}}
%\email{dawn@physics.ucsb.edu}

\slugcomment{Accepted for publication in ApJ}

\shorttitle{A YOUNG, UNREDDENED, LOW METALLICITY GALAXY AT HIGH REDSHIFT}
\shortauthors{ERB ET AL.}

\altaffiltext{1}{Based on data obtained at the 
W.M. Keck Observatory, which is operated as a scientific partnership
among the California Institute of Technology, the University of
California, and NASA, and was made possible by the generous financial
support of the W.M. Keck Foundation.}  
\altaffiltext{2}{Department of Physics, University of California Santa Barbara, Santa Barbara, CA 93106-9530}
\altaffiltext{3}{Spitzer Fellow}
\altaffiltext{4}{Institute of Astronomy, Madingley Road, Cambridge CB3 0HA, UK}
\altaffiltext{5}{International Centre for Radio Astronomy Research, University of Western Australia, 35 Stirling Highway, Crawley, WA 6009, Australia}
\altaffiltext{6}{Department of Astronomy, University of California Los Angeles, Los Angeles, CA 90095-1547}
\altaffiltext{7}{Department of Astronomy, California Institute of Technology, MS 249--17, Pasadena, CA 91125}
\altaffiltext{8}{Hubble Fellow}
\altaffiltext{9}{National Optical Astronomy Observatory, 950 N Cherry Ave, Tucson, AZ 85719}

\begin{abstract}
Increasingly large samples of galaxies are now being discovered at redshifts $z\sim5$--6 and higher.  Many of these objects are inferred to be young, low in mass, and relatively unreddened, but detailed analysis of their high quality spectra will not be possible until the advent of future facilities.  In this paper we shed light on the physical conditions in a plausibly similar low mass galaxy by presenting the analysis of the rest-frame optical and UV spectra of Q2343-BX418, an $L^*$ galaxy at $z=2.3$ with a very low mass-to-light ratio and unusual properties: BX418 is young ($<100$ Myr), low mass ($M_{\star}\sim10^9$ \msun), low in metallicity ($Z\sim 1/6$ \zsun), and unreddened ($E(B-V)\simeq0.02$, UV continuum slope $\beta=-2.1$).  We infer a metallicity $12+\log({\rm O/H})=7.9\pm0.2$ from the rest-frame optical emission lines. We also determine the metallicity via the direct, electron temperature method, using the ratio \Othreeuv\ $\lambda\lambda$1661, 1666/\Othree\ $\lambda$5007 to determine the electron temperature and finding  $12+\log({\rm O/H})=7.8\pm0.1$.  These measurements place BX418 among the most metal-poor galaxies observed in emission at high redshift.  The rest-frame UV spectrum, which represents $\sim12$ hours of integration with the Keck telescope, contains strong emission from \lya\ (with rest-frame equivalent width 54 \AA), \hetwo\ $\lambda1640$ (both stellar and nebular), \Cthree\ $\lambda\lambda1907$, 1909 and \Othreeuv\ $\lambda\lambda1661$, 1666.  The \ion{C}{4}/\Cthree\ ratio indicates that the source of ionization is unlikely to be an AGN.  Analysis of the \hetwo, \Othreeuv\ and \Cthree\ line strengths indicates a very high ionization parameter $\log U\sim-1$, while \lya\ and the interstellar absorption lines indicate that outflowing gas is highly ionized over a wide range of velocities. It remains to be determined how many of BX418's unique spectral features are due to its global properties, such as low metallicity and dust extinction, and how many are indicative of a short-lived phase in the early evolution of an otherwise normal star-forming galaxy.
\end{abstract}


\section{Introduction}
Many of the key questions in the study of galaxy formation and evolution are also the most difficult to address, because they center on the properties of the very faint galaxies in the early universe.  Significant numbers of such galaxies are now being confirmed at $z\sim5$--6 (e.g.\ \citealt{yds+05,mcd+06,aoi+08,vgd+09,seb+09,sec+10}), and the recent installation of the Wide Field Camera 3 (WFC3) on the {\it Hubble Space Telescope} has resulted in a flurry of candidate galaxies at $7<z<10$ \citep{bio+10b,obi+10,ywh+10,bwe+10,mdc+10}.  The general properties of these galaxies can be inferred from their broadband spectral energy distributions, but as a result of their faint continuum magnitudes, spectral information---when available at all---is limited to strong \lya\ emission or low signal-to-noise composite spectra.  Because of this flux limit, significant progress in the understanding of the detailed properties of these galaxies is unlikely to occur before the era of JWST and the extremely large ground-based telescopes.

An alternative and immediately accessible approach is to identify galaxies at somewhat lower redshifts that are likely to be similar to these high redshift objects.  Galaxies at \ztwo\ are still bright enough for detailed study with current technology, and have the further advantage that the full set of rest-frame optical emission lines are observable in the near-IR from the ground.  If we wish, then, to identify \ztwo\ galaxies that are most likely to illuminate the properties of galaxies in the very early universe, what features should we look for? Intuitively, we expect that such galaxies will be young, low in mass and metallicity, and relatively unreddened.  Recent results bear this out.  At increasing redshifts, the UV-continuum slope $\beta$ ($f_{\lambda} \propto \lambda^{\beta}$) of galaxies becomes bluer;  the most luminous galaxies have $\beta \sim -1.5$ at \ztwo, and $\beta \sim -2.4$ at $z\sim6$ \citep{bif+09}.  At a given redshift, fainter galaxies also show bluer slopes.  At $z\sim7$--8, galaxies are found to have $\beta \sim -2$ to $-3$ \citep{bio+10a,fpg+10}.  Given the limited data available on these galaxies, the origin of such blue slopes is difficult to determine.  \citet{bio+10a} suggest that the bluest slopes of the faintest galaxies cannot be produced by normal stellar populations at nonzero metallicities, while \citet{fpg+10} argue that recourse to Population III stars, a top-heavy initial mass function, or other unusual mechanisms is not necessary to account for the blue color.  Instead, they find that the galaxies' observed SEDs are best explained by a young ($<100$ Myr), low metallicity ($Z < 0.1$ \zsun\ at 1 $\sigma$ confidence) stellar population with little to no dust extinction and stellar masses of $10^8$--$10^9$ \msun.  We are unlikely to find galaxies dominated by Population III stars at \ztwo, but the set of properties found by \citet{fpg+10} may well be approximated by some unusual lower redshift objects.

In this paper we present the analysis of the rest-frame ultraviolet and optical spectra of one such highly unusual galaxy.  Q2343-BX418 is an $L^*$ galaxy at $z=2.3$ selected by the standard rest-frame UV color criteria \citep{ssp+04}, but it is nevertheless among the youngest and lowest stellar mass continuum-selected \ztwo\ galaxies identified to date, with rest-frame $B$-band mass-to-light ratio $M/L=0.03$.  More uniquely, BX418 also shows extremely low dust obscuration and low metallicity.  We use the UV and optical line ratios in combination with photoionization modeling to show that the properties of this galaxy are significantly different from those of more typical \ztwo\ galaxies, and from local starburst galaxies as well.  Our study relies on an extremely deep UV spectrum: given spectra of sufficient quality, observations in the rest-frame UV are sensitive to dust content, age, galaxy-scale outflows, metallicity, ionization parameter, and the initial mass function.  In order to obtain the relatively high signal-to-noise necessary for this analysis, detailed studies of the UV spectra of high redshift galaxies have so far been restricted primarily to the composite spectra of many galaxies \citep{ssp+03} or to galaxies whose luminosities have fortuitously been boosted by gravitational lensing \citep{psa+00,prs+02,cvl08,qpss09,qsp+10,dds+09}.  The result is that these studies have primarily served to illuminate the properties of typical galaxies at \ztwo--3.  Here we take advantage of $\sim15$ hours of integration with the Keck telescope to study an object that is considerably younger, bluer, and lower in mass than these more typical galaxies.

The paper is organized as follows.  We describe the observations and data reduction in Section \ref{sec:obs}, the construction of a composite spectrum for comparison in Section \ref{sec:comp}, and the global properties of BX418 in Section \ref{sec:global}, including the stellar population properties (Section \ref{sec:sed}), the star formation rate (Section \ref{sec:sfr}), and the kinematics and gas mass (Section \ref{sec:kin_gas}).  In Section \ref{sec:cloudy} we discuss the suite of photoionization models used to match the observed line ratios.  We focus on the rest-frame optical spectrum in Section \ref{sec:optical}, including constraints on metallicity and ionization parameter.  The rest-frame UV spectrum is the subject of Section \ref{sec:uv}:  Section \ref{sec:outflows} deals with the interstellar absorption and \lya\ emission, and Section \ref{sec:he2} describes the \hetwo\ emission. In Section \ref{sec:nebularuv} we discuss nebular emission from carbon and oxygen and resulting constraints on the C/O abundance ratio.  The stellar wind lines are described in Section \ref{sec:windlines}, and the weak fine structure emission lines in Section \ref{sec:fs}.  We discuss all of these results in Section \ref{sec:disc}, and summarize our conclusions in Section \ref{sec:summary}.  We use a cosmology with $H_0=70\;{\rm km}\;{\rm s}^{-1}\;{\rm Mpc}^{-1}$, $\Omega_m=0.3$, and $\Omega_{\Lambda}=0.7$, and assume a \citet{c03} initial mass function and a solar oxygen abundance of $12+\log({\rm O/H})=8.69$ \citep{agss09}.

\section{Observations and Data Reduction}  
\label{sec:obs}
\subsection{Optical Spectroscopy}
The $z=2.3048$ galaxy Q2343-BX418 ($\alpha=23$:46:18.571, $\delta=+12$:47:47.379; J2000) was observed in the course of the \ztwo\ galaxy survey described by \citet{ssp+04}.  In September and October 2003, we created four slitmasks in the Q2343 field for deep observation with the Low Resolution Imaging Spectrometer (LRIS; \citealt{occ+95}) on the Keck I telescope.  BX418 appeared on two of these masks, and was observed for 5.5 hrs on 24--26 September 2003 and for 6.1 hrs on 26--28 October 2003, for a total of 11.6 hrs of integration.  The observations were conducted as described by \citet{ssp+04}.  We used the 400 groove mm$^{-1}$ grism blazed at 3400 \AA\ for the LRIS blue channel, and divided the beam with a dichroic at 6800 \AA.  In the red channel we used the 400 groove mm$^{-1}$ grating blazed at 8500 \AA, though in the case of BX418 all of the scientifically useful information falls in the blue channel.   Each mask was observed at the parallactic angle, for approximately two hours each night.  The deep masks contained a total of 77 objects, receiving integrations ranging from 5.5 to 22.8 hrs.  Inspection of the spectra showed many to be of high quality but similar to the average UV spectra of galaxies at \ztwo--3 (e.g.\ \citealt{ssp+03}); BX418, however, contains several distinctive spectral features.

Data reduction was done according to standard procedures described by \citet{sas+03}.  The spectra from each of the two masks containing BX418 were reduced and extracted independently, and then these 1D spectra were combined to produce the final spectrum.  Flux calibration was checked by integrating the spectrum over the $G$ bandpass using the IRAF task ``sbands" and comparing with the broadband magnitude.  The two measurements differed by only 0.03 mag, so no correction to the flux calibration was applied.  The data reduction procedures also included the production of a 2D variance spectrum for each slit on each mask observed.  These were scaled by the number of exposures of each mask, extracted to 1D using the same apertures used for the science spectra, and combined using standard error propagation to construct a 1$\sigma$ error spectrum which was used to determine uncertainties.  The spectrum of BX418 covers wavelengths $\sim3400$--6400 \AA, or $\sim1000$--1930 \AA\ in the rest frame.  The slit width was 1.2\arcsec, but because the galaxy is compact and the seeing was good, the spectral resolution is primarily determined by the angular size of the galaxy, which was $\sim0.85$\arcsec.  The resulting resolution is $\sim400$ \kms, and ranges from $\sim550$ \kms\ at the blue end of the spectrum to $\sim300$ \kms\ at the red end.  

\subsection{Near-IR Spectroscopy}
BX418 was also observed with NIRSPEC \citep{mbb+98} on the Keck II telescope, in order to obtain infrared spectra of the rest-frame optical emission lines.  The $K$-band observation of \Ha\ was included in the survey of \citet{ess+06mass}; these data were obtained on 13 June 2004, and consisted of five 15 minute integrations using the NIRSPEC-6 filter.  We also observed BX418 in the $H$ (for \Othree\ $\lambda\lambda$5007, 4959 and \Hb) and $J$ (for \Otwo\ $\lambda$3727) bands in September 2004.  The $H$-band observations were conducted on 6 September 2004 using the NIRSPEC-5 filter, and consisted of four 15 minute integrations.  Similarly, we obtained four 15 minute $J$-band integrations using the NIRSPEC-3 filter on 7 September 2004.  As measured from the sky lines, the spectral resolution is 15 \AA\  ($\sim 200$ \kms) in the $K$ band, 10 \AA\ ($\sim180$ \kms) in $H$, and 10 \AA\ ($\sim240$ \kms) in $J$.  The data were reduced using the standard procedures described by \citet{ess+03}.  

When comparing observations taken at different times and with different spectral configurations, relative (and absolute) flux calibration is always a concern.  As discussed by \citet{ess+06mass}, NIRSPEC flux calibration is subject to significant uncertainties, primarily because of slit losses.  In the case of BX418, however, we have reason to believe that these uncertainties are minimal.  As described below, \Ha\ emission from BX418 has also been observed with OSIRIS, the near-IR integral field spectrograph with laser guide star adaptive optics on the Keck II telescope.  The two flux measurements agree within 3\%, well below the uncertainties on both observations.  The OSIRIS observations also show that the nebular emission arises from a very small region with radius 0.8 kpc, or 0.1\arcsec\ at $z=2.3$.  Thus the spatial extent of the galaxy in the NIRSPEC observations is determined almost entirely by the seeing, which was $\sim0.5$\arcsec\ for all observations.  The slit width was 0.76\arcsec, and we therefore expect slit losses to be minimal.   Conditions were photometric for all observations.  As noted above, the flux calibration of the rest-frame UV spectrum was checked by comparison with the broadband $G$ magnitude and found to agree; we are therefore confident in comparisons between the LRIS and NIRSPEC spectra as well.

\subsection{Optical and IR Imaging}
We also make use of deep optical, near-IR and mid-IR images.  The Q2343 field was imaged in the $U_n$, $G$ and $R_s$ bands with the Large Format Camera (LFC) on the Palomar 200-inch telescope in August 2001, as described by \citet{ssp+04}.  Near-IR $J$ and $K$ band observations were conducted with the Wide-Field Infrared Camera (WIRC, \citealt{wirc}), also on the Palomar telescope.  These observations are described by \citet{ess+06mass}.  Finally, the Q2343 field was observed at 4.5 and 8.0 \micron\ with the IRAC \citep{irac} instrument on the {\it Spitzer Space Telescope}, and at 24 \micron\ with MIPS.  We use only the 4.5 \micron\ IRAC data, as BX418 was not detected at 8 \micron.  The 4.5 \micron\ image had an integration time of 18000 sec and a 3$\sigma$ depth of $\sim$0.16 $\mu$Jy.  The IRAC image was reduced as described by \citet{sse+05}.  BX418 was also undetected in the MIPS 24 \micron\ image, which had an effective integration time of 9.6 hrs and a 3$\sigma$ depth of $\sim$15 $\mu$Jy.  The MIPS data were reduced as described by  \citet{rsf+06}.  

A portion of the Q2343 field including BX418 was observed using the F160W filter ($\lambda \sim 1.54 \micron$, tracing rest-frame 4700 \AA\ at the redshift of BX418) of the WFC3/IR camera on board the {\it Hubble Space Telescope} ({\it HST}) on UT 15 June 2010, as part of a survey of star-forming galaxies in deep extragalactic fields (GO-11694).  The integration time was 9$\times$900 sec, for a total of 8100 sec.  The details of the {\it HST} observations and data reduction are described in full by Law et al. (in preparation); in brief, the {\it HST} images were drizzled to a pixel scale of 0.08 arcsec pixel$^{-1}$ with a point spread function of FWHM $= 0.19$\arcsec\ (corresponding to 1.6 kpc at the redshift of BX418), and reach a $5\sigma$ limiting magnitude of 27.9 AB in a 0.4\arcsec\ diameter aperture.  

\section{Comparison Sample} 
\label{sec:comp}
In order to compare the rest-frame UV spectrum of BX418 with that of more typical galaxies at \ztwo, we have constructed a composite spectrum of all of the galaxies in the \ztwo\ sample for which we have determined stellar population parameters, and which have spectra of high enough signal-to-noise for reliable redshift determination.  The composite is the average of 1186 spectra of 966 galaxies (because some galaxies were observed more than once).  It was constructed in the same fashion as the \zthree\ Lyman break galaxy composite spectrum described by \citet{ssp+03}, and is very similar to that spectrum in appearance.  The galaxies included in the \ztwo\ composite have the following medians and semi-interquartile ranges:  redshift $z_{\rm med} =2.2\pm0.3$, stellar mass $M_{\star, \rm med}  = 9.7\pm8.9\times10^{9}$ \msun, and age $571\pm538$ Myr, assuming constant star formation.  The median reddening is $E(B-V)_{\rm med}  = 0.17\pm0.06$.  The properties of BX418 are given in Section \ref{sec:sed} for comparison.

\begin{deluxetable*}{c c c c c c c}
\tablewidth{0pt}
\tabletypesize{\footnotesize}
\tablecaption{Photometry\label{tab:phot}}
%\rotate
\tablehead{
\colhead{$U_n$} & 
\colhead{$G$} & 
\colhead{${\cal R}$} & 
\colhead{$J_{\rm AB}$} & 
\colhead{$H_{\rm AB}$\tablenotemark{a}} & 
\colhead{$K_{\rm AB}$\tablenotemark{b}} &
\colhead{4.5 $\mu$m} 
}
\startdata
24.3$\pm$0.1 & 23.9$\pm$0.1 & 24.0$\pm$0.1 & 24.5$\pm$0.4 & 23.04$\pm$0.04 & 23.7$\pm$0.4 & 23.8$\pm$0.5
\enddata
\tablecomments{All magnitudes are AB.}

\tablenotetext{a}{\textit{HST} F160W magnitude, with no correction for
  [O~{\sc iii}] $\lambda\lambda$4959, 5007 and \Hb\ emission.  The
  corrected magnitude is 23.49$\pm$0.08.}
\tablenotetext{b}{Uncorrected for \Ha\ emission, which accounts for
  $0.4\pm0.1$ mag.}

\end{deluxetable*}

\begin{figure}[htbp]
\plotone{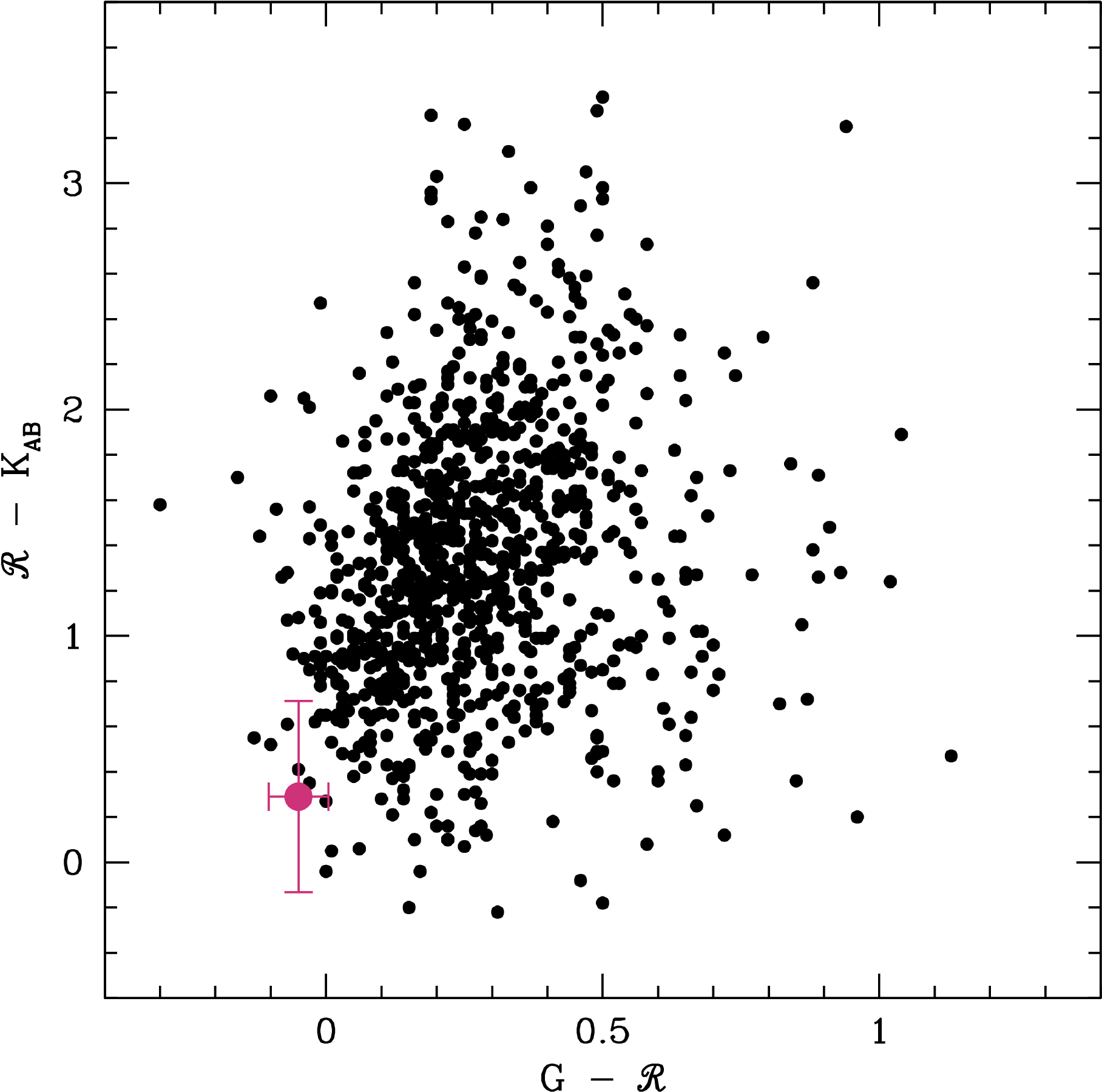}
\caption{The $G-{\cal R}$ vs.\ ${\cal R} -K_{AB}$ color-color diagram, in which BX418 is compared with 1116 spectroscopically confirmed \ztwo\ galaxies selected by the BX, BM and MD color criteria \citep{sas+03,ssp+04}.}
\label{fig:gmr_rmk}
\end{figure}

\section{Global Properties}
\label{sec:global}
Broadband magnitudes for BX418 are given in Table \ref{tab:phot}, from $U_n$ to 4.5 \micron.  Note that the spectral energy distribution is nearly flat, with AB magnitudes $m_{\rm AB} \simeq 24\pm0.5$ in all observed bands after correction for line emission.  These colors are highly unusual, as shown in Figure \ref{fig:gmr_rmk}.  Here we present the $G-{\cal R}$ vs.\ ${\cal R}-K$ color-color diagram for 1116 spectroscopically confirmed \ztwo\ galaxies selected by the BX, BM and MD color criteria \citep{sas+03,ssp+04}.  BX418 is shown by the large circle in the lower left corner of the diagram.  There are few objects in this corner, and BX418 lies at the extreme such that galaxies bluer in $G-{\cal R}$ are redder in ${\cal R}-K$, while galaxies bluer in ${\cal R}-K$ are redder in $G-{\cal R}$.  In physical terms, the blue $G-{\cal R}$ color is indicative of low extinction, but not necessarily a young stellar population, while the blue ${\cal R}-K$ color implies a low stellar mass and a young age as well, given the high star formation rates of these galaxies.  Young, low mass, and unreddened galaxies can therefore be expected to fall in the lower left corner of the color-color diagram; they are few in number because young galaxies usually have redder UV slopes \citep{ssa+01,sse+05,rep+10}.  From the $G-{\cal R}$ color, we find a UV continuum slope $\beta=-2.1$, placing BX418 at the extreme blue end of the $\beta$ distribution of luminous \ztwo\ galaxies \citep{rsf+06,rep+10}.  Note that at $z=2.3$ \lya\ falls between the $U_n$ and $G$ bands, so the observed colors are not affected by the strong \lya\ emission.  However, the $H$ and $K$ magnitudes are affected by the strong rest-frame optical emission lines; combined \Othree $\lambda\lambda$4959,5007 and \Hb\ emission contributes 0.45 mag to the observed $H$-band, while \Ha\ emission contributes 0.4 mag to the $K$-band magnitude.

BX418 is undetected in deep 8 and 24 \micron\ imaging, with 3$\sigma$ upper limits of $m_{AB}>22.5$ at 8 \micron\ and $f_{24\;\micron}<34$ $\mu$Jy.  Most AGN in the UV-selected \ztwo\ galaxy sample have a rising, power law SED at 8 and 24 \micron, with $f_{24\;\micron}>100$ $\mu$Jy \citep{rse+06}; BX418's nondetection in these bands therefore argues against the presence of an obscured AGN.   

\begin{figure}[htbp]
\plotone{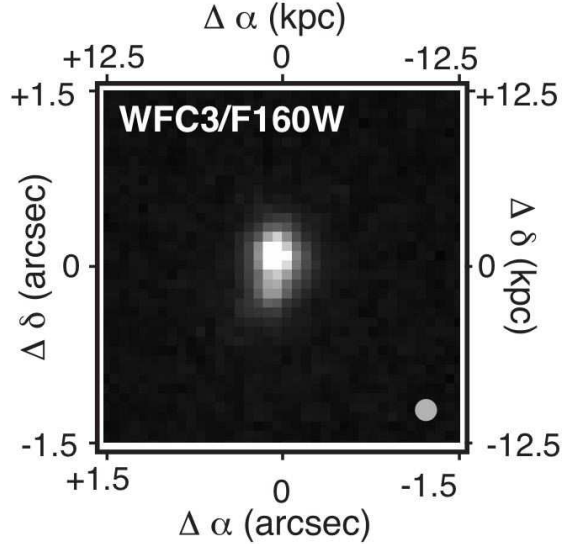}
\caption{Rest-frame optical image of BX418, taken with the Wide Field Camera 3 on the \textit{Hubble Space Telescope}.  The area shown is 3\arcsec $\times$ 3\arcsec, and the grey circle at lower right shows the FWHM of the 0.19\arcsec\ PSF.  BX418 is compact, with a half light radius of 1.5 kpc.}
\label{fig:wfc3}
\end{figure}

In Figure \ref{fig:wfc3} we present a high resolution rest frame optical image of BX418, taken with the Wide Field Camera 3 on the \textit{Hubble Space Telescope}, using the F160W filter.  This demonstrates that the galaxy is very compact, with a half light radius of 1.5 kpc; for comparison, the average half light radius of galaxies in the UV-selected sample is $\sim3$ kpc.  BX418 shows a slight elongation along the N-S axis (also seen in the OSIRIS observations of \citealt{lse+09} described in Section \ref{sec:kin_gas} below), but otherwise has no distinctive morphological features.

\subsection{Stellar Population Modeling}
\label{sec:sed}
Stellar population properties are determined by modeling of the broadband SED in the usual fashion, employing the \citet{bc03} population synthesis models and following methods previously described by \citet{sse+05} and \citet{ess+06mass}.  We quantify the uncertainties in the modeling using a series of 10,000 Monte Carlo simulations, as described by \citet{sse+05}.  These simulations perturb the galaxy's photometry in accordance with the photometric errors, and compute the best fitting model for the perturbed photometry, including a variety of star formation histories.  We find that 91\% of the 10,000 trials favor a young, low mass galaxy with age $<150$ Myr; the remaining 9\% allow a more massive and significantly older stellar population.  Based on the strong spectral evidence that the galaxy is young and metal poor, we adopt the younger and less massive solution, taking as the stellar mass, age, and $E(B-V)$ the mean values of these quantities in the 91\% of trials which give this general solution.  Uncertainties are then given by the standard deviation.   With the exception of a somewhat younger age (38 Myr vs.\ 100 Myr), these values are not significantly different from the best fit constant star formation model given by the observed photometry.  We determine the stellar population parameters in this way in order to obtain uncertainties which are not skewed by the tail of older solutions disallowed by the spectrum.   We have also checked these results using a newer set of population synthesis models which offer an improved description of thermally pulsating AGB stars (Charlot \& Bruzual 2007;  S. Charlot, private communication), with the SMC rather than \citet{cab+00} extinction law, and with lower metallicity models of 0.4 $Z_{\odot}$ rather than solar.  None of these variations significantly change the results.  We find $M_{\star}=9\times10^8$ \msun, an age of 38 Myr, and $E(B-V)=0.02$.  These values and their associated uncertainties are also given in Table \ref{tab:sed}.  

In addition we have assessed the probability of a hidden older stellar population by fitting two component stellar population models, as described by \citet{ess+06mass}.  A general two component model, in which the age and mass of both components are allowed to vary arbitrarily, provides a good fit to the data but increases the total mass by at most a factor of two.  A maximally massive model, in which the optical and IR light is fit by a maximally old stellar population and the UV residuals are then fit with an episode of current star formation, increases the total stellar mass by a factor of ten.  However, this model is a significantly poorer fit to the observed photometry, and overpredicts the star formation rate by a factor of $>2$ relative to that predicted by independent observations of \Ha\ with both NIRSPEC (long slit spectroscopy) and OSIRIS (integral field spectroscopy).  We therefore consider a significant underlying older stellar population to be unlikely.

\begin{deluxetable*}{c c c c c c}
\tablewidth{0pt}
\tabletypesize{\footnotesize}
\tablecaption{Stellar Population, Gas and Kinematic Properties\label{tab:sed}}
%\rotate
\tablehead{
\colhead{$M_{\star}$\tablenotemark{a}} & 
\colhead{Age\tablenotemark{a}} & 
\colhead{$E(B-V)$\tablenotemark{a}} & 
\colhead{SFR$_{\Ha}$\tablenotemark{b}} & 
\colhead{$M_{\rm gas}$\tablenotemark{c}} &
\colhead{$M_{\rm dyn}$\tablenotemark{d}} \\
\colhead{($10^9 $M$_{\odot}$)} &
\colhead{(Myr)} &
\colhead{} &
\colhead{(M$_{\odot}$ yr$^{-1}$)} &
\colhead{($10^9 $M$_{\odot}$)} &
\colhead{($10^9 $M$_{\odot}$)} 
}
\startdata
% 1.1$\pm$1.0 & 101$\pm$137 & 0.03$\pm$0.03 & 15$\pm$1 & 2.5 & 3 \\  these are CSF best fit
$0.9\pm0.4$ & $38\pm23$ & $0.02\pm0.03$ & $15\pm2$ & 2.5 & 3    % from MC sims
\enddata

\tablenotetext{a}{Stellar mass, age and $E(B-V)$ from SED modeling as described in the text.}
\tablenotetext{b}{Star formation rate from \Ha.  No extinction correction has been applied, but uncertainties include both random flux error and the effects of extinction up to $E(B-V)<0.05$.}
\tablenotetext{c}{Gas mass inferred from the Kennicutt-Schmidt law.}
\tablenotetext{d}{\citet{lse+09}}

\end{deluxetable*}

\subsection{Star Formation Rates and Timescales}
\label{sec:sfr}
We use the NIRSPEC $\Ha$ luminosity to determine a star formation rate of $15\pm2$ \msunyr, where no extinction correction has been applied but the uncertainties include both statistical flux errors and a range of extinction $0<E(B-V)<0.05$.  This is a factor of $\sim2$ lower than the average \Ha-determined star formation rate of \ztwo\ galaxies \citep{ess+06stars}.  We also calculate the SFR from the 1500 \AA\ luminosity, finding SFR$_{\rm UV}= 9$ \msunyr, again with no extinction correction applied.  For both calculations we use the prescriptions of \citet{k98}.   It is well known that \Ha\ emission and the UV continuum luminosity probe different timescales; \Ha\ emission arises in \ion{H}{2} regions around the most massive stars, and therefore traces the nearly instantaneous star formation rate, while the UV luminosity continues to rise for the first $\sim100$ Myr.  In a very young galaxy, we would therefore expect \Ha\ to give a higher star formation rate than the UV continuum.  This is indeed the case in BX418, where SFR$_{\rm UV}$/SFR$_{\Ha}=0.65$.  Using a  Starburst99 \citep{sb99} constant star formation model matched as closely as possible to the properties of BX418, we predict the UV continuum luminosity as a function of time, and find that it reaches $\sim65$\% of its final value at an age of $\sim15$ Myr.  This is lower than our estimated age, but within the uncertainties.  It is also comparable to the estimated dynamical time of $\sim25$ Myr (see below).  Applying our best fit $E(B-V)=0.02$ to both the \Ha\ and UV fluxes, we find SFR$_{\rm UV}$/SFR$_{\Ha}\simeq 0.8$, which occurs at an age of $\sim40$ Myr; this is well-matched to our best fit age.  However, we caution that while all of these numbers are consistent with the young age we derive for BX418, uncertainties in the extinction mean that the ratio of SFRs alone is insufficient to reliably constrain the age to $<100$ Myr.  Applying a correction of $E(B-V)=0.05$, still within the uncertainties, to both the \Ha\ and UV fluxes results in SFRs that are equal within the photometric and spectroscopic uncertainties.

We can also place a lower limit on the SFR from the \lya\ emission, from which we find SFR$_{\lya}=6$ \msunyr.  A comparison of the SFRs derived from \Ha\ and \lya\ suggests a \lya\ escape fraction of $\sim40$\%, as also discussed in Section \ref{sec:lya}.

\subsection{Kinematics, Gas Mass and Gas Fraction}
\label{sec:kin_gas}
The \Ha\ emission from BX418 has also been observed with the integral field spectrograph OSIRIS on the Keck II telescope using laser guide star adaptive optics, as part of the survey described by \citet{lse+09}.  The measured \Ha\ flux was in excellent agreement with that derived from our NIRSPEC observations.  The OSIRIS observations indicate that the \Ha\ emission is very compact, covering $2.2\pm0.6$ kpc$^{2}$ with effective radius $r=0.8\pm0.1$ kpc.   BX418 also shows a low mean velocity dispersion $\sigma_{\rm mean}=61\pm17$ \kms\ (where  $\sigma_{\rm mean}$ is the flux-weighted mean of the velocity dispersions of individual spatial pixels) and a slight velocity gradient given by $v_{\rm shear}=23\pm12$ \kms\ over a distance of $\sim1$ kpc; $v_{\rm shear}$ is defined as half of the maximum difference in velocity between any two postions along a pseudo-slit oriented to maximize the velocity gradient.  Thus like most low to average mass \ztwo\ galaxies studied to date, BX418 appears to be kinematically dominated by random motions rather than rotation \citep{lse+09,fgb+09}.  From the \Ha\ kinematics and size, \citet{lse+09} derive a dynamical mass $M_{\rm dyn}=3\times10^9$ \msun.  We also use the size and velocity dispersion to estimate the dynamical time, $t_{\rm dyn} \sim 2r/\sigma \simeq$ 25 Myr.   This is again comparable to the age derived from SED fitting, and provides an approximate lower limit to the age of the galaxy.

The flux and spatial extent of the \Ha\ emission observed by OSIRIS can be used to estimate the mass of cold gas associated with star formation.  We use the inversion of the Kennicutt-Schmidt star formation law \citep{k98schmidt} given by \citet{ess+06mass}, and find $M_{\rm gas}=2.5\times10^9$ \msun.  We combine this with the stellar mass $M_{\star}=9\times10^8$ \msun\ to find baryonic mass $M_{\rm bar}=3.4\times10^9$ \msun, in good agreement with the dynamical mass, and gas fraction $\mu=M_{\rm gas}/(M_{\star}+M_{\rm gas})=0.7$.  This high gas fraction is consistent with the galaxy's young age and low metallicity, and is comparable to the high gas fractions inferred for other young galaxies at \ztwo\ \citep{ess+06mass}.

\section{Photoionization Modeling}
\label{sec:cloudy}
In the following sections we compare the rest-frame optical and UV emission line ratios with a suite of photoionization models in order to obtain constraints on metallicity, the ionization parameter and ionization correction factors, and the relative abundances of C and O.  The models were constructed with version 08.00 of Cloudy, last described by \citet{cloudy}.  We employ an input ionizing spectrum from Starburst99 \citep{sb99}, assuming continuous star formation at 15 \msunyr, an age of 50 Myr, and a metallicity of $Z=0.004$, or $\sim0.3$ \zsun, the closest available metallicity to that measured for BX418.  We use the Pauldrach/Hillier stellar atmospheres, which have ionizing fluxes discussed by \citet{snc02}.  Uncertainties associated with these ionizing fluxes are discussed in more detail in Section \ref{sec:he2}.  We consider metallicities of 0.05, 0.1, 0.2, 0.3 and 0.5 Z$_{\odot}$.  We use Cloudy's default \ion{H}{2} region abundance set, and in addition reduce the abundances of C and N by factors of 0.85, 0.65, 0.45, 0.25, 0.15, and 0.05 for each of the metallicities adopted.  The default C/O and N/O ratios are 0.75 and 0.17 respectively; the resulting reduced C/O ratios are 0.65, 0.49, 0.34, 0.19, 0.11 and 0.04, and the reduced N/O ratios are 0.15, 0.11, 0.079, 0.044, 0.026 and 0.0088.  

The ionization parameter represents the ratio of ionizing photon to particle density, and is defined in dimensionless form as 
\begin{equation}
U\equiv\frac{Q(H)}{4\pi r_o^2\, n(H) c},
\end{equation}
where $Q(H)$ is the number of ionizing photons emitted per unit time, $r_o$ is the assumed distance from the ionization source to the cloud, and $n(H)$ is the total hydrogen density.   Because neither the particle density nor the ionization parameter can be measured independently for BX418, we instead construct a set of models with reasonable values of $n(H)$ and $r_o$ that result in ionization parameters ranging from $\log U=-3.3$ to $\log U=-0.2$.  We examine the variation of the line ratios of interest as a function of ionization parameter, metallicity, and C/O and N/O abundance, and choose the models best matching all of the observations.  As described in detail in the following sections, the properties of BX418 can only be reproduced by models with low metallicity, a significantly reduced C/O ratio, and an extremely high ionization parameter.

\begin{figure*}[htbp]
\plotone{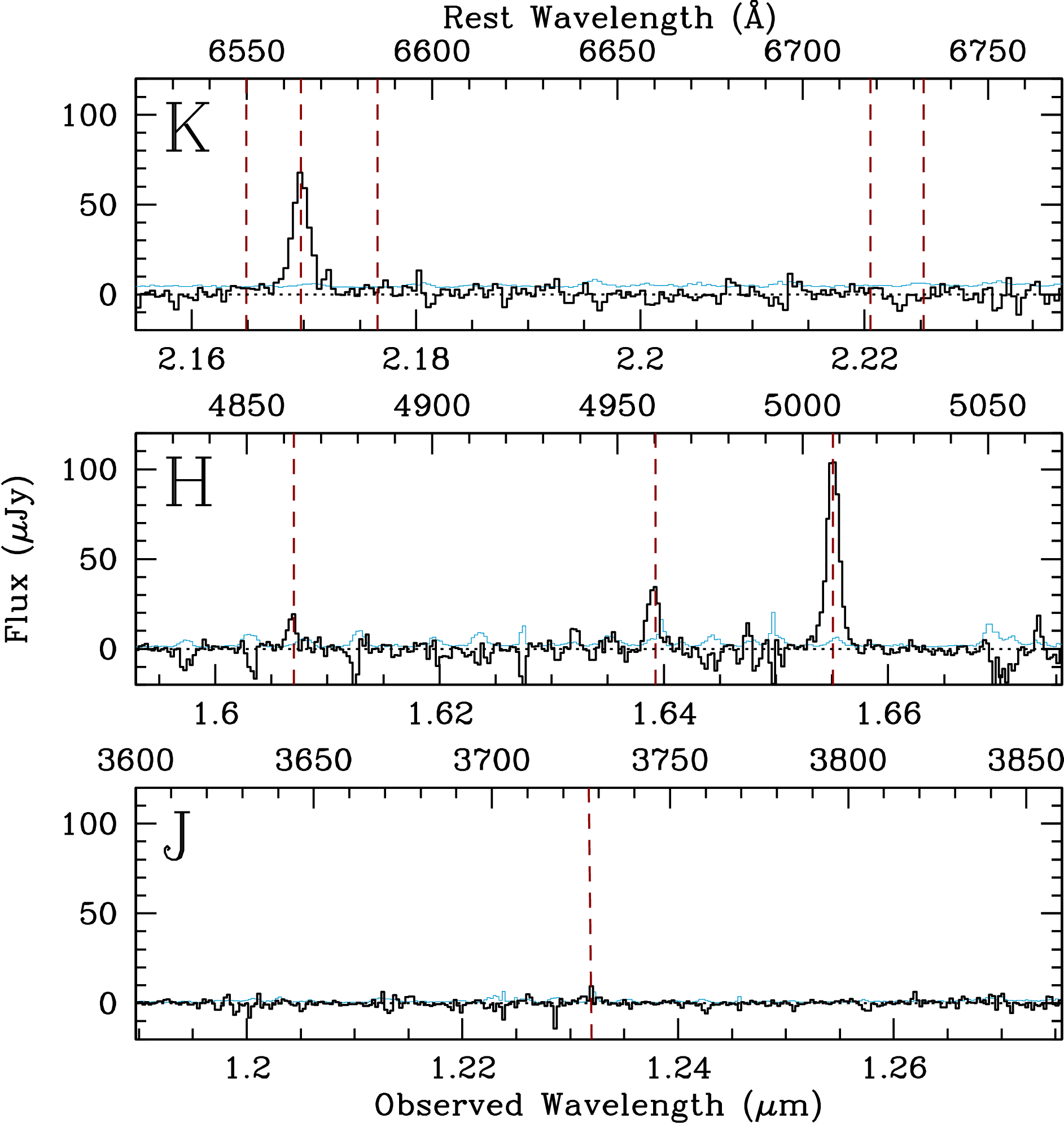}
\caption{Rest-frame optical spectra in the K, H and J bands, taken as three separate observations with NIRSPEC.  The thin blue lines show the 1$\sigma$ error spectra, and the vertical dashed lines mark the locations of the strong emission lines.  From left to right in the K band: [\ion{N}{2}] $\lambda$6548 (undetected), \Ha, [\ion{N}{2}] $\lambda$6583 (undetected), and [\ion{S}{2}] $\lambda\lambda$6716, 6731 (undetected); in the H band, \Hb, [\ion{O}{3}] $\lambda$4959, and [\ion{O}{3}] $\lambda$5007; and in the J band,  [\ion{O}{2}] $\lambda$3727 (undetected).}
\label{fig:irspec}
\end{figure*}

\begin{figure}[htbp]
\plotone{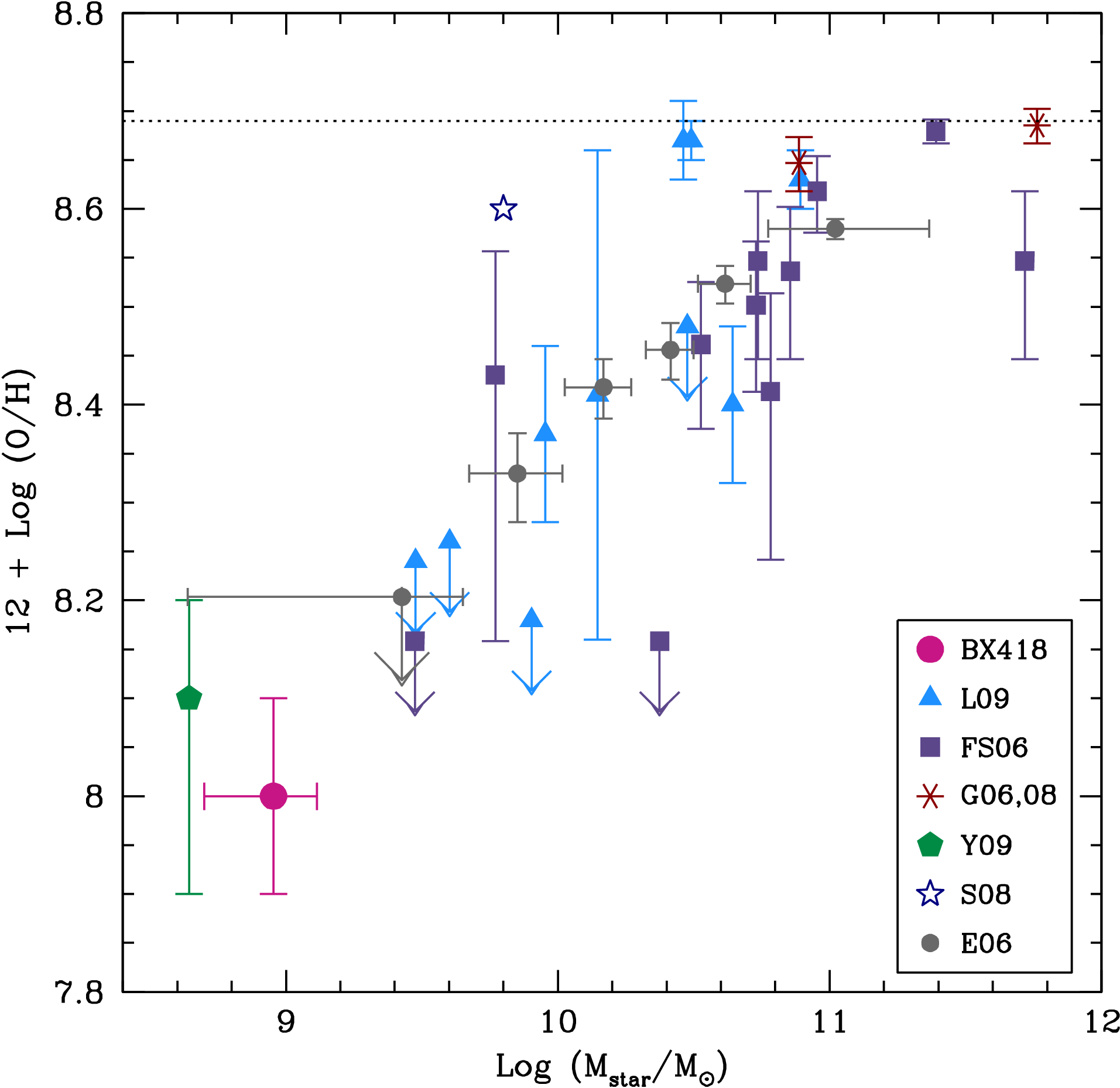}
\caption{BX418 and the mass-metallicity relation for \ztwo\ galaxies.  All measurements are either from the \Ntwo/\Ha\ calibration of \citet{pp04} or have been converted to this method using the prescriptions of \citet{ke08}.  BX418 is indicated by the large magenta circle at lower left.  Other data points are from \citealt{lse+09} (L09), \citealt{fgl+06} (FS06), \citealt{gte+06,gbb+08} (G06, 08), \citealt{yk09} (Y09), \citealt{sse+08} (S08), and the composite spectra of \citealt{esp+06} (E06).  The horizontal dotted line shows the solar oxygen abundance.}
\label{fig:mz}
\end{figure}

\section{The Rest-Frame Optical Spectrum}  
\label{sec:optical}
The $J$, $H$ and $K$ band observations described in Section \ref{sec:obs} cover the approximate rest frame wavelength ranges 3450--4200, 4550--5400, and 5900--7000 \AA, respectively, and therefore include the lines   \Otwo\ $ \lambda$3727, \Hb, \Othree\ $\lambda\lambda$4959, 5007, \Ha, and \Ntwo\ $\lambda$6583.  Upper limits on \Otwo\ and \Ntwo\ and fluxes of the remaining lines are reported in Table~\ref{tab:lines}, and the spectra are shown in Figure \ref{fig:irspec}.  We use the strongest of these lines, \Ha\ and \Othree\ $\lambda$5007, to determine the systemic redshift of BX418, $z_{\rm sys}=2.3048$.  In this section we use the ratios of all of these lines to place constraints on metallicity, ionization parameter, and extinction.  

\begin{deluxetable*}{l l l c l l l}
\tablewidth{0pt}
\tabletypesize{\footnotesize}
\tablecaption{Emission  Lines\label{tab:lines}}
%\rotate
\tablehead{
\colhead{Ion} & 
\colhead{$\lambda_{\rm rest}$\tablenotemark{a}} & 
\colhead{$W_0$\tablenotemark{b}} & 
\colhead{Flux} & 
\colhead{FWHM\tablenotemark{c}} &
\colhead{$\Delta v$\tablenotemark{d}} &
\colhead{Instrument} \\
\colhead{} &
\colhead{(\AA)} &
\colhead{(\AA)} &
\colhead{($10^{-17}$ erg s$^{-1}$ cm$^{-2}$)} &
\colhead{(km s$^{-1}$)} &
\colhead{(km s$^{-1}$)} &
\colhead{}
}
\startdata
Ly$\alpha$ & 1215.6701  & 54.0$\pm$1.2 & 29.3$\pm$0.4 & 840$\pm$17 & 307$\pm$3 & LRIS-B \\
Si {\sc ii}* & 1264.738 & 0.4$\pm$0.1 & 0.2$\pm$0.06 & ...  & 220$\pm$72 & LRIS-B \\
Si {\sc ii}* & 1309.276 & 0.5$\pm$0.2 & 0.2$\pm$0.07 & ... & 206$\pm$71 & LRIS-B \\
C {\sc ii}* & 1335.708  & 0.7$\pm$0.1 & 0.3$\pm$0.06 & ... & 197$\pm$51 & LRIS-B \\
Si {\sc ii}* & 1533.431 & $<0.4$       & $<0.2$     & ... &  ... & LRIS-B \\
He {\sc ii} & 1640.42    & 2.7$\pm$0.2 & 0.8$\pm$0.1 & 612$\pm$64 &  $-17\pm$55 & LRIS-B \\
O {\sc iii}] & 1660.809 & 1.0$\pm$0.2 & 0.3$\pm$0.07 & $<276$ &  193$\pm$57 & LRIS-B \\
O {\sc iii}] & 1666.150 & 1.3$\pm$0.2 & 0.4$\pm$0.08 & 235$\pm$93 & $-2\pm$41 & LRIS-B \\
C {\sc iii}] & 1906.683, 1908.734\tablenotemark{d} & 7.1$\pm$0.4 & 1.4$\pm$0.1 & 225$\pm$36 & $81\pm$21 & LRIS-B \\
$[$O {\sc ii}] & 3727.09, 3728.79    & ...  & $<1.9$    & ...  &  ... & NIRSPEC \\
H$\beta$     & 4862.721  & 44$\pm$9         & 2.6$\pm$0.3 & $<102$ & $-46\pm$19 & NIRSPEC \\
$[$O {\sc iii}$]$ & 4960.295 & 93$\pm$13         & 5.4$\pm$0.3 & 120$\pm$23 &  $-47\pm$10 & NIRSPEC \\
$[$O {\sc iii}] & 5008.239 & 285$\pm$26          & 16.7$\pm$0.2 & 161$\pm$7 &  $-9\pm$4 & NIRSPEC \\
H$\alpha$ & 6564.614     & 450$\pm$247   & 8.0$\pm$0.2 & 155$\pm$11 & ... & NIRSPEC \\
$[$N {\sc ii}] & 6585.27   & ...          & $<0.3$     & ... &  ... & NIRSPEC
\enddata
\tablenotetext{a}{Vacuum rest wavelength}
\tablenotetext{b}{Rest frame equivalent width, computed from the
  spectrum for the LRIS-B observations and from the comparison of the
  line flux and broadband magnitude for the rest-frame optical
  emission lines.  No
  broadband magnitude is available for the other lines observed with
  NIRSPEC.  Where the lines are
  undetected, 3$\sigma$ limits are given.}
\tablenotetext{c}{Full width at half maximum, corrected for
  instrumental resolution.  If the uncorrected line width minus its
  error is less than the instrumental resolution, we report a
  1$\sigma$ upper limit of the corrected FWHM plus its 1$\sigma$
  error.  We do not report widths for the weak UV fine structure
  lines, which are unresolved.}
\tablenotetext{d}{Velocity offset relative to H$\alpha$
  emission. Velocity differences of \Hb\ and [O {\sc iii]} $\lambda$4959
with respect to [O {\sc iii}] $\lambda$5007 are due to sky line residuals. }
\tablenotetext{e}{We assume a wavelength of 1907.709 \AA\ for the
  blended C {\sc iii}] doublet.}
\end{deluxetable*}

\subsection{Metallicity and Ionization Parameter}  
\label{sec:metal}
The gas phase metallicity of BX418 can be determined from the ratios of the rest-frame optical emission lines, using the various familiar strong-line metallicity diagnostics.\footnote{Unfortunately the $T_e$-sensitive auroral line [\ion{O}{3}]$\lambda4363$ falls outside of our observational windows, though given the faintness of this line we would be unlikely to detect it even in the low metallicity case.}  We begin by obtaining an upper limit on the metallicity using the ratios of [\ion{N}{2}]/\Ha\ (N2) and  ([\ion{N}{2}]/\Ha)/([\ion{O}{3}]/\Hb) (O3N2) as calibrated by \citet{pp04}.  We obtain 12~+~log(O/H) $<8.1$ from N2, and 12~+~log(O/H) $<8.0$ from O3N2.  These upper limits  place BX418 on the lower branch of the double-valued $R_{23}$ metallicity calibration, as does the \Ntwo/\Ha\ ratio itself \citep{ke08}.  Given this constraint, we can then use the $R_{23}$ method to determine metallicity.  This is the ratio of the oxygen lines \Othree\ $\lambda\lambda4959$, 5007 and \Otwo\ $\lambda$3727 to \Hb, and is perhaps the most commonly used strong line metallicity diagnostic.  We use the calibration of \citet{m91}, in the analytical form given by \citet{kkp99}.  This form incorporates the dependence of $R_{23}$ on the ionization parameter via the inclusion of the ionization parameter-sensitive ratio \Othree/\Otwo.  Our lower limit $\log$(\Othree/\Otwo)$>1.1$ indicates a high ionization parameter of at least $\log U\sim-2$.  In determining the metallicity, we consider the range  $1.1<$~log({\Othree/\Otwo)~$<2$, where the upper limit comes from the maximum value of this ratio in our Cloudy photoionization models, corresponding to a low metallicity $Z=0.05$ \zsun\ and a very high ionization parameter $\log U\sim-1$.  We do not apply an extinction correction, which would be negligible given the very low reddening implied by the SED modeling.  Incorporating both the uncertainty in the \Otwo\ flux and random error in the fluxes of the other lines, we find 12~+~log(O/H)~$=7.9\pm0.2$, or $Z=1/6$ \zsun.  This places BX418 among the most metal-poor star-forming galaxies known at high redshift (much lower abundances are found among damped \lya\ absorption systems, which show a wider range in metallicity;  e.g.\ \citealt{pskh97,pw99}).  This is shown in Figure~\ref{fig:mz}, which places BX418 on the mass-metallicity relation at \ztwo.

In Section \ref{sec:nebularuv} we also derive the metallicity via the direct, electron temperature method, using the ratio \Othreeuv\ $\lambda\lambda$1661, 1666/\Othree\ $\lambda$5007 to determine the electron temperature.  We find 12~+~log(O/H)~$=7.8\pm0.1$, in excellent agreement with the strong line metallicity derived above.  From later work on the UV spectrum (Section \ref{sec:he2}), we also find that reproducing all of the line ratios requires a very high ionization parameter of $\log U\sim-1$.  We discuss this extreme ionization parameter in Section \ref{sec:disc_logu}.  

\subsection{Extinction and the \Ha/\Hb\ Ratio}
\label{sec:balmerdec}
We have a further estimate of dust extinction in BX418 in the \Ha/\Hb\ ratio.  In principle this provides a more robust measurement than the SED fitting, but in practice we are limited by the low S/N of the \Hb\ line.  For case B recombination, $\Ha/\Hb=2.8$; from our measured \Ha\ and \Hb\ fluxes, we find $\Ha/\Hb=3.1\pm0.4$.  This is consistent with zero extinction and with the low value of $E(B-V)$ estimated from the SED fitting, but allows for greater reddening as well.  We conclude that the S/N of the  \Ha/\Hb\ ratio is too low to provide additional constraints on the amount of dust extinction.

\begin{figure*}[htbp]
\plotone{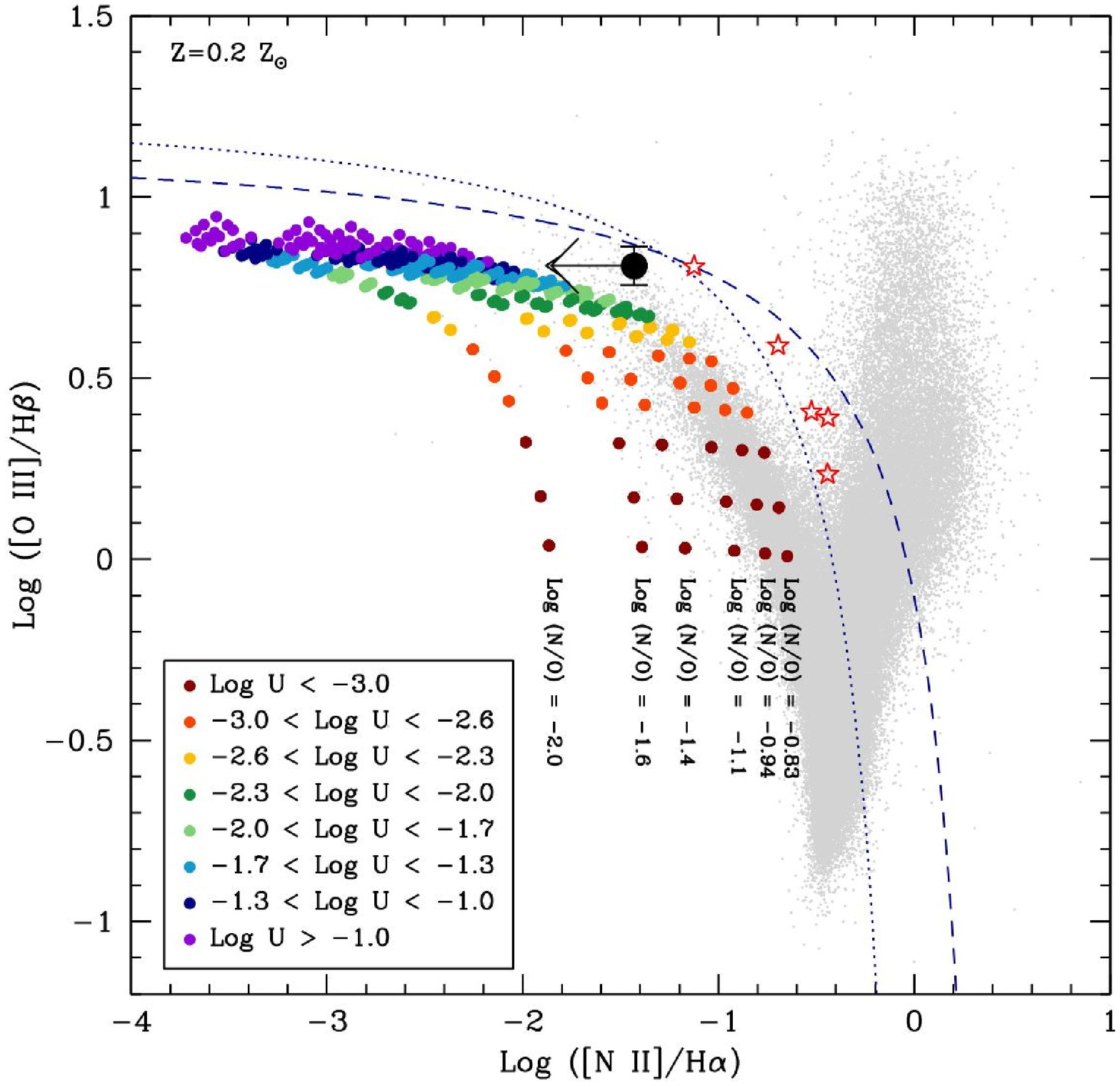}
\caption{The \Ntwo/\Ha\ vs.\ \Othree/\Hb\ diagnostic diagram.  BX418 is indicated by the large black circle, and the small grey points show $\sim$96,000 objects from the Sloan Digital Sky Survey.  Colored points are predictions from photoionization models, coded by ionization parameter as labeled at lower left.  The ionization parameter increases upward and to the left along the curves.  All models shown have 20\% of the solar oxygen abundance, and curves are shown for six different values of N/O, illustrating the effect of decreasing nitrogen abundance on the \Ntwo/\Ha\ ratio.  For higher overall oxygen abundances, the curves move up and to the right.  The open red stars show other galaxies at \ztwo\ from \citet{esp+06}; only objects with detections of all four lines are shown.  The dotted and dashed lines show empirical and theoretical divisions between galaxies and AGN, from \citet{kht+03} and \citet{kds+01} respectively.}
\label{fig:bpt}
\end{figure*}

\subsection{The \Ntwo/\Ha\ vs.\ \Othree/\Hb\ Diagnostic Diagram and the \Ntwo/\Ha\ Ratio}
The \Ntwo/\Ha\ vs.\ \Othree/\Hb\ line ratio diagram is one of several diagnostic tools commonly used to gain insight into the physical conditions in galaxies and distinguish between star-forming galaxies and AGN \citep{bpt81,vo87}.  In this section we use the upper limit on \Ntwo\ and our measurements of the other three lines to place BX418 on this diagram, and our suite of photoionization models to interpret the results.  The diagram is shown in Figure \ref{fig:bpt}, where the large black circle shows the limit on the position of BX418.  For context, the small grey points show $\sim97,000$ objects from the Sloan Digital Sky Survey; the left branch shows the location of star-forming galaxies, while AGN extend upward and to the right.  The dotted and dashed lines show empirical and theoretical divisions between galaxies and AGN (from \citealt{kht+03} and \citealt{kds+01} respectively).  The open red stars show other galaxies at \ztwo\, as discussed by \citet{esp+06}.

Next we plot the line ratios predicted by the Cloudy models, using filled circles color-coded by ionization parameter.  We restrict the models shown to the set with $Z=0.2$ \zsun\ for clarity; for higher metallicities, the curves move up and to the right.  We see that the \Othree/\Hb\ ratio increases with increasing ionization parameter (because the fraction of oxygen in the O$^{+2}$ ionization state increases with ionization parameter, while the strength of \Hb\ depends on the star formation rate).  \Ntwo/\Ha, on the other hand, decreases with increasing ionization parameter (as the N$^+$/N$^{+2}$ ratio decreases with increasing  ionization parameter).  Unsurprisingly, the \Ntwo/\Ha\ ratio also depends on the nitrogen to oxygen abundance ratio N/O.  While we have no constraints on N/O in BX418, it is expected that the nitrogen abundance will be depressed relative to oxygen in young galaxies, because N is produced by intermediate mass stars as well as by the massive stars that produce O and is therefore released into the ISM more slowly.   This effect has been seen in the lensed $z=2.7$ galaxy MS1512-cB58, in which N is underabundant with respect to O by a factor of $\sim3$ \citep{prs+02}, as well as in local galaxies and in damped \lya\ systems \citep{pzsc08}.  We therefore plot curves for several values of N/O in Figure \ref{fig:bpt}.  The rightmost curve uses the default N/O ratio of Cloudy's H II region abundance set; proceeding to the left, each curve further reduces the abundance of N (for cB58, \citet{prs+02} find $\log\; ({\rm N/O}) = -1.89$).  

We use this diagram to make several points.  First, the \Ntwo\ flux is likely to be well below our lower limit; we are unlikely to detect it even with significantly deeper data, especially if the nitrogen abundance is low with respect to oxygen, as is likely.  We also note that under these extreme conditions of very high ionization parameter and (possibly) low N/O ratio, the N2 metallicity calibration would provide a significant underestimate of the oxygen abundance.  This is not unexpected, as the local sample from which the relation is calibrated contains no galaxies with such an extreme ionization parameter.\footnote{The $R_{23}$ diagnostic we use to determine the metallicity in Section \ref{sec:metal} does not suffer from the same issue, as it incorporates the ionization parameter; we have used the best-fitting Cloudy models to verify that the $R_{23}$ index reproduces the input metallicity of the models within our uncertainties.}  

This very low \Ntwo/\Ha\ ratio places BX418 far from the region of the diagram occupied by AGN, suggesting that significant AGN contribution to the ionizing flux of BX418 is unlikely.  However, most of the AGN shown in Figure \ref{fig:bpt} are considerably higher in metallicity than BX418. \citet{ghk06} have studied emission line diagnostics in lower metallicity AGN, finding that the \Ntwo/\Ha\ ratio is much more metallicity-dependent than \Othree/\Hb, and that therefore low metallicity AGN move to the left on the diagram, with relatively little vertical shift.  Given BX418's low metallicity and probable low nitrogen abundance, its position on the diagnostic diagram alone is not sufficient to rule out the presence of an AGN.  However, as we have argued elsewhere, the lack of strong \ion{C}{4} emission indicates a radiation field softer than that seen in AGN (Section \ref{sec:uv}), and the nondetection of BX418 in deep 8 and 24 \micron\ images argues against the presence of an obscured AGN.

Next we observe that the few \ztwo\ galaxies for which all four emission lines have been detected lie significantly to the right of the probable location of BX418 (and from the sequence of local star-forming galaxies; see \citealt{lsc+08} and \citealt{bpc08}).  This is true by definition, given our limited sensitivity to \Ntwo, but stacking of spectra shows that most star-forming \ztwo\ galaxies have log (\Ntwo/\Ha) $\gtrsim-1$ \citep{esp+06}.  This much higher \Ntwo/\Ha\ ratio confirms that average \ztwo\ galaxies (while still having ionization parameters higher than are seen in typical star-forming galaxies in the local universe)
have a lower ionization parameter and higher metallicity than BX418, in agreement with other observations \citep{hsk+09,esp+06}.

\section{The Rest-Frame UV Spectrum}
\label{sec:uv}
Rest-frame UV spectra of galaxies contain a myriad of features sensitive to dust content, age, galaxy-scale outflows, metallicity, ionization parameter, and the initial mass function (e.g.\ \citealt{ssp+03}).  BX418 is not unusual among \ztwo\ galaxies in showing strong \lya\ emission and strong absorption from high ionization interstellar lines, but it is rare in that the low ionization absorption lines are weak or undetected, and in its extremely strong emission from \hetwo\ $\lambda1640$, \Cthree\ $\lambda\lambda1907$, 1909 and \Othreeuv\ $\lambda\lambda1661$, 1666.  The spectrum is shown in full in Figure \ref{fig:fullspec}, and with an expanded vertical scale for detail in Figure \ref{fig:specpanels}. 

\begin{figure*}[htbp]
\plotone{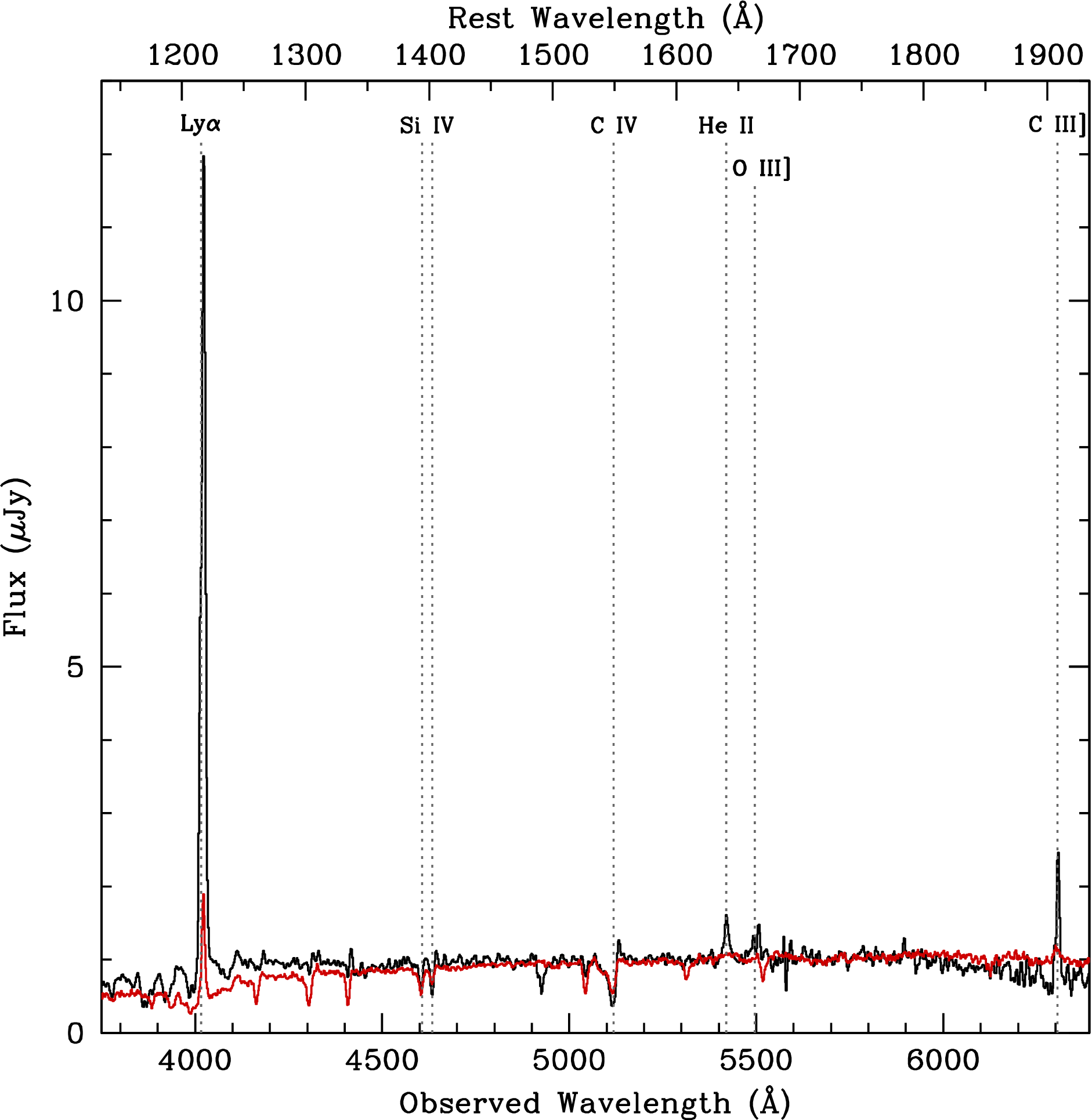}
\caption{The rest-frame UV spectrum of Q2343-BX418, shown in full to emphasize the strong \lya\ emission.  The strongest features are marked with vertical dotted lines: from left, \lya\ emission, the interstellar high ionization absorption lines \ion{Si}{4} $\lambda$1394, \ion{Si}{4} $\lambda$1402, and \ion{C}{4} $\lambda$1549, \ion{He}{2}  $\lambda$1640 emission, \ion{O}{3}]  $\lambda$1663 emission, and \ion{C}{3}]  $\lambda$1909 emission.  The spectrum is not normalized; note the flatness of the UV slope.  Overplotted for comparison (thin red line) is the composite spectrum of 966 galaxies with a mean redshift $\langle z\rangle=2.2$.}
\label{fig:fullspec}
\end{figure*}

\begin{figure*}[htbp]
\plotone{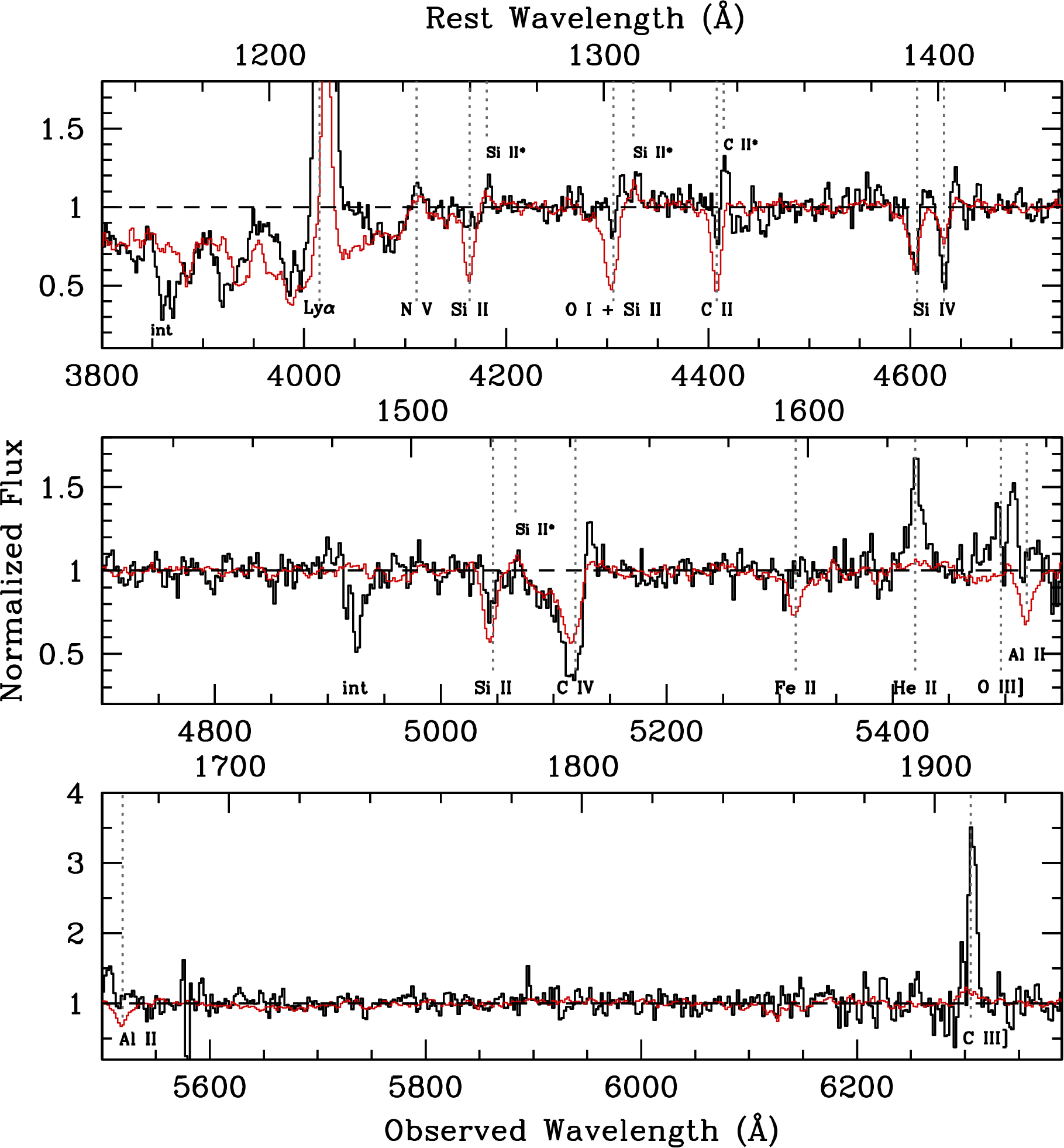}
\caption{The normalized UV spectrum shown on an expanded vertical scale.  We overplot the normalized \ztwo\ composite in red for comparison.  Lines in both spectra are identified.  The features marked ``int" at 3859 and 4926 \AA\ are intervening \lya\ and C~{\sc iv} absorption from the $z=2.176$ foreground galaxy Q2343-BX442, with an impact parameter of 140 kpc.  This system is discussed in Section \ref{sec:intervene}.}
\label{fig:specpanels}
\end{figure*}

\begin{deluxetable*}{l l l l l}
\tablewidth{0pt}
\tabletypesize{\footnotesize}
\tablecaption{Interstellar Absorption Lines\label{tab:abs}}
%\rotate
\tablehead{
\colhead{Ion} & 
\colhead{$\lambda_{\rm rest}$\tablenotemark{a}} & 
\colhead{$f$\tablenotemark{b}} & 
\colhead{$W_0$\tablenotemark{c}} & 
\colhead{$\Delta v$\tablenotemark{d}} \\
\colhead{} &
\colhead{(\AA)} &
\colhead{} &
\colhead{(\AA)} &
\colhead{(km s$^{-1}$)}
}
\startdata
Si {\sc ii} & 1260.4221 & 1.18 & $>-0.5$ & ...\\
O {\sc i}\tablenotemark{e} & 1302.1685 & 0.04887 & $>-0.3$ & ...\\
Si {\sc ii}\tablenotemark{e} & 1304.3702 & 0.094 & $>-0.3$ & ...\\
C {\sc ii} & 1334.5323 & 0.1278 & $>-0.5$ & ...\\
Si {\sc iv} & 1393.7602 & 0.5140 & $-1.1\pm0.1$ & $-153$\\
Si {\sc iv} & 1402.7729 & 0.2553 & $-1.3\pm0.1$ & $-161$\\
Si {\sc ii} & 1526.7070 & 0.133 & $-0.7\pm0.1$ & $-91$\\
C {\sc iv}\tablenotemark{f} & 1548.204 & 0.1908 & $-5.1\pm0.2$ & $-583$\\
C {\sc iv}\tablenotemark{f} & 1550.781 & 0.09522 & $-5.1\pm0.2$ & $-583$\\
Fe {\sc ii} & 1608.4509 & 0.058 & $>-0.6$ & ...\\
Al {\sc ii}\tablenotemark{g} & 1670.7886 & 1.833 & $>-0.7$ & ...
\enddata 

\tablenotetext{a}{Vacuum rest wavelength.}
\tablenotetext{b}{Oscillator strength, from \citet{morton03}.}
\tablenotetext{c}{Rest frame equivalent width.  Where the lines are
  undetected, 3$\sigma$ limits are given.}
\tablenotetext{d}{Velocity offset relative to H$\alpha$ emission.}
\tablenotetext{e}{Blended.}
\tablenotetext{f}{Blended; $W_0$ is given for the blend, and $\Delta
  v$ assumes that the rest wavelength of the blend is 1549.479 \AA.}
\tablenotetext{g}{Blended with O {\sc iii}] $\lambda1666$ emission.}

\end{deluxetable*}

Before proceeding with detailed discussion of the spectrum, we ask whether these strong emission lines indicate the presence of an AGN.  We compare BX418's emission line ratios (see Table \ref{tab:lines}) with those of the composite spectrum of 16 narrow-line AGN presented by \citet{shs+02}.  These AGN have a mean redshift $\langle z \rangle = 2.67$, and were selected in the course of the \zthree\ Lyman break galaxy survey \citep{sas+03}.  The composite AGN spectrum shows many of the same emission lines as BX418, but is notably different in its much stronger \ion{C}{4} emission.  The hard AGN ionizing spectrum is evident in the ratio \ion{C}{4}/\Cthree~$\sim2$; in contrast, BX418 has \ion{C}{4}/\Cthree~$\sim0.3$, indicative of a significantly softer radiation field.  The composite AGN spectrum also has relatively strong \ion{C}{4} in comparison with \lya, with \ion{C}{4}/\lya~$\sim0.25$, while BX418 has \ion{C}{4}/\lya~$\sim0.01$.  Given the weakness of the \ion{C}{4} emission, which can be fully accounted for by nebular and P~Cygni stellar emission (see Section \ref{sec:windlines}), significant AGN contribution to BX418's ionizing flux is unlikely.

\subsection{Outflowing Gas}
\label{sec:outflows}
The strongest features in the UV spectrum of a typical high redshift galaxy---a variety of interstellar absorption lines, and \lya\ emission and/or absorption---reflect the kinematics of powerful, galaxy-scale outflows of gas \citep{ssp+03}.  In nearly all such spectra, the interstellar absorption lines are blueshifted with respect to the systemic redshift of the galaxy, while \lya\ emission (if present) is redshifted.  The standard model invoked to explain this pattern is of a nearly spherical outflow; the blueshifted absorption lines arise from the gas approaching the observer, on the near side of the galaxy, while the resonantly scattered \lya\ photons are most likely to escape the galaxy in the direction of the observer when they are shifted away from their resonant frequency by scattering off the receding gas on the far side of the galaxy, resulting in a redshift.  The interstellar absorption lines therefore provide a reasonably straightforward probe of the kinematics of the outflowing gas, subject to the uncertainties imposed by limited spectral resolution and partial covering fraction.  \lya\ emission also reflects the kinematics of the outflow, but the complexities of \lya\ radiative transfer make its interpretation much more difficult.  Several recent models explore the connections between \lya\ (and interstellar) line profiles and the kinematics of the gas (e.g. \citealt{vsm06,vsat08,ses+10}).

\subsubsection{Interstellar Absorption}
\label{sec:interstellar}
A detailed comparison of the UV spectrum of BX418 and the composite spectrum of $\sim1000$ \ztwo\  galaxies is shown in Figure \ref{fig:specpanels}.  The composite spectrum is characterized by strong low-ionization absorption lines from Si, O, C, Fe and Al, originating in outflowing neutral and low ionization gas as described above.  In BX418 these lines are weak or absent.  We adopt a 3$\sigma$ threshold for detection of these lines; applying this threshold, the only low ionization line with a statistically significant detection is \ion{Si}{2} $\lambda1527$.  Other lines that appear to be weakly present in Figure \ref{fig:specpanels}, such as \ion{Si}{2} $\lambda$1260, \ion{O}{1}~+~\ion{Si}{2} $\lambda$1303 and \ion{C}{2} $\lambda$1334, fall just below this threshold.

In typical galaxies at \ztwo--3, the low ionization lines are strongly saturated, indicating that the equivalent width is determined by the velocity range and covering fraction of the absorbing gas.  Although the lines are barely present in BX418, we can nevertheless obtain interesting constraints by comparing the two  \ion{Si}{2} lines at 1260 and 1527 \AA.  On the linear, optically thin part of the curve of growth, the equivalent width of the line $W\propto N\lambda^2 f$, where $N$ is the column density, $\lambda$ is the wavelength of the transition, and $f$ is the oscillator strength.  For the two \ion{Si}{2} lines, the oscillator strengths of the blue and red transitions are 1.18 and 0.133 respectively \citep{morton03}, so that in the optically thin case $W(1260)/W(1527)=6.0$.  We measure $W_0=-0.7$ \AA\ for \ion{Si}{2} $\lambda1527$, while the 1260 line is undetected with a 3$\sigma$ limit $W_0>-0.5$ \AA.  In the optically thin case we would find $W(1260)\simeq-4$ \AA, strongly ruled out by the data.  We also use the Cloudy models to assess the strength of any predicted emission from the \ion{Si}{2} lines, finding that such emission is at most $\sim10$\% of the strength of the observed absorption, too weak to be detected in the spectrum or affect the measured line ratios.  We therefore conclude that in spite of their weakness, the low ionization lines are saturated, as in other galaxies at \ztwo--3.  
 
The spectrum also contains high ionization absorption lines from \ion{Si}{4} and \ion{C}{4}.  In contrast to the low ionization lines, these are as strong or stronger than their counterparts in the composite spectrum.  As with the \ion{Si}{2} lines discussed above, the \ion{Si}{4} doublet at $\lambda\lambda$1393, 1402 \AA\ can be used to obtain constraints on the optical depth of the absorbing gas.  The oscillator strengths of the members of the doublet (see Table \ref{tab:abs}) imply $W(1393)/W(1402)=2.0$ in the optically thin case; the line strengths in the composite spectrum are consistent with this ratio, as they are in the \zthree\ LBG composite studied by \citet{ssp+03}.  In BX418, however, the two \ion{Si}{4} lines are of approximately equal strength, indicating that the high ionization lines are saturated.  Although the composite spectra show that this is not the usual case, BX418 is not the only such example; the high ionization lines are also saturated in several of the lensed galaxies as well, though these objects also have strong low ionization lines \citep{psa+00,qpss09,dds+09}.   We have also checked the predicted strength of nebular emission from the two \ion{Si}{4} transitions: this emission is expected to be stronger than the \ion{Si}{2} emission considered above, but for the range of ionization parameters BX418 requires (log $U\gtrsim -2$ to log $U\sim-1$), emission from the two members of the doublet is approximately equal, leaving the absorption line ratio unaffected.

The equivalent width of a strongly saturated line is determined by the velocity range and covering fraction of the absorbing gas; if the line is resolved, the transition describes the covering fraction of gas as a function of velocity.  Given the fairly low resolution and limited S/N of the BX418 spectrum, it is difficult to distinguish between these two effects, and detailed modeling of the kinematics of the outflowing gas is not feasible.  However, the available information suggests that the range of outflow velocities is typical of \ztwo--3 galaxies.  The \ion{Si}{4} lines have a mean blueshift of $\sim150$ \kms, and show asymmetric absorption extending to approximately $-700$ \kms\ relative to the systemic velocity defined by the nebular emission lines (Figure \ref{fig:velprofiles}, left two panels); both the centroid and range are characteristic of interstellar absorption in \ztwo\ galaxies (\citealt{psa+00,qpss09,qsp+10,ses+10}).  The S/N of the low ionization lines is too low to make a reliable measurement of the velocity range, but the mean blueshift of the \ion{Si}{2} $\lambda$1527 line is similar to that of the \ion{Si}{4} lines, and the low and high ionization lines typically show similar kinematics.  The \lya\ emission line (discussed below) also shows a very wide range of velocities.  This would suggest that covering fraction is the primary factor driving the relative strengths of the low and high ionization lines, and that the covering fraction of gas producing the high ionization lines is significantly higher than that responsible for the low ionization lines.  Given the extreme ionization parameter (see Section \ref{sec:he2}), this result is not surprising.  

\begin{figure*}[htbp]
\plotone{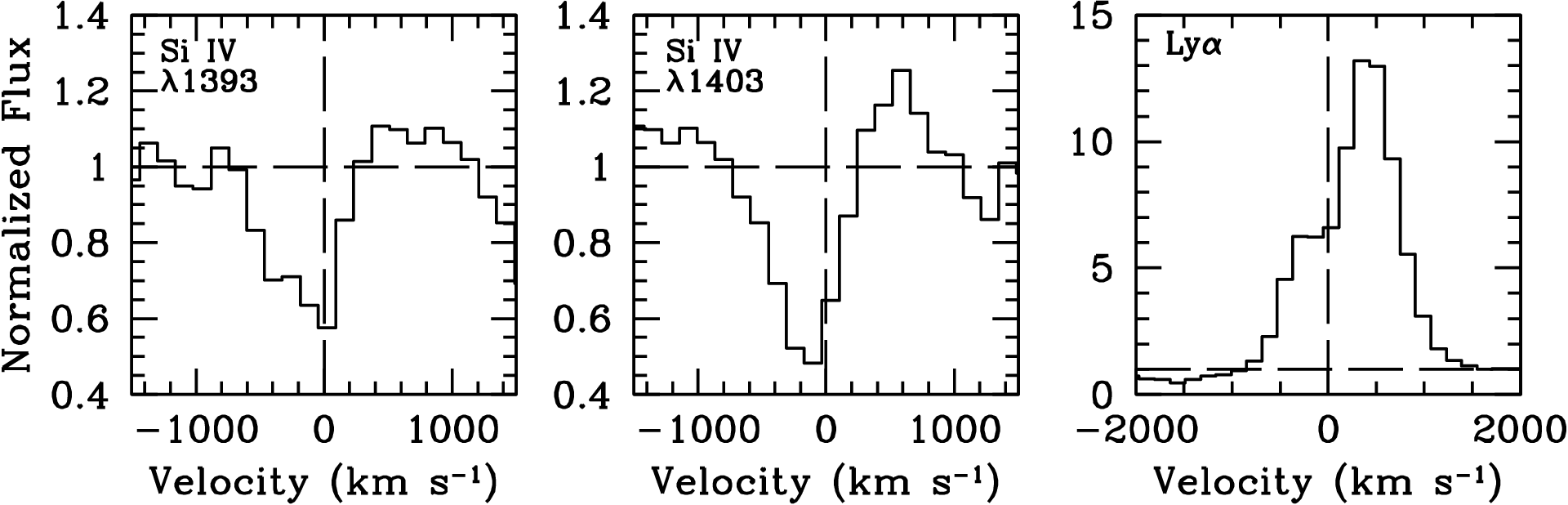}
\caption{Absorption or emission line profiles as a function of velocity, relative to the systemic velocity defined by \Ha\ and \Othree\ $\lambda$5007 emission.  The two left panels show the \ion{Si}{4} $\lambda\lambda 1393$, 1402 lines, and the right panel shows \lya.}
\label{fig:velprofiles}
\end{figure*}

\subsubsection{\lya\  Emission} 
\label{sec:lya}
BX418 is a strong \lya\ emitter, with rest-frame equivalent width $W_{\lya}=54$ \AA.  The \lya\ line is also unusually broad, with FWHM~$\sim850$ \kms, compared to the average of $\sim650$ \kms\ seen in \ztwo--3 galaxies \citep{ses+10}.  The reason for the unusually large line width is clear from the line profile shown in the right panel of Figure \ref{fig:velprofiles}: while typical \lya\ profiles of high redshift galaxies are purely redshifted and show a sharp blue edge, the \lya\ emission in BX418 shows a significant blueshifted component as well.  The total range of velocities spanned by the line is large, ranging from approximately $-800$ to $+1500$ \kms.  The line centroid is redshifted with $\Delta v=300$ \kms, somewhat lower than the \ztwo\ average of $\sim450$ \kms; this lower than average value is clearly due to the significant blueshifted emission, rather than to a lower characteristic outflow velocity.

Given the ubiquitous nature of outflows in high redshift galaxies, a great deal of energy has been devoted to modeling the radiative transfer of \lya\ photons in an expanding medium.  This is a complex problem, and even the simplest models suffer from significant degeneracies in recovering a unique kinematic model from a given line profile.  Given these degeneracies, and the fact that our resolution is too low to resolve any detailed structure that may be present in the profile, we do not attempt rigorous modeling.  Nevertheless it is interesting to consider BX418 in the context of these models.  According to the work of \citet{vsm06,vsat08}, if the outflow is considered to be a shell with velocity $V_{\rm exp}$, the redshifted peak of the \lya\ emission falls at $v\sim 2\times V_{\rm exp}$; this represents emission backscattered off the far side of the receding shell. There may also be secondary emission peaks at $v\sim0$ and $v\lesssim-V_{\rm exp}$, corresponding to photons escaping from the blue and red wings of the approaching shell.  In a more realistic model with a broad range of outflow velocities, the emission will span a broader range, with peaks at the velocities of lowest optical depth.  The significant blueshifted emission in BX418 therefore suggests that the optical depth to \lya\ photons is unusually low in the foreground outflowing material.  This is consistent with our conclusion in Section \ref{sec:interstellar} above that the covering fraction of neutral gas is low, probably because a higher than average fraction of the outflow is highly ionized.  The low dust content undoubtedly also contributes to the apparent ease with which \lya\ photons escape the outflowing gas.

\citet{ses+10} have recently applied simple analytic models to the kinematics of outflowing gas, and are able to reproduce observed interstellar absorption and \lya\ emission profiles with two gas components, one outflowing and one at or near the systemic velocity of the galaxy.  These results suggest that the observed line profiles may be more dependent on bulk gas velocities than on detailed radiative transfer effects.  We have not attempted to use these schematic models to reproduce the \lya\ line profile of BX418 in detail, but the models do demonstrate that significant blueshifted emission can be produced if the optical depth in the outflow is sufficiently low.

Although BX418 is a sample of only one, it is interesting to compare with the results of \citet{vsat08}, who model the \lya\ emission in a sample of 11 galaxies at $2.8<z<5$.  They suggest that the equivalent width and FWHM of \lya\ emission are correlated with the column density of neutral hydrogen $N_{\rm HI}$ in opposite senses, and that therefore systems with both high equivalent width and large FWHM should not be observed.  They note that observed \lya\ emitters (LAEs, with $W_{\lya}>20$ \AA) have FWHM less than $\sim500$ \kms.  BX418 clearly does not follow this trend.  With $W_{\lya}=54$ \AA, it is well above the threshold to be selected as an LAE, and yet it has FWHM~$\sim850$ \kms.  The \citet{vsm06,vsat08} expanding shell models are unable to fit broad lines with the low column density required to produce a high equivalent width; however, these models assume unity covering fraction, which is unlikely to be the case in BX418, and it is probably the patchy distribution of optically thick gas which allows significant escape of \lya\ photons from the blueshifted approaching outflow, thus adding to both the equivalent width and FWHM of the line.  This situation is unusual, but BX418 is not the only example; as noted in Section \ref{sec:disc_hiz}, galaxies with significant Lyman continuum emission also show blueshifted \lya\ emission (Steidel et al., in preparation).  Models with a more realistic treatment of gas covering fraction may be more successful in reproducing the observed diversity of \lya\ profiles.

Given this strong \lya\ emission and BX418's low dust content, we might expect that a large fraction of the \lya\ photons are able to escape the galaxy.  We assess this by comparing the \lya\ and \Ha\ fluxes: for case B recombination, we expect $\lya/\Ha=8.3$ \citep{fo85}.  We observe $\lya/\Ha=3.7$, a factor of $\sim2$ lower than the case B prediction, or escape fraction $f_e=0.4$.  \citet{vsat08} suggest that dust extinction is the primary factor in determining $f_e$, finding that their sample is well fit by 
\begin{equation}
f_e=10^{-7.71\times {\rm E(B-V)}}.
\label{eq:fesc}
\end{equation}
The correlation between dust content and escape fraction is also seen in local galaxies by \citet{aks+09}, and in a larger high redshift sample by \citet{kse+10}, who find that the data scatter about the \citet{vsat08} model, with the majority of points having a lower escape fraction than predicted by Equation \ref{eq:fesc} (all calculations assume the \citealt{cab+00} extinction law).  From this relation and our inferred $E(B-V)=0.02$ we would predict $f_e=0.7$, higher than the observed value; like most of the \citet{kse+10} sample, BX418 falls below the \citet{vsat08} prediction.  However, given the significant uncertainties in $E(B-V)$, the predicted $f_e$ is within our uncertainties.  This comparison also assumes that \Ha\ and \lya\ emission have the same spatial extent, which may not be the case because \lya\ is subject to greater scattering.  Our observed $f_e$ is therefore a lower limit, since the spectroscopic slit may be missing a larger fraction of the \lya\ photons.

\subsubsection{Intervening Absorption and the Physical Extent of Outflows}
\label{sec:intervene}
One of the stronger absorption features in the spectrum, with equivalent width $W=2$ \AA, falls at 4926 \AA\ (see Figure \ref{fig:specpanels}).  There is no intrinsic absorption feature at this wavelength in the rest frame of BX418, as shown by the composite spectrum.\footnote{This absorption feature prevents the detection of any possible N {\sc iv}] $\lambda$1487 emission; see Section \ref{sec:nebularuv}.}  We identify this line, and a corresponding feature at 3859 \AA, as intervening \ion{C}{4} and \lya\ absorption from the foreground galaxy Q2343-BX442, with $z=2.176$.  BX442, identified in the \Ha\ survey of \citet{ess+06mass}, is separated from BX418 by an impact parameter of 17\arcsec, or 140 kpc.   Other galaxies we find near BX418 are known or likely to be at lower redshifts, given their colors and magnitudes.  %more?

Absorption systems along the line of sight to background galaxies and QSOs can be used to assess the spatial extent of galactic outflows (e.g.\ \citealt{assp03,ass+05,ssrb06,ses+10}).   Close pairs of background QSOs and foreground galaxies are rare but offer high S/N; galaxy-galaxy pairs are much more common but are more difficult because of the faintness of the background object.  \citet{ses+10} have recently stacked spectra of galaxy-galaxy pairs as a function of impact parameter, to determine the average absorption of various transitions with distance; they find significant absorption in both \ion{C}{4} and \lya\ at an impact parameter of $\sim60$ kpc, and at $\sim100$ kpc the average spectrum still shows clear absorption in \lya.   Individual pairs can show much stronger absorption; for example, \citet{ass+05} and \citet{ssrb06} identify strong absorption from \ion{C}{4}, \lya, \ion{O}{6} and other ions in a QSO spectrum, and associate it with a massive galaxy at an impact parameter of 115 kpc.  With equivalent width $\sim2$ \AA, the absorption from BX442 is similarly well above the average value (absorption of this strength would have been easily detectable in the composite spectra of \citealt{ses+10}).  The sample of individual pairs of galaxies with detected foreground absorption is still too small to test for any correlations between the properties of the foreground galaxies and the strength of the observed absorption, but we note that BX442 is among the oldest and most massive galaxies in the sample, with an estimated age of 2.8 Gyr and stellar mass $M_{\star}=1.1\times10^{11}$ \msun; it has therefore had considerable time in which to drive material to large distances.   Given BX442's large stellar mass, it probably occupies a dark matter halo more massive than the \ztwo\ average, $\langle M_{\rm halo}\rangle=9\times10^{11}$ \msun\ \citep{cst+08}; for a halo of mass $\sim2\times10^{12}$ \msun, the virial radius is $r_{\rm vir}\simeq125$ kpc (see calculations by \citealt{ses+10}).  Thus it appears that BX442 has driven significant quantities of gas to the virial radius and beyond.

\subsection{\ion{He}{2} Emission}
\label{sec:he2}
One of the most striking differences between the UV spectrum of BX418 and the average \ztwo\ galaxy spectrum is BX418's strong \hetwo\ $\lambda 1640$ emission, which has equivalent width $W_0=2.7$ \AA\ and flux relative to \Hb\ $F_{\rm He II}/F_{\Hb}\sim0.3$.  The line is shown in Figure~\ref{fig:he2}, with the \ztwo\ composite spectrum overplotted for comparison.  Broad \hetwo\ $\lambda 1640$ emission is characteristic of high redshift galaxies; the weak emission in the composite shown in Figure~\ref{fig:he2} has equivalent width $W_0\sim0.6$ \AA, and \hetwo\ emission is clearly seen in the composite spectrum of \zthree\ Lyman break galaxies analyzed by \citet{ssp+03}, with $W_0=1.3\pm0.3$ \AA\ (as measured by \citealt{bpc08}) and FWHM $\sim1500$ \kms.  Because of its broad width, the line is attributed to the fast, dense stellar winds of Wolf-Rayet stars.  Broad \hetwo\ emission with $W_0=2.45\pm0.22$ \AA\ is also seen in the spectrum of the $z=2.7$ lensed galaxy known as the 8 o'clock arc \citep{dds+09}, and at similar strength in the spectrum of the lensed $z=3.8$ galaxy observed by \citet{cvl08}, an object which otherwise appears similar to cB58.

While the \hetwo\ emission in BX418 is clearly broad, it also has a narrow peak not seen in the composite spectra or in the high resolution spectrum of the 8 o'clock arc.  As shown in the right panel of Figure~\ref{fig:he2}, the line is well fit by the superposition of two Gaussian components:  a broad line with FWHM $\sim1000$ \kms\ accounts for $\sim75$\% of the line flux, while a narrow, unresolved component makes up the remaining $\sim25$\%.  Compared to a single Gaussian, the superposition of these two components provides a better fit to the line at the 99\% confidence level; a single Gaussian underpredicts both the broad wings and the narrow peak of the line.  As with the composite spectra, the width of the broad component indicates an origin in the winds of W-R stars, while the narrow component is likely to be nebular emission.  We discuss the stellar and nebular \hetwo\ emission separately below.

\begin{figure*}[htbp]
\plotone{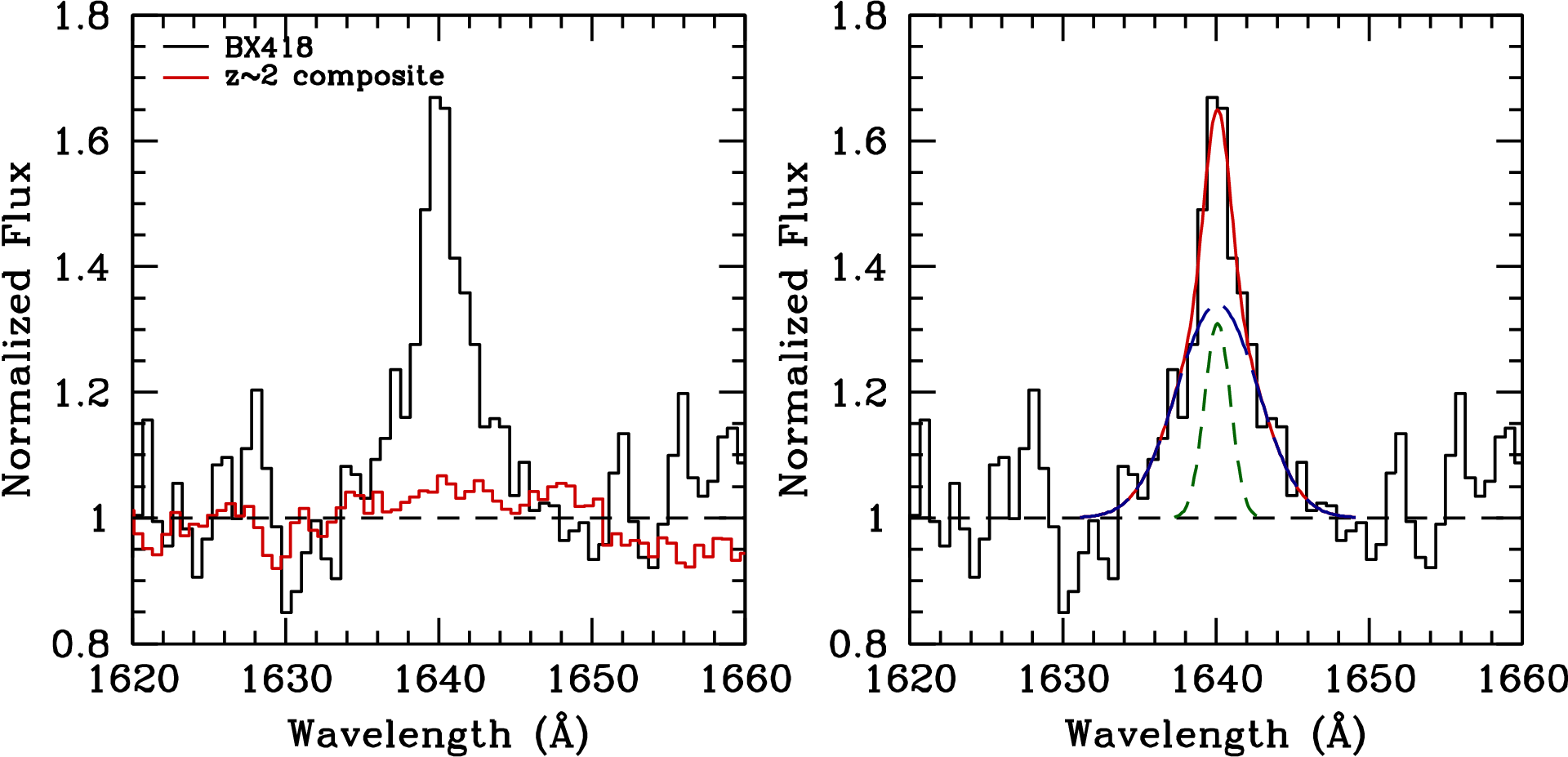}
\caption{{\it Left:} The strong \hetwo\ $\lambda$1640 emission line (black), compared to the broad and far weaker emission in the composite spectrum of 966 \ztwo\ galaxies (red).  {\it Right:}  The line is well fit by a superposition of two Gaussian components (solid red line).  The broad component (long-dashed dark blue line) has FWHM $ \sim1000$ \kms, while the narrow component (short-dashed green line) is unresolved.  We attribute the broad emission to W-R stellar winds, and the narrow component to nebular \hetwo\ emission.}
\label{fig:he2}
\end{figure*}

\subsubsection{Stellar \hetwo\ Emission}
The strength of the stellar \hetwo\ line undoubtedly depends on the number of W-R stars producing the emission, and probably depends on the type and metallicity of these stars as well.  Because the W-R phase is very short, age and star formation history are also crucial.  There has been considerable recent work addressing these issues, both locally and at high redshift, but progress is hampered by the lack of local observations of individual W-R stars of low metallicity.

Although detailed observations are still needed, current models indicate that line luminosities of W-R stars decrease with decreasing metallicity, due to the reduced density of stellar winds \citep{snc02,ch06}.  Studies of lensed galaxies and composite spectra are broadly consistent with this trend, as the only strong stellar \hetwo\ emission seen in a lensed galaxy with a determination of metallicity belongs to the massive and metal-rich 8 o'clock arc (\citealt{dds+09}; metallicity measurements are not available for the galaxy observed by \citealt{cvl08}).  This emission is also stronger than than seen in the \zthree\ composite spectrum, which probably has a lower average metallicity.

The precise form of the line scaling with metallicity is unknown.  \citet{bpc08} combine models of \hetwo\ $\lambda1640$ luminosities with a variety of star formation histories in an effort to account for the \hetwo\ emission seen by \citet{ssp+03}; they find that, while \hetwo\ emission peaks a few Myr after a burst of star formation, even for more extended star formation histories the line should be detectable with an equivalent width of 1--2 \AA, at metallicities above $Z\sim0.5$ \zsun.  While this model may account for the \hetwo\ emission seen in the \zthree\ and \ztwo\ composite spectra and in the 8 o'clock arc,  it is more problematic in the case of BX418, with $Z=0.2$ \zsun.  At this low metallicity, the \hetwo\ emission is never predicted to exceed $W_0\sim0.5$ \AA, even immediately after a burst of star formation.  More recent stellar population synthesis models which include binary evolution \citep{es09} offer better agreement with our observations, however.  These models so far include only instantaneous bursts rather than more extended star formation histories, but they find that with binary evolution taken into account, W-R features are both stronger and occur at later times.  For a metallicity $Z=0.004$, close to the metallicity we find for BX418, their models predict an equivalent width $W_0\simeq3$ \AA\ at an age of $\sim10$ Myr.  The equivalent width is stronger at lower metallicities because the W-R phase lasts longer; WNL stars form late, through binary evolution, and have a longer lifetime at lower metallicities because it takes longer for the weak stellar winds to remove the stars' hydrogen envelopes.  Thus the lower metallicity stars make a greater contribution to the integrated spectrum.

While these models may account for the \hetwo\ emission seen in BX418, the situation remains complicated.  Comparison of BX418 with the \ztwo\ composite spectrum supports the scenario in which low metallicity stars make a greater contribution to the \hetwo\ equivalent width.   We infer from the \ztwo\ mass-metallicity relation that the average metallicity of the galaxies making up the \ztwo\ composite is about 0.5 \zsun\ \citep{esp+06}, and the \citet{es09} models predict significantly weaker \hetwo\ emission at this metallicity.  The observations of lensed galaxies described above suggest the opposite picture, however, since the only detection of \hetwo\ emission is in the most metal-rich object observed.  In addition, the effect of varying star formation histories still needs to be taken into account; the instantaneous burst model is certainly not appropriate for typical galaxies in the \ztwo\ sample and is probably not the best model for BX418 either.  Even for BX418, it remains to be determined how much of the \hetwo\ line strength is due to metallicity, and how much may be due to observation at a particular, brief stage of evolution.  Both additional spectra of low metallicity galaxies and models of the line strength for a variety of star formation histories are needed to address this question.

\begin{figure*}[htbp]
\plotone{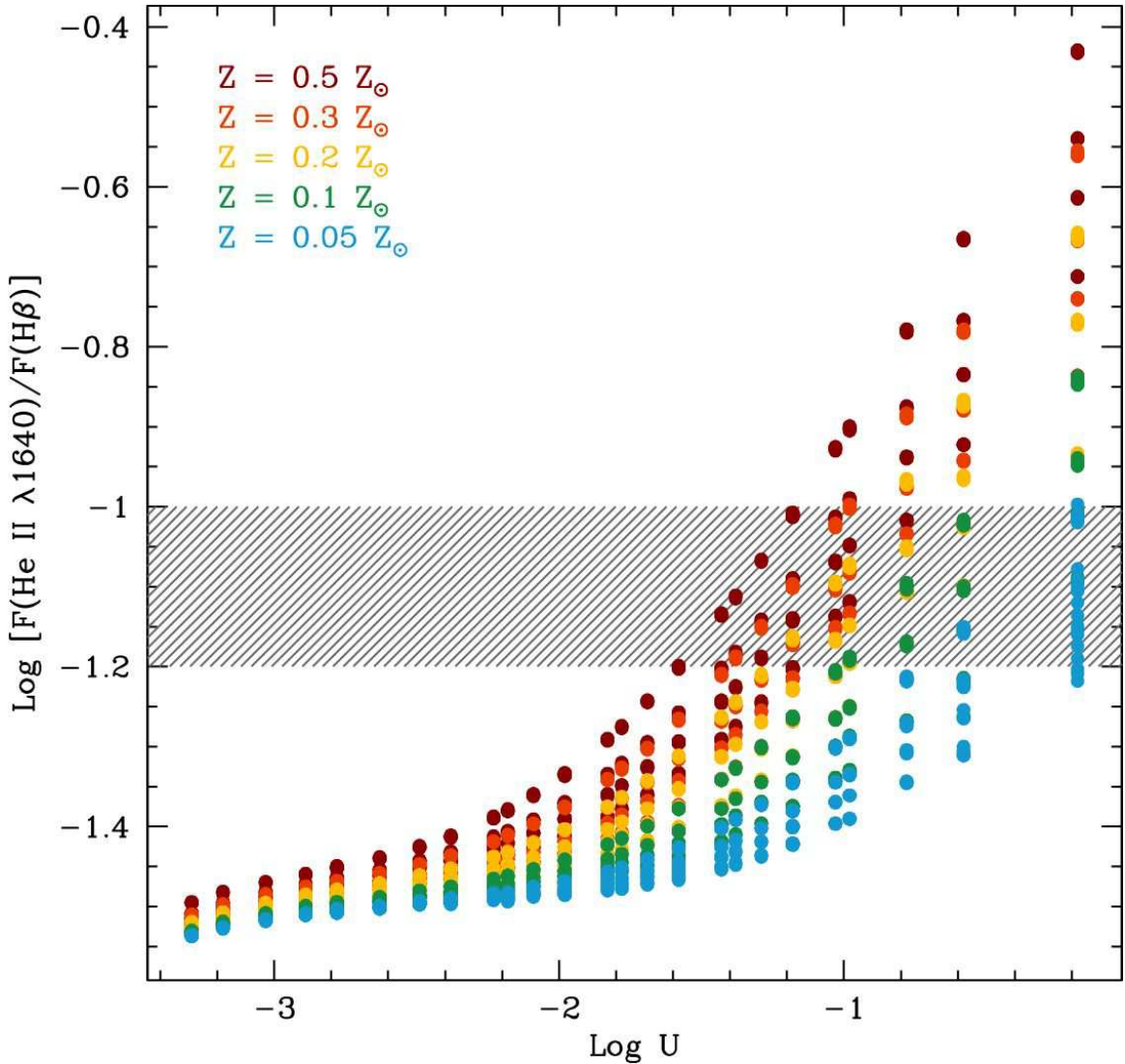}
\caption{Predicted nebular \hetwo\ emission as a function of metallicity and ionization parameter, from Cloudy photoionization modeling.  The colors represent different metallicities as labeled at upper left, and the shaded region shows the estimated \ion{He}{2}/\Hb\ ratio of BX418, including uncertainties.  For a metallicity $Z\sim0.2$ \zsun, an ionization parameter $\log U\sim-1$ is required to match observations.}
\label{fig:cloudy_he2}
\end{figure*}

\subsubsection{Nebular \hetwo\ Emission}
Decomposition of the \hetwo\ line profile indicates that about 25\% of the flux arises from unresolved nebular emission, for which there is no evidence in the composite spectra at either \ztwo\ or \zthree.  This is an immediate indication of an unusually hard ionizing spectrum, with significant numbers of photons with energies greater than 54.4 eV.  Nebular \hetwo\ emission is occasionally observed in local W-R galaxies, and appears to occur primarily in lower metallicity systems \citep{ctm86,gkc+91,clt04,bkd08}.  Even locally, however, explanations for the source of the ionizing photons differ.  Models of the ionizing fluxes of W-R stars \citep{snc02} show that flux above 54 eV is a strong function of metallicity; in the weaker, less dense winds of metal-poor W-R stars, He$^{++}$ does not recombine as it does in a denser wind, and significant flux is emitted at energies above the He$^+$ edge.  The models of \citet{snc02} also predict that nebular \hetwo\ emission will be seen only immediately after an instantaneous burst of star formation; for continuous star formation, the fraction of sufficiently energetic photons is so small that nebular \hetwo\ emission is not expected to be detectable at any metallicity.  \citet{bkd08} also investigate the origin of nebular \hetwo\ emission, using a large sample of galaxies showing W-R features from the Sloan Digital Sky Survey.  Using timescales derived from \Hb\ equivalent widths, they argue that nebular \hetwo\ emission arises earlier than would be expected if its primary source were W-R stars, and that therefore low metallicity O stars must produce significant numbers of photons above 54 eV.  At least qualitatively speaking, either of these explanations can probably account for the nebular emission in BX418.  Its low metallicity should allow the production of He-ionizing photons from low metallicity stars, and the unusual strength of the stellar \ion{He}{2} emission suggests that we are observing a strong burst of star formation.  

We further investigate the conditions needed to produce significant nebular \hetwo\ emission with photoionization models from Cloudy.  As described in Section \ref{sec:cloudy}, we use an input ionizing spectrum from Starburst99, incorporating the ionizing fluxes of \citet{snc02} and using a metallicity $Z=0.004$.   We caution that in interpreting the Cloudy results, it is important to remember that the model output is only as reliable as the input, and the input ionizing spectrum above 54.4 eV should be considered highly uncertain.  \citet{snc02} note that the flux above 54.4 eV depends sensitively on the wind density and on the form of the law used to scale the winds with metallicity; this scaling law isn't known precisely even for O stars, and for W-R stars it is not known at all.  They also point out that the grid of models excludes the rare, hot W-R stars with weak winds that have significant ionizing fluxes above 54 eV; these are very rare in the Galaxy and LMC, but could be more significant at lower metallicities.  It is clear that a more thorough understanding of the source of these energetic photons locally is required in order to interpret results at high redshift.

With these cautions in mind, we proceed with the modeling assuming that 25\% of the \hetwo\ emission is nebular, for a flux relative to \Hb\ of log ($F_{\rm He II}/F_{\Hb})\simeq-1.1$.   The Cloudy models then predict the nebular \hetwo\ flux as a function of metallicity and ionization parameter.  The results of the modeling are shown in Figure~\ref{fig:cloudy_he2}: the \hetwo/\Hb\ ratio increases with both metallicity and ionization parameter, and for the metallicity of BX418, an extremely high ionization parameter $\log U\sim-1$ is required to account for the observed emission.   Interestingly, the increase of \hetwo/\Hb\  with metallicity and ionization parameter is {\it not} primarily driven by an increase in \hetwo\ luminosity.  Rather, at very high ionization parameters (log $U\gtrsim-2$) and in the presence of dust grains, hydrogen-ionizing photons are selectively depleted by energy-specific grain opacities peaking near 912 \AA, resulting in a decreasing \Hb\ flux with both ionization parameter and metallicity \citep{blm+98}.  The metallicity dependence arises because higher metallicity models are assumed to also have more grains.  The implications of this effect are discussed further in Section \ref{sec:disc_logu}; for now, we note that entirely dust-free models do not successfully reproduce the observed line ratios of BX418, and that in spite of the inferred low extinction such models seem unrealistic given the metallicity of $\sim1/6$ \zsun\ and the known correlation between metallicity and dust content.   Also recall that log $U\gtrsim-2$ is required by the lower limit on the \Othree/\Otwo\ ratio (Section \ref{sec:metal}).

Log $U\sim-1$ is much higher than the ionization parameters inferred for more typical \ztwo\ galaxies (which are in turn higher than those inferred for local star-forming galaxies); for example, \citet{hsk+09} use the \Othree/\Otwo\ ratio of two lensed and apparently representative \ztwo\ galaxies to infer $\log U\sim-2.5$.  While unusual, this high ionization parameter is not inconsistent with other observations, and may be expected from a compact and energetic ionizing source such as a cluster of massive, low metallicity stars.  We discuss this high ionization parameter further in Section \ref{sec:disc_logu}, and emphasize in conclusion that regardless of the uncertainties associated with the input ionizing spectrum and the photoionization modeling, the fact remains that the \hetwo\ emission seen in BX418 is exceptionally strong and indicative of an ionizing spectrum that must be significantly harder than that seen in more typical \ztwo--3 galaxies. 

\subsection{Nebular Emission Lines, Direct Metallicity and C/O Abundance}
\label{sec:nebularuv}
The UV spectrum of BX418 is also notable for its strong nebular emission from \Othreeuv\ $\lambda\lambda$1661, 1666 \AA\ and \Cthree\ $\lambda\lambda$1907, 1909 \AA.  Inspection of Figure~\ref{fig:specpanels} shows that the \Cthree\ emission is far stronger than that observed in the composite \ztwo\ spectrum, while \Othreeuv\ emission is not present in the composite at all (it is, however, detected in the \zthree\ composite of \citealt{ssp+03}, and is noticeably stronger in their subset of galaxies with strong \lya\ emission; see Section \ref{sec:z23comp}).  We use the Cloudy models to investigate the conditions under which such strong \Othreeuv\ and \Cthree\ emission arises.

The photoionization models show that \Othreeuv\ emission increases strongly with ionization parameter, and decreases with metallicity; at $Z=0.5$ \zsun\ and $\log U=-2.5$,  the combined strength of the \Othreeuv\ doublet is roughly $\sim5$\% of the flux of \Hb, while for $Z=0.2$ \zsun\ it reaches the observed flux of $\sim25$--30\% of \Hb\ for ionization parameters ranging from $\log U \sim -1.6$ to $-0.8$.   This is only partially dependent on the decrease in \Hb\ at very high ionization parameters discussed above; much of the increase in the \Othreeuv\ flux with ionization parameter occurs well before the photon depletion effect becomes significant.  This result is consistent with the high ionization parameter inferred from the \hetwo\ emission, and accounts for the lack of \Othreeuv\ emission in the \ztwo\ composite as well.  We also note that the detection of the \Othreeuv\ line is complicated by the presence of \ion{Al}{2} absorption at 1670 \AA; the line is undoubtedly more easily detected when the low ionization absorption lines are weak, as in BX418, or in higher resolution spectra.

The ratio \Othreeuv\ $\lambda\lambda$1661, 1666/\Othree\ $\lambda$5007 is sensitive to the electron temperature.  Using the calibration of \citet[][see their Figure 1]{vcd04}, we find $T_e\simeq15,000^{+500}_{-700}$ K, in agreement with the best fitting photoionization models, which also have $T_e\simeq15,000$ K.  This determination of the electron temperature allows a direct measurement of the metallicity via the $T_e$ method; following \citet{ism+06}, we find $12+\log({\rm O/H})=7.8\pm0.1$, in agreement with the $R_{23}$-derived metallicity of $12+\log({\rm O/H})=7.9\pm0.2$ (Section \ref{sec:metal}).

The modeling also shows that \Cthree\ emission is strongly dependent on both metallicity and ionization parameter, with the line strength increasing with ionization parameter at all metallicities, peaking at $Z\sim0.2$ \zsun, and decreasing at both higher and lower metallicities.  Thus BX418 appears to have the optimal conditions for strong \Cthree\ emission, with an extremely high ionization parameter and a metallicity that is low but not too low. Further investigation of the models shows that the metallicity dependence of \Cthree\ emission is largely a dependence on the abundance of carbon, which we study in more detail below.

Other high ionization nebular emission lines are occasionally seen in high redshift galaxy spectra.  Weak \ion{N}{4}] $\lambda$1487 emission is seen in cB58 \citep{psa+00}, but because of the intervening absorption feature discussed in Section \ref{sec:intervene} we are unable to tell whether or not it is present in BX418.  The Cloudy photoionization models predict weak \ion{N}{4}] emission, at the level of a few percent of \Hb, depending on the N/O ratio.  We have also checked for the presence of \ion{N}{3}] $\lambda1750$ emission, finding an upper limit of F(\ion{N}{3}]/\Hb)$<0.1$.  Unfortunately this is not sensitive enough to constrain the nitrogen abundance; for all metallicities and all values of N/O considered, the Cloudy models predict F(\ion{N}{3}]/\Hb)$<0.1$.  We discuss the possibility of nebular \ion{C}{4} emission in Section \ref{sec:windlines}.

\begin{figure*}[htbp]
\plotone{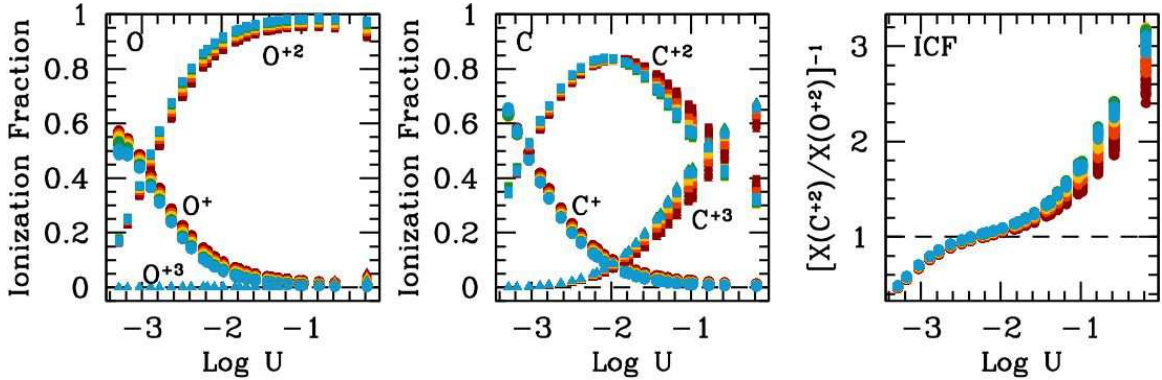}
\caption{The ionization states of oxygen (left panel) and carbon (middle panel) as a function of ionization parameter, from the Cloudy models.  The right panel shows the ionization correction factor (see Equation \ref{eq:icf}) as a function of ionization parameter.  The colors represent metallicities ranging from $Z=0.05$ \zsun\ (blue) to $Z=0.5$ \zsun (red), as labeled in Figure \ref{fig:cloudy_he2}.}
\label{fig:icf}
\end{figure*}

 \subsubsection{The C/O Abundance Ratio and the Origin of Carbon}
The origin of carbon is a subject of some controversy.  It is produced primarily through He burning via the triple $\alpha$ reaction, which may occur in both massive ($M>8\; \msun$) and low to intermediate mass ($1<M<8$ \msun) stars; however, there is considerable disagreement over the mass range of the stars responsible for most carbon production.  The C/O ratio is observed to increase with increasing O/H in both individual stars and in galaxies and H II regions.  This trend has been studied by many authors, using a variety of chemical evolution models and stellar yields.  Some find that the data are best explained by C arising almost exclusively from massive stars, in which case the trend in C/O vs.\ O/H is due to metallicity-dependent stellar winds, as mass loss and ISM enrichment are greater at higher metallicities \citep{hek00}.  Others \citep{crm03} reach the opposite conclusion, finding that carbon production is dominated by low to intermediate mass ($1<M<3$ \msun) stars, and that the C/O vs.\ O/H trend is therefore an evolutionary effect due to the delayed release of carbon relative to oxygen (which is produced almost exclusively by massive stars) in younger and less metal-rich systems.  \citet{cpeg05} find that massive and low to intermediate mass stars contribute roughly equal amounts of carbon in the solar vicinity, while \citet{acn+04} find that the metallicity-dependent mass loss of high mass stars is the primary factor in producing the C/O vs.\ O/H trend, with some secondary contribution from delayed C production in lower mass stars.

Measurement of the C/O ratio in a very young starburst can provide valuable discrimination between these models, since any time delay in carbon production will be most pronounced at early times and with high star formation rates.  We estimate the C/O abundance of BX418 via the ratio of \Othreeuv\ $\lambda\lambda$1661, 1666 \AA\ to \Cthree\ $\lambda\lambda1907$, 1909 \AA, following the method described by \citet{gsd+95} and \citet{ssp+03}, who derive the C$^{+2}$/O$^{+2}$ ionic abundance ratio from the emission-rate coefficients for collisionally excited emission lines:
\begin{equation}
\frac{\rm C^{+2}}{\rm O^{+2}} = 0.15\; e^{-1.1054/t}\, \frac{I(\rm C\; \textsc{iii}] \;\lambda\lambda 1907, 1909)}{I(\rm O\; \textsc{iii}]\;\lambda\lambda 1661, 1666)},
\label{eq:co}
\end{equation}
where $t=T_e/10^4$ K.  Note that this expression differs slightly from that of \citet{gsd+95}, which includes only \Othreeuv\ $\lambda 1666$; we include both members of the \Othreeuv\ doublet.  This ionic abundance ratio will be equal to the true C/O abundance only if C and O have the same relative ionization.  This may be approximately true in many circumstances,  but is not necessarily the case.  O$^+$ and O$^{+2}$ have higher ionization potentials than C$^+$ and C$^{+2}$ (35.1 and 54.9 eV vs.\ 24.4 and 47.9 eV respectively), so in regions ionized by a hard ionizing spectrum, as is apparently the case in BX418, a significant amount of carbon may be in the form of C$^{+3}$, and the C$^{+2}$/O$^{+2}$ ionic abundance ratio would then underestimate the true C/O abundance.  In order to correct for this effect it is necessary to apply an ionization correction factor (ICF) as described by \citet{gsd+95}:
\begin{eqnarray}
\frac{\rm C}{\rm O} &=& \frac{\rm C^{+2}}{\rm O^{+2}} \left[ \frac{X(\rm C^{+2})}{X(\rm O^{+2})} \right]^{-1} \nonumber\\
&=& \frac{\rm C^{+2}}{\rm O^{+2}} \times \rm ICF ,
\label{eq:icf}
\end{eqnarray}
where $X(\rm C^{+2})$ and $X(\rm O^{+2})$ are the C$^{+2}$ and O$^{+2}$ volume fractions, respectively.

We use the Cloudy models to examine the ionization states of C and O, and estimate the ionization correction factor as a function of ionization parameter.  The results are shown in Figure \ref{fig:icf}.  The two left panels show the ionization fractions of O and C as a function of ionization parameter; the difference due to oxygen's higher ionization potential is apparent, as for $\log U \gtrsim -2$, nearly all the oxygen is in the form of O$^{+2}$, while the fraction of C$^{+2}$ declines at $\log U > -2$, becoming equal to the fraction of C$^{+3}$ around $\log U \sim -1$.  We translate these fractions into the ionization correction factor in the right panel of Figure \ref{fig:icf}.  For moderate ionization parameters around $\log U \sim -2$, the ICF $\sim 1$ (with little dependence on metallicity), and C$^{+2}$/O$^{+2}$ will be a good approximation for C/O.  Indeed, in the sample of dwarf galaxies studied by \citet{gsd+95}, the range of ICF found is only 1.06--1.33.  On the other hand, with increasing ionization parameter the ICF becomes more significant, and the metallicity dependence increases as well.  

There is considerable uncertainty associated with the derivation of the ICF for BX418.  Our primary constraint on the ionization parameter comes from the \hetwo\ flux; this determination of $\log \;U$ relies heavily on the uncertain input ionizing spectrum used by the models, as discussed in Section \ref{sec:he2}.  For $Z=0.2$ \zsun, the best-fitting models require approximately $\log U = -1.0 \pm 0.15$.  The corresponding ionization correction factor is therefore $\rm ICF = 1.7 \pm 0.2$.  For the electron temperature we adopt $T_e=15,000^{+500}_{-700}$ K as indicated by the \Othreeuv\ $\lambda\lambda$1661, 1666/\Othree $\lambda$5007 ratio.  Including these uncertainties in the ICF, in $T_e$, and the flux uncertainties in the \Cthree\ and \Othreeuv\ lines, we find $\rm{C/O} = 0.24 \pm 0.05$, and $\log \rm{(C/O)} = -0.62^{+0.08}_{ -0.1}$.  This is in agreement with the input C/O ratio of the best-fitting Cloudy models, all of which have $\rm{C/O} = 0.19$.

We show the $\log \rm{(C/O)}$ vs.\ $\log \rm{(O/H)}$ diagram in Figure~\ref{fig:co_oh}, using the oxygen abundance determined by the electron temperature method for consistency with local observations.  BX418 is shown by the large magenta circle, and the remaining symbols show local galaxies as detailed in the caption.  The shaded region shows the range of C/O determined from the LBG composite spectrum by \citet{ssp+03}, $\log \rm{(C/O)} = -0.68\pm0.13$; there is no reliable O/H measurement for this set of galaxies, and no ionization correction was applied.  \citet{ssp+03} also found a somewhat lower value of C/O for their subset of galaxies of galaxies with strong \lya\ emission, $\log \rm{(C/O)} = -0.74\pm0.14$, again with no ionization correction applied (for comparison, we find $\log \rm{(C^{+2}/O^{+2})} = -0.87$ without the ionization correction for BX418).

\begin{figure}[htbp]
\plotone{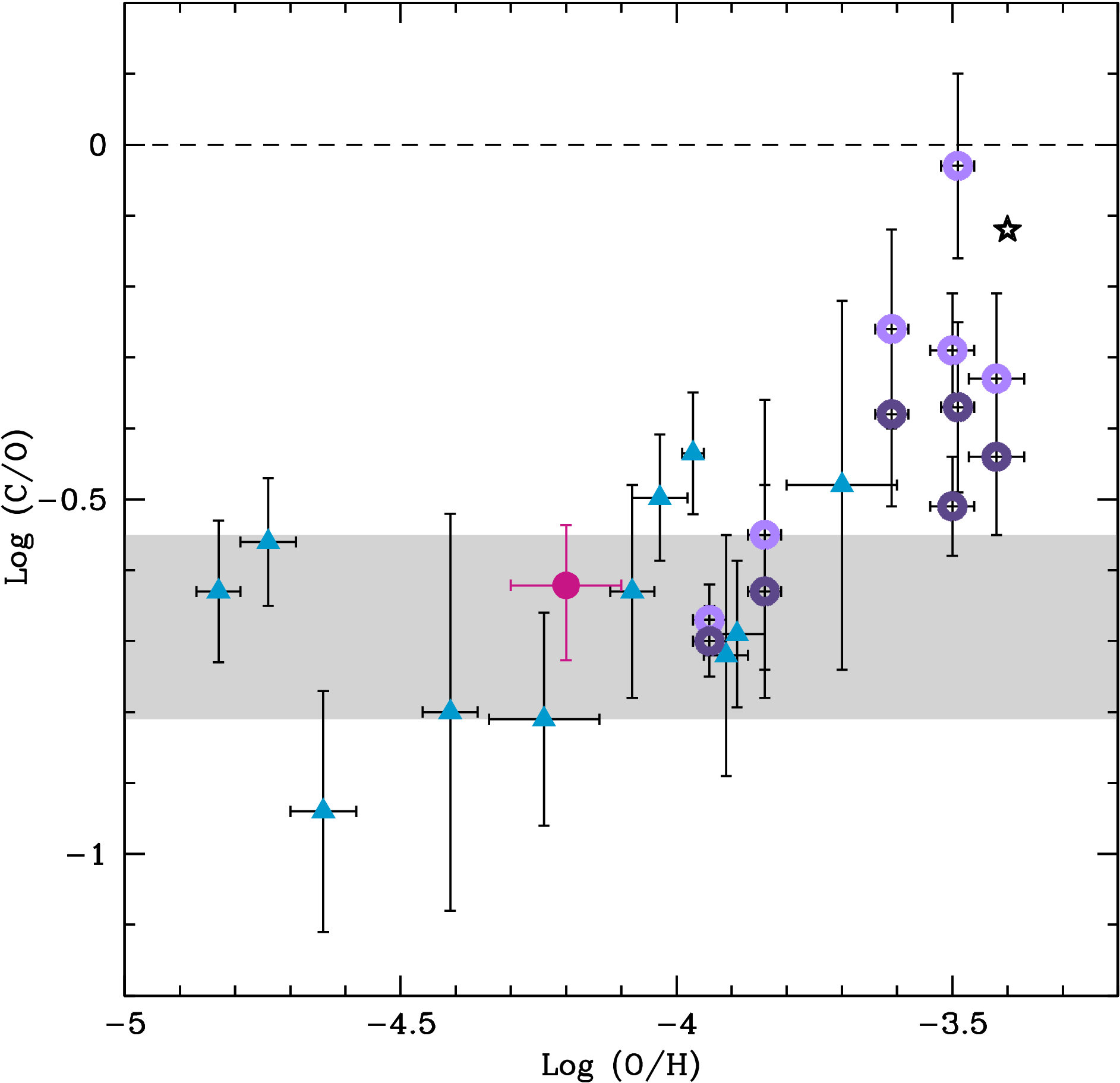}
\caption{The carbon-to-oxygen ratio vs.\ the oxygen abundance.  BX418 is shown by the large magenta circle.  The other points show values for local galaxies.  Blue triangles are local dwarf irregular galaxies from \citet{gsd+95,gss97} and \citet{ks98}, and the open light and dark purple circles show local spirals from \citet{gsp+99}, with two different extinction corrections.  The black star shows the default value of the \ion{H}{2} region abundance set used by Cloudy, corresponding to the Orion nebula;  the most recent estimate of this ratio in Orion is log (C/O) $ = -0.21$ \citep{epg+04}.  The shaded region shows the value determined from the \zthree\ LBG composite spectrum by \citet{ssp+03}.}
\label{fig:co_oh}
\end{figure}

BX418 shows excellent agreement with the local trend in C/O vs.\ O/H, in spite of its differences with respect to the local sample.  The local objects it most resembles in abundance ratios are \ion{H}{2} regions of dwarf emission line galaxies \citep{gsd+95}. While a detailed comparison with such objects is difficult, local \ion{H}{2} galaxies probably have more extended star formation histories than BX418, extending to at least 1 Gyr and consisting of short bursts alternating with quiescent periods (\citealt{tsrt04,wctk04,pghd10}; we assess the possibility of an earlier episode of star formation in BX418 in Section \ref{sec:sed} and find that it is a poorer fit to the observed SED than a single stellar population, but we cannot rule out this possibility entirely).  The star formation activity of these galaxies is also less extreme than that seen in BX418, as indicated by their lower ionization parameters \citep{pd03,pghd10}.  If a delay in carbon production by low mass stars were the primary cause of the increase in C/O with O/H, we would likely see a significantly lower C/O ratio in a very young, extreme starburst.  Furthermore, the best-fit age for BX418, $\sim38$ Myr, allows only massive stars with $M\gtrsim10$ \msun\ to have left the main sequence; an age of 100 Myr allows for carbon enrichment from stars with $M\gtrsim6$ \msun.  It is therefore unlikely that carbon production is dominated by $1<M<3$ \msun\ stars as suggested by \citet{crm03}.  Rather, we concur with \citet{acn+04}, who conclude that the apparent universality of the metallicity dependence of C/O in objects of widely varying star formation histories is more likely to be due to variable stellar yields than to evolutionary effects.

\subsection{Stellar Wind Lines}
\label{sec:windlines}
The most prominent stellar features in UV spectra of high redshift galaxies are generally the wind lines from massive stars; the high ionization lines of \ion{N}{5} $\lambda1240$, \ion{Si}{4} $\lambda\lambda1394$, 1402 and \ion{C}{4} $\lambda\lambda1548$, 1551 show distinctive P Cygni profiles indicative of stars with masses of at least 30 \msun\ \citep{lrh95}.  These lines depend sensitively on the properties of the massive stellar population, so in principle they may provide a great deal of information.  In practice, however, their interpretation is difficult.  The strength of these features depends on metallicity, the age of the starburst, the slope and upper mass cutoff of the IMF, and possibly on time-dependent dust extinction as well \citep{lcm02}.  The lines themselves are complex blends of stellar wind and photospheric features and interstellar absorption; nebular emission may also contribute to the line profiles.  The situation is further hampered by our lack of detailed understanding of the scaling of stellar winds with metallicity, and the limited number of observations of massive stars at low metallicities.  Given these substantial uncertainties, and the relatively low resolution of the spectrum which precludes accurate separation of the stellar and interstellar lines, we cannot draw robust conclusions from the stellar wind features in the UV spectrum of BX418.  Nevertheless these lines show intriguing differences from typical high redshift galaxy spectra, as we show below.

Figure \ref{fig:pcyg_comp} shows the \ion{N}{5} $\lambda1240$, \ion{Si}{4} $\lambda\lambda1394$, 1402 and \ion{C}{4} $\lambda\lambda1548$, 1551 lines in the spectrum of BX418, with the composite spectrum of $\sim1000$ \ztwo\ galaxies overplotted (in red) for comparison.  For all  three lines, the most notable difference between BX418 and the composite spectrum is the strength of P Cygni emission, which is stronger in BX418 in each case (the emission occurs at 1244, 1405 and 1553 \AA\ in the \ion{N}{5}, \ion{Si}{4} and \ion{C}{4} lines respectively).  The broad, blueshifted absorption, most apparent in \ion{N}{5} and \ion{C}{4}, is very similar in BX418 and the composite (the differences in the narrower interstellar absorption lines of \ion{Si}{4} and \ion{C}{4} are discussed in Section \ref{sec:interstellar}).  

\begin{figure*}[htbp]
\plotone{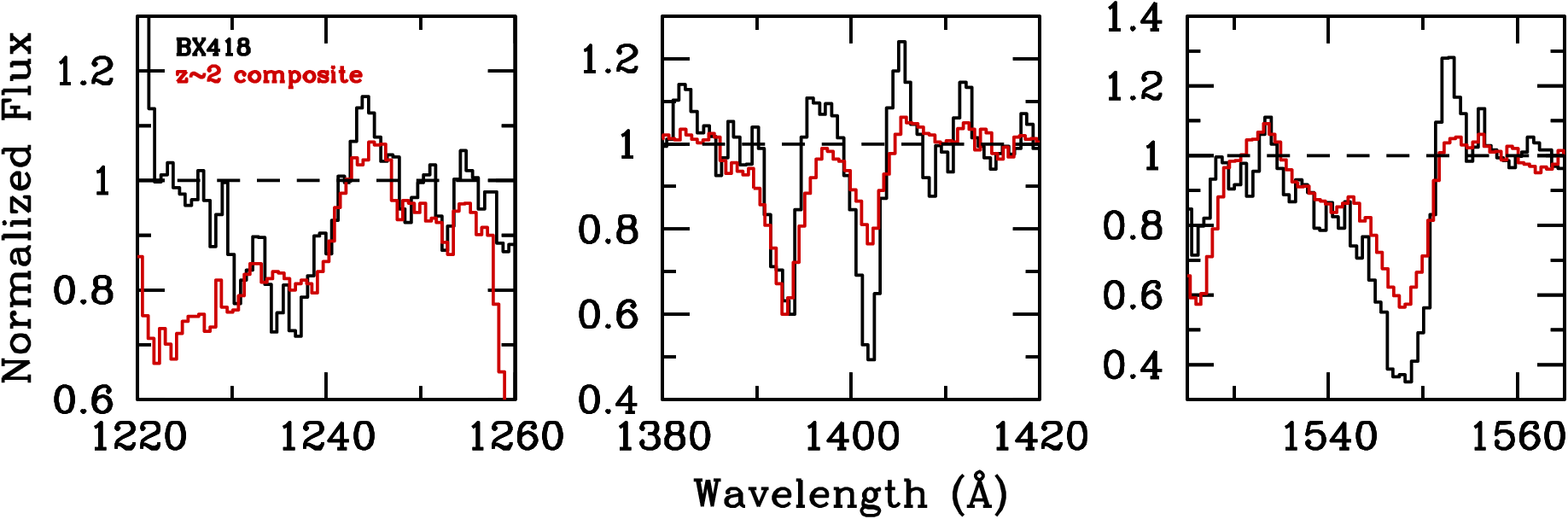}
\caption{Profiles of the stellar wind lines \ion{N}{5} $\lambda1240$ (left panel), \ion{Si}{4} $\lambda\lambda1393$, 1402 (middle panel), and \ion{C}{4} $\lambda\lambda1548$, 1551 (right panel).  The spectrum of BX418 is shown in black, and the composite spectrum of $\sim1000$ \ztwo\ galaxies is shown in red for comparison.}
\label{fig:pcyg_comp}
\end{figure*}

\begin{figure*}[htbp]
\plotone{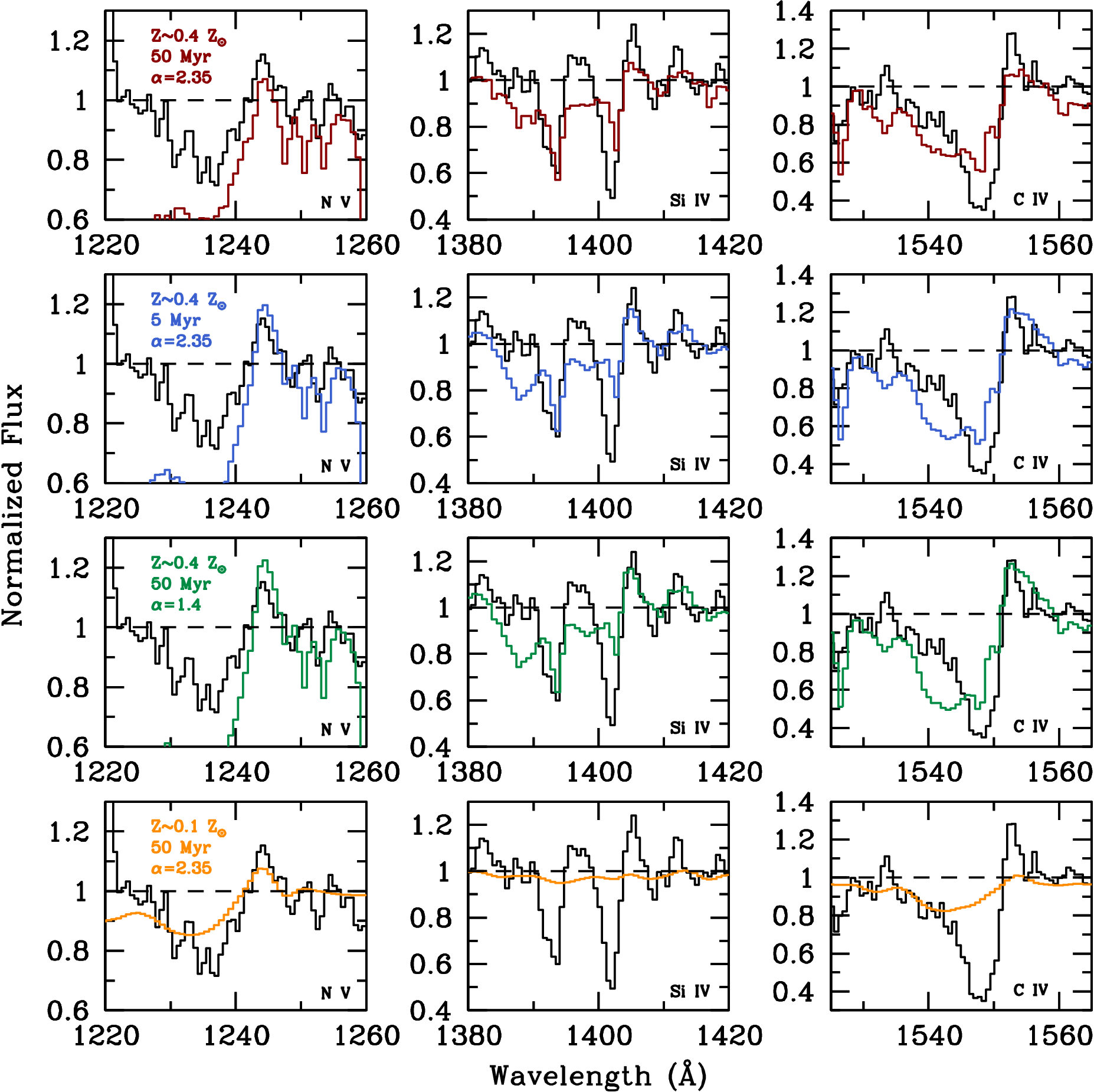}
\caption{A comparison of the profiles of the wind lines \ion{N}{5} $\lambda1240$ (left panels), \ion{Si}{4} $\lambda\lambda1393$, 1402 (middle panels), and \ion{C}{4} $\lambda\lambda1548$, 1551 (right panels) with synthetic spectra.  The top three rows are continuous star formation Starburst99 models using the composite LMC/SMC metallicity of $Z\sim0.4$ \zsun.  The top row (dark red model) shows a 50 Myr old starburst and an IMF with Salpeter slope and upper mass cutoff of 100 \msun.  The second row (blue) shows a 5 Myr old starburst and an IMF with Salpeter slope and upper mass cutoff of 100 \msun, and the third row (green) shows a 50 Myr old starburst with a flatter IMF slope of $\alpha=1.4$.  The bottom row (orange) shows a model from \cite{rpl+04} with $Z\sim0.1$ \zsun, an age of 50 Myr, and a Salpeter IMF; this model was chosen to match the depth of the broad absorption in the \ion{C}{4} profile. The spectrum of BX418 is shown in black in all panels.}
\label{fig:pcyg_mod}
\end{figure*}

Broadly speaking, the behavior of these lines with varying stellar population properties is understood: the P Cygni emission is strongest in a young stellar population of 3--5 Myr and stabilizes by $\sim20$ Myr, increases with any variation of the IMF that increases the number of massive stars (a flatter slope or higher upper mass cutoff), and decreases at lower metallicities.  The strength of absorption depends primarily on the density of the stellar wind and therefore on the mass loss rate.  Mass loss rates are lower at lower metallicities, and thus the absorption troughs are weaker at lower metallicities.  The absorption profiles also show a mild age dependence, particularly in the case of \ion{C}{4}, with stronger absorption in younger starbursts.  However, significant work remains to be done in quantifying these dependencies and in creating a wide range of models for comparison with observation.  The flexible Starburst99 models \citep{sb99} allow the creation of synthetic UV spectra with a variety of ages and IMFs, but include only two metallicities, solar and the average LMC/SMC metallicity of $Z\sim0.4$ \zsun.  More recently, \citet{rpl+04} combined Starburst99 with the model atmosphere code WM-BASIC to create a suite of synthetic spectra covering a wider range of metallicities; however, this work was focused primarily on the stellar photospheric lines, and the authors caution that a more detailed study of the wind lines is still needed.  Because of the focus on photospheric lines, processes such as shock emission that affect only the high ionization wind lines were not included in the models.  These limitations have been addressed to some extent in the latest generation of the WM-BASIC code \citep{lob+10}. However, the wind line profiles generated by running Starburst99 with this latest version of WM-BASIC do not differ appreciably from those used in the present
work, and none of the discussion below is altered by the use of these latest models.

We compare the profiles of BX418's stellar wind lines with various models in Figure \ref{fig:pcyg_mod}.  The top three rows are Starburst99 LMC/SMC metallicity models with $Z\sim0.4$ \zsun; this is a factor of $\sim2$ higher than the metallicity of BX418, and these models may be expected to have stronger emission and absorption than comparable models of lower metallicity.  Nevertheless they are illustrative of the behavior of the wind lines with variations in age and IMF.  The top panels of Figure \ref{fig:pcyg_mod} (dark red lines) show the standard model which has been shown previously to give generally good agreement with high redshift galaxy spectra; this model has a Salpeter slope and an upper mass cutoff of 100 \msun, and an age of 50 Myr (other studies have assumed an age of 100 Myr; there is almost no difference between the two models).   The second set of panels (blue lines) show a model with the same IMF but a younger age of 5 Myr, and the third row (green lines) shows a 50 Myr old spectrum with a flatter slope of $\alpha=1.4$, resulting in more massive stars.  The primary effect of these two variations is to strongly increase the predicted emission.  The bottom row shows an interpolation between the $Z=0.05$ \zsun\ and $Z=0.2$ \zsun\ models of \citet{rpl+04}, to create a $Z\sim0.1$ \zsun\ model in which the depth of the broad  \ion{C}{4} absorption profile approximately matches observations.  Note that this model predicts no emission, except in the \ion{N}{5} line.  As we discuss further below, none of the models considered simultaneously match the strengths of the absorption and emission components of the P Cygni features.

As noted above and shown in Figure \ref{fig:pcyg_comp}, the depth of absorption is almost identical in BX418 and in the \ztwo\ composite spectrum; this is surprising given that the latter almost certainly has a significantly higher average metallicity.  The absorption profiles do also depend on age, but models indicate that the strength of the absorption is much more sensitive to metallicity than to age.  It is possible that differences in the average age of the stellar populations cancel out the differences in metallicity to create nearly identical absorption profiles, but this would be a surprising coincidence.  More likely, the fact that the  $Z\sim0.1$ \zsun\ \citet{rpl+04} model matches the composite spectrum as well as BX418 is an indication that our understanding of the age and metallicity dependence of these lines remains incomplete.

The $Z\sim0.1$ \zsun\ model matching the depth of the P Cygni absorption predicts no emission in either the \ion{C}{4} or \ion{Si}{4} lines, clearly at odds with observations.   Standard models with a Salpeter slope and an age of at least 50 Myr also underpredict the emission component of the lines.  Models matching the strong emission require either a very young age (Figure \ref{fig:pcyg_mod}, blue model in second row), or an IMF with more massive stars (green model, third row; note that both of these models overestimate the metallicity of BX418, so even more extreme variations may be required to match the observed emission).  As discussed in Section \ref{sec:he2}, a young burst superimposed on a somewhat older population may be required to produce the strong stellar \ion{He}{2} emission; therefore a young age such as that seen in the second row of Figure \ref{fig:pcyg_mod} may not be unreasonable.  Perhaps the simplest explanation, however, and the one most likely to account for both the absorption and emission, is that the \ion{C}{4} and \ion{Si}{4} emission is at least in part nebular.  The best fitting Cloudy models  predict \ion{C}{4} emission at the level of $\sim8$\% of \Hb\ and $\sim15$\% of \Cthree\ (see also \citealt{lcm02}).  This is consistent with observations, although we may expect nebular emission to be difficult to detect when superimposed over the strong interstellar and stellar absorption; on the other hand, as resonance lines, these transitions are subject to multiple scatterings, and like \lya, they may be most likely to escape the galaxy when they are redshifted away from their resonant wavelength.  This would effectively mimic the stellar P Cygni emission profile.  No nebular \ion{N}{5} emission is predicted by the models, but the \citet{rpl+04} models predict stellar \ion{N}{5} emission even at metallicities when there is none from \ion{C}{4} or \ion{Si}{4}.  

Finally, we note that the failure of models to reproduce all of the features of the wind line profiles is not unique to BX418.  Most previous studies of UV spectra of high redshift galaxies have noted either an overprediction of the absorption \citep{psa+00, ssp+03} or an underprediction of the emission \citep{qpss09}.  Metallicity effects and nebular emission have been invoked to account for the discrepancies.  Future observations of massive stars over a wide range of metallicities and improved models of stellar winds across this metallicity range will shed further light on the interpretation of these complex features.

\subsection{Fine Structure Emission Lines}
\label{sec:fs}
One of the puzzles of the \zthree\ LBG composite spectrum studied by \citet{ssp+03} was the relatively strong emission from the excited fine structure lines \ion{Si}{2}*, at 1265, 1309, and 1533 \AA.  These lines are clearly visible in the spectrum of BX418 and in the \ztwo\ composite; both spectra are shown in Figure \ref{fig:specpanels}, where the lines are marked.  BX418 also shows significant emission from \ion{C}{2}* $\lambda$1335, which \citet{ssp+03} did not detect, probably because of strong \ion{C}{2} $\lambda$1334 absorption.  Similarly, we see strong \ion{C}{2} $\lambda$1334 absorption in the \ztwo\ composite, but no \ion{C}{2}* emission.  Presumably the weakness of the low ionization interstellar absorption lines in BX418 allows its detection here.  These lines are not detected in the UV spectra of local starbursts, as discussed by \citet{ssp+03} and shown more recently by \citealt{smc+06} (see their Figure 13, which shows a clear comparison with the LBG composite spectrum).  Although the lines are weak, they are marginally stronger in BX418 than in the composite; this may be due to the stronger absorption in the composite, or it may suggest that the processes that give rise to these lines in high redshift  but not local starbursts are even stronger in BX418. 

\citet{ssp+03} attempted to reproduce these lines with Cloudy photoionization models but were unable to do so, finding that all models which satisfactorily reproduced other observed lines underpredicted these lines by at least an order of magnitude.  They concluded from this result that the lines are unlikely to originate in photoionized \ion{H}{2} regions.  We also attempt to use our Cloudy models to reproduce these lines, but similarly find that none of the models considered (whether they match the other properties of BX418 or not) come close to the observed line strengths.  Even assuming that all of the emission from the \ion{Si}{2} and \ion{C}{2} multiplets arises at the excited fine structure wavelength, models which reproduce other observations underpredict the fine structure emission by factors of 5--10 depending on the line.  We conclude that if these lines do arise from the \ion{H}{2} regions, either we are not modeling the regions correctly or some factor affecting these lines is not treated accurately by the models.

The relative velocity of the lines may offer additional clues to their origin.  In the \zthree\ composite, these lines are redshifted by $\sim100$ \kms, though they may be biased to the red by absorption on the blue side of the lines.  The \ion{Si}{2}* $\lambda$1265, $\lambda$1309 and \ion{C}{2}* $\lambda$1335 lines are also detected in the higher resolution spectrum of the lensed galaxy cB58, where they are again redshifted by $\sim200$ \kms\ \citep{psa+00}.  The lines are consistently redshifted in BX418 as well, by $\sim200$ \kms, and in this case the weakness of the absorption makes significant biasing of the line centers less likely.  This consistent redshift suggests an origin in the outflowing gas, perhaps through resonant scattering of photons originating in the \ion{H}{2} regions (though, as noted above, the predicted emission from \ion{H}{2} regions is not strong enough).  \citet{ssp+03} also considered the possibility of the lines arising in the outflow, but concluded that this explanation is not fully satisfactory because of the narrowness of the lines.  The outflowing gas is apparently optically thin in these transitions, 
and therefore the lines should span the full range of velocities in the outflow, $\sim1000$ \kms\ or more.  Instead they are quite narrow.  Similar arguments apply to BX418, where the \lya\ emission spans a range of over 2000 \kms, but the fine structure lines are unresolved.  However, it is also possible that the line strength varies as a function of velocity, and limited S/N prevents us from detecting emission except where it is strongest.  The origin of these lines remains a puzzle.

\section{Discussion}
\label{sec:disc}
\subsection{Ionization Parameter}
\label{sec:disc_logu}
A very high ionization parameter of $\log U \sim-1$ is suggestive of large numbers of massive, low metallicity stars forming in a small volume, in a configuration that must be considerably more extreme than anything observed locally.  The source of the ionizing photons is unlikely to be an AGN, as indicated by the low \ion{C}{4}/\Cthree\ ratio and the nondetection of BX418 in deep 8 and 24 \micron\ images.   

The factors controlling the ionization parameter of an \ion{H}{2} region are discussed in detail by \citet{dfs+06}.  The ionization parameter increases significantly with decreasing metallicity, because of the harder ionizing spectra and weaker winds of low metallicity stars.  It also increases as the density of the \ion{H}{2} region increases relative to that of the surrounding medium, and has a weak inverse coupling with the pressure of the surrounding medium and a weak positive coupling with the mass of the star cluster.  \citet{dfs+06} model the probability distributions of ionization parameters in local \ion{H}{2} regions, finding a maximum value nearly an order of magnitude lower than the log $U\sim-1$ we infer for BX418.  Similarly, \citet{kds+01} model the metallicities and ionization parameters of starburst galaxies, and consider a maximum value of $\log U =-2$.  \citet{fgl+01} have estimated the ionization parameter in the star-forming regions of the starburst galaxy M82, finding $\log U \approx -2.3$ on spatial scales of tens to hundreds of kpc.

Although measurements of the emission lines needed to determine the ionization parameter are few at high redshift, the values inferred similarly do not approach that of BX418, although they are higher than the ionization parameters of local galaxies of similar metallicities \citep{pss+01, mng+08,hsk+09}.  There is an exception, however, in the unusual emission line spectrum of the gravitationally lensed Lynx arc at $z=3.36$ \citep{fvh+03,vcd04}.  This object shows strong, narrow high ionization emission lines in the UV spectrum, including all of the lines observed in emission in BX418 and several others as well.  Unlike BX418 it shows little to no continuum, so the interstellar absorption features we observe are not present.  It is interpreted as a low mass \ion{H}{2} galaxy undergoing a very young burst of low metallicity star formation, with an ionization parameter $\log U \sim-1$.  Some of the line ratios are unlike those of BX418, however; in particular, \ion{C}{4} emission in the Lynx arc is 3.7 times that of \Hb\ (though this relies on a significant absorption correction), whereas we observe $<10$\% of \Hb\ at most.  The spectrum of BX418 appears to be intermediate between that of the Lynx arc and more typical high redshift galaxies.

The difference in ionization parameter between BX418 and that of more typical objects is more likely to be due to geometry than metallicity, since local galaxies with metallicities even lower than BX418 do not have ionization parameters nearly as high.  We clearly have no data capable of constraining the geometry of star formation in BX418, so any conclusions we draw must be purely speculative.  Given the scalings of the factors contributing to the ionization parameter, however, the most likely scenario may be one in which the galaxy's massive stars are both numerous and densely packed enough that the galaxy is essentially a single large \ion{H}{2} region with an extremely intense radiation field.  The integrated spectra of more typical galaxies at \ztwo\ would then represent the superposition of individual \ion{H}{2} regions, with properties closer to those of the regions modeled locally.  Clearly both additional data and additional modeling are needed to confirm or refute such a scenario.

A potentially important implication of the extreme ionization parameter was raised in Section \ref{sec:he2}.   The Cloudy photoionization models indicate that at very high ionization parameters and in the presence of dust grains, hydrogen-ionizing photons are selectively depleted by energy-specific grain opacities peaking near 912 \AA, resulting in a decreasing flux in the hydrogen recombination lines with both increasing ionization parameter and increasing metallicity, assuming that galaxies with higher metallicities also have higher dust content \citep{blm+98}.   Modeling suggests that the effect begins to be significant at log $U\sim-2$, the minimum ionization parameter required by our measurements of \Cthree\ and \Othree/\Otwo\ (these constraints do not depend on the decreasing flux of \Hb, unlike that obtained from \hetwo).  The magnitude of the effect depends on both the amount and the size distribution of the dust grains, about which we know very little.  We have not corrected the observed line ratios for extinction, given the low value of $E(B-V)$; any correction applied would increase the fluxes of the UV lines relative to \Hb, thereby requiring an ionization parameter even higher than that which we infer already.  We have, however, allowed grains to be included in the photoionization models (Cloudy includes Orion nebula grains as part of the \ion{H}{2} region abundance set, and we have reduced them by varying factors to account for the low dust content of BX418).  We have not attempted to use the models to constrain the dust properties of BX418, given the strong dependence on geometry, but we cannot reproduce the observed line ratios with entirely dust-free models.  Such a situation is unlikely given the metallicity of $\sim1/6$ \zsun\ and the known correlation between metallicity and dust content.

For the grains included in the models, the hydrogen recombination line flux is reduced by a factor of $\sim2$ at log $U\sim-1$.  Such a reduction, while highly uncertain, would have significant implications.  Most obviously, the true star formation rate would be a factor of $\sim2$ higher than the value we have inferred from \Ha, and thus a factor of $\sim3$ higher than the SFR obtained from the UV continuum.  This is marginally inconsistent but not impossible; dust allowed by both the SED modeling and the \Ha/\Hb\ ratio combined with a still rising UV continuum flux due to a young age could produce such observed SFRs, especially when combined with the systematic and observational uncertainties.  The generally good agreement of SFRs derived from \Ha\ with those derived from all other available diagnostics \citep{ess+06stars,rep+10} indicates that this effect cannot be widespread at \ztwo, and indeed we do not expect it to be, given that typical \ztwo\ galaxies appear to have significantly lower ionization parameters.  For BX418, more precise determinations of the star formation rate from a variety of indicators are needed to determine the magnitude of this effect. 

The reduction in \Ha\ and \Hb\ flux would also affect any metallicity indicators employing those lines.  In the case of the $R_{23}$ indicator used here (which takes ionization parameter into account) the Cloudy models suggest that the effect is mild, since the input metallicity of the best-fitting models is in reasonable agreement with the $R_{23}$ metallicity calculated from the predicted line ratios of those models.  

It is primarily the nebular \hetwo\ emission which requires an ionization parameter as high as $\log U \sim -1$.  This is uncertain; not only is the \hetwo/\Hb\ ratio dependent on the diminution of the \Hb\ flux by grains, which are extremely difficult to constrain, the detection of nebular emission relies on deconvolution with the broader stellar emission.  As described in Section \ref{sec:he2}, the two component fit we have adopted provides a better fit to the line at the 99\% confidence level compared to a single, broad line, and it is obviously different in shape from purely stellar \hetwo\ emission of similar strength \citep{dds+09}.  Nevertheless we cannot rule out the possibility that the emission is purely stellar, with an unusual profile.  However, even if the \hetwo\ emission is stellar, there is significant additional evidence for an ionization parameter nearly as high: the lower limit on the  \Othree/\Otwo\ ratio indicates $\log U \gtrsim -2$, as does the strength of the \Cthree\ emission, while the strong \Othreeuv\ emission indicates $\log U \gtrsim -1.6$, with only mild dependence on the diminution of \Hb.  We therefore consider the high ionization parameter to be a robust result, subject to the (not inconsiderable) uncertainties of the input ionizing spectrum used for the photoionization modeling.

\subsection{Comparisons with Other Galaxies at \ztwo--3}
\label{sec:z23comp}
We have shown through comparison with the composite spectrum of $\sim1000$ \ztwo\ galaxies that BX418 is quite different from typical galaxies at this redshift.  We have also constructed a composite spectrum of the youngest galaxies in the \ztwo\ sample for comparison; this spectrum is generally quite similar to the composite of the full \ztwo\ sample, and the distinctive features of BX418 are not appreciably stronger in the young subset of \ztwo\ galaxies, suggesting that age alone is not enough to account for BX418's uniqueness.  A full study of the variation of the spectra of \ztwo\ galaxies as a function of stellar population properties or particular spectral features is beyond the scope of this paper and will be presented elsewhere.  However, it may be instructive to compare BX418 with the composite spectra of \zthree\ Lyman break galaxies.  \citet{ssp+03} divided the \zthree\ sample into quartiles based on the strength of \lya\ emission, and found that, compared to the subset of galaxies with the strongest absorption, the strongest \lya\ emitters have bluer UV slopes, weaker low ionization absorption lines, slightly stronger \hetwo\ emission, stronger emission from \Othreeuv, and stronger emission in the \ion{C}{4} stellar wind line.  These are all characteristics of BX418 (though the \hetwo\ emission of BX418 is much stronger than that seen in any of the \zthree\ composites), suggesting that BX418 lies at the extreme end of a continuum of galaxy properties that may be parameterized by the strength of \lya\ emission.  

Much has been written about the relationship of \lya\ emission to other galaxy properties and much remains to be understood, but perhaps the most reliable conclusion is that stronger \lya\ emission is associated with lower dust content (e.g.\ \citealt{vsat08,kse+10} and references therein),  and perhaps more specifically with dust geometry.  \citet{kse+10} find that strong \lya\ emission is more common in older galaxies at \zthree, and propose that this is because supernova-driven outflows have cleared out channels through which the \lya\ photons can escape in the older galaxies.  

Other recent results are suggestive of the importance of dust geometry as well.  \citet{rep+10} have recently investigated the dust properties of galaxies at \ztwo, using MIPS 24 \micron\ imaging; they find that while galaxies with young best-fit ages have redder UV continuum slopes, this does not necessarily mean that they are dustier.  Rather, they appear to have less dust than their UV slopes would imply, and are therefore better described by a steeper extinction law than more typical galaxies at \ztwo\ (BX418 has a bluer slope than most of these galaxies, and is inferred to have little dust content no matter which extinction law is applied).  The reasons for this are not fully understood; younger galaxies may have a larger dust covering fraction, more consistent with a foreground screen, or the size distribution of dust grains may be different in younger objects.  BX418 is both young and blue, and we have inferred a low covering fraction of neutral gas and dust; this may explain at least some of its differences from more typical young galaxies.  While much remains to be clarified regarding the processes contributing to all of the features we have observed in BX418, it does seem clear that the covering fraction of gas and dust is a major factor in modulating the emergent UV spectrum of galaxies at high redshifts.

\subsection{Implications for Higher Redshifts}
\label{sec:disc_hiz}
We have suggested that because of its low mass, young age, low metallicity and low dust content, BX418 is likely to be similar to many of the galaxies now being discovered at $5\lesssim z \lesssim 8$.  Detailed observations from future ground and space-based facilities are needed to confirm this claim, but if true the implications for high redshift galaxy formation and reionization may be significant.  The \lya\ profile of BX418 shows substantial blueshifted emission, which is a signature of \zthree\ galaxies with a significant Ly continuum escape fraction (Steidel et al., in preparation).  If BX418 is indeed similar to typical galaxies at higher redshifts, the implication is that the escape fraction may significantly increase with increasing redshift, as already shown between $z\sim1$ and \zthree\ \citep{stf+10} and predicted at higher redshifts \citep{wc09,rs10}.

The detection of strong \lya\ and \hetwo\ $\lambda1640$ emission in a high redshift object is suggestive of the presence of extremely low metallicity Population III stars, as discussed in detail by \citet{jh06}, and by \citet{fvh+03} in the context of the Lynx arc.  Given the many metal lines, our measured abundance of $\sim\frac{1}{6}$ \zsun, and the fact that we are able to reproduce the observed line ratios with a normal stellar population, we do not believe this scenario applies to BX418.  Indeed, models of Population III stars predict much higher \hetwo\ equivalent widths of at least $5\lesssim W_{\rm He\;II} \lesssim 80$ \AA, and \lya\ equivalent widths of several hundred \AA\ as well \citep{s03}.  

It is also worth noting that after \lya, the strongest emission line in the rest-frame UV spectrum of BX418 is not \hetwo\ but \Cthree\ $\lambda$1909.  This line is observable from the ground to $z\sim10$ and even beyond, and therefore future spectroscopic searches for galaxies at very high redshift may wish to target this line for unambiguous redshift confirmation.  Finally, study of BX418 may provide insight into the issue of correction of broadband magnitudes to account for nebular line emission; we find that \Ha\ emission contributes $0.4\pm0.1$ mag to the observed K-band magnitude, while combined \Othree\ $\lambda\lambda$4959,5007 and \Hb\ emission account for $0.45\pm0.4$ mag in the H-band.  Such corrections are important to consider when modeling the stellar populations of $z\sim7$--8 galaxies, for which \Othree\ $\lambda\lambda$4959,5007 and \Hb\ fall in the {\it Spitzer} IRAC bands (e.g.\ \citealt{lgb+10}).  Given the wider bandpasses of the IRAC bands, however, the correction for a similar galaxy at $z\sim7$--8 would not be as large as those we find for the $H$ and $K$ bands.

\subsection{Remaining Questions and Further Work}
Perhaps the most important question we are unable to address with current observations is how many of the unusual spectral properties of BX418 are characteristic of galaxies with similar colors and metallicities, and how many are due to a very short-lived phase in its evolution.  A larger sample of high quality observations of young, blue galaxies is needed to address this question, although this is a daunting prospect given their rarity and the substantial observing time required to obtain a spectrum of the S/N presented here.

Local observations and improved modeling will also help to address this question.  The expected strength of \hetwo\ emission from W-R stars as a function of metallicity and age is still not well known, and it is primarily the strong stellar \hetwo\ emission that suggests that we may be observing BX418 at an unusual phase in its evolution.  Similarly, the age and metallicity dependence of the P Cygni emission and absorption lines from massive stars is not well known; an improved understanding of these lines will be of great benefit to high redshift work, since they are among the stronger features in most spectra of distant galaxies.

\section{Summary}
\label{sec:summary}
We have presented the analysis of the rest-frame UV and optical spectra of Q2343-BX418, a young ($<100$ Myr), low mass ($M_{\star}\sim10^9$ \msun), low metallicity ($Z\sim 1/6$ \zsun), and unreddened ($E(B-V)\simeq0.02$, UV continuum slope $\beta=-2.1$) galaxy at $z=2.3$.  In spite of its low mass, BX418 is an $L^*$ galaxy at \ztwo, with a very low mass-to-light ratio $(M/L)_{\rm B}=0.03$.  Our primary conclusions are as follows:

\begin{enumerate}
\item{Using ratios of the strong optical emission lines, we measure an oxygen abundance of $12+\log({\rm O/H})=7.9\pm0.2$, or $Z\sim1/6$ \zsun.  We measure lower limits on the ratio of \Othree/\Otwo, indicating a high ionization parameter, and of \Ntwo/\Ha, indicating a low metallicity.  We also determine the metallicity via the direct, electron temperature method, using the ratio \Othreeuv\ $\lambda\lambda$1661, 1666/\Othree\ $\lambda$5007 to determine the electron temperature and finding  $12+\log({\rm O/H})=7.8\pm0.1$.} 
\item{The source of ionization in BX418 is unlikely to be an AGN.  The \ion{C}{4}/\Cthree\ ratio is significantly lower than that seen in AGN, indicating a softer radiation field, and BX418 is undetected in deep imaging at 8 and 24 \micron, with a 24 \micron\ upper limit well below the luminosities seen in most AGN of comparable redshifts and optical magnitudes.}
\item{The rest-frame UV spectrum contains strong high ionization interstellar absorption lines from outflowing gas, while the low ionization lines are extremely weak.  We compare the equivalent widths and velocity profiles of the interstellar lines, and conclude that the covering fraction of high ionization gas is significantly higher than that of the low ionization gas, and thus that a larger than average fraction of the outflow is highly ionized.}
\item{The \lya\ emission line (with rest-frame $W_{\lya}=54$ \AA) is extremely broad, with FWHM~$\sim850$ \kms, and shows significant blueshifted emission as well as the usual asymmetric redshifted profile.  This is likely due to a low optical depth in the outflowing gas, which allows substantial escape of \lya\ photons from the approaching gas on the near side of the galaxy.  From the \lya/\Ha\ ratio we find a \lya\ escape fraction $f_e=0.4$, assuming that \lya\ and \Ha\ have the same spatial distribution.  The true escape fraction may be higher than this, if \lya\ emission is more extended because of scattering.}
\item{The UV spectrum contains strong \hetwo\ $\lambda1640$ emission, which is well fit by the superposition of a broad stellar component and unresolved nebular emission.  Photoionization modeling indicates that the nebular emission requires a very high ionization parameter $\log U\sim-1$.  The strong stellar emission may require the presence of significant numbers of very young W-R stars.}
\item{Strong nebular emission from \Cthree\ $\lambda\lambda1907$, 1909 and  \Othreeuv\ $\lambda\lambda1661$, 1666 is also present, and requires $Z\sim0.2$ \zsun\ and $\log U\gtrsim-1.6$.  We use the \Cthree/\Othreeuv\ ratio to estimate the C/O abundance ratio, finding that it is consistent with the trend in C/O vs.\ O/H observed locally.  Given BX418's young age and extreme star formation, this implies that the trend of increasing C/O with O/H is more likely to be due to metallicity-dependent stellar yields than to time delay effects, and that carbon production is not dominated by low mass stars.}
\item{We observe strong P Cygni emission from the stellar wind lines \ion{C}{4} and \ion{Si}{4}, but are unable to reproduce both the emission and the absorption profiles with models.  The strong emission requires a significant contribution from nebular emission, a stellar population age of only $\sim5$ Myr, or a top-heavy initial mass function, while the absorption suggests a low metallicity of $Z\sim0.1$ \zsun.  However, the similarity of the absorption profile to that of the higher metallicity composite spectrum of \ztwo\ galaxies suggests that our understanding of this feature is incomplete.}
\end{enumerate}

The properties of BX418 are broadly similar to those inferred for the young, low mass galaxies now being discovered at much higher redshifts, and therefore its detailed spectral properties may shed light on the likely physical conditions in these higher redshift objects, well before facilities are available to measure them directly.  Important questions remain, however, that can be addressed by observations both locally and at \ztwo.  A larger sample of high quality spectra of young, very blue galaxies at \ztwo\ is required in order to assess how many of BX418's unusual features are due to a short-lived evolutionary phase, and improved data and models of massive stars at low metallicity in the local universe are needed to aid in the interpretation of these spectra.

\acknowledgements
We thank the anonymous referee for a constructive report; Gary Ferland for valuable assistance with the interpretation of the Cloudy results; Paul Crowther, Kristian Finlator, Jay Gallagher, Samantha Rix and Elizabeth Stanway for useful conversations; and the staff of Keck Observatory for assistance with the observations. DKE is supported by the Spitzer Fellowship Program of the National Aeronautics and Space Administration under Award No. NAS7-03001 and the California Institute of Technology.   Additional support comes from the US National Science Foundation through grants AST-0606912 and AST-0908805, and the David and Lucile Packard Foundation (AES, CCS).  CCS acknowledges additional support from the John D. and Catherine T. MacArthur Foundation.  We wish to extend thanks to those of Hawaiian ancestry on whose sacred mountain we are privileged to be guests.


\begin{thebibliography}{}

\bibitem[{Adelberger} {et~al.}(2005){Adelberger}, {Shapley}, {Steidel},  {Pettini}, {Erb}, \& {Reddy}]{ass+05}
{Adelberger}, K.~L., {Shapley}, A.~E., {Steidel}, C.~C., {Pettini}, M., {Erb},  D.~K., \& {Reddy}, N.~A. 2005, \apj, 629, 636

\bibitem[Adelberger {et~al.}(2003)Adelberger, Steidel, Shapley, \&  Pettini]{assp03}
Adelberger, K.~L., Steidel, C.~C., Shapley, A.~E., \& Pettini, M. 2003, \apj,  584, 45

\bibitem[{Akerman} {et~al.}(2004){Akerman}, {Carigi}, {Nissen}, {Pettini}, \&  {Asplund}]{acn+04}
{Akerman}, C.~J., {Carigi}, L., {Nissen}, P.~E., {Pettini}, M., \& {Asplund},  M. 2004, \aap, 414, 931

\bibitem[{Ando} {et~al.}(2007){Ando}, {Ohta}, {Iwata}, {Akiyama}, {Aoki}, \&  {Tamura}]{aoi+08}
{Ando}, M., {Ohta}, K., {Iwata}, I., {Akiyama}, M., {Aoki}, K., \& {Tamura}, N.  2007, \pasj, 59, 717

\bibitem[{Asplund} {et~al.}(2009){Asplund}, {Grevesse}, {Sauval}, \&  {Scott}]{agss09}
{Asplund}, M., {Grevesse}, N., {Sauval}, A.~J., \& {Scott}, P. 2009, \araa, 47,  481

\bibitem[{Atek} {et~al.}(2009){Atek}, {Kunth}, {Schaerer}, {Hayes},  {Deharveng}, {{\"O}stlin}, \& {Mas-Hesse}]{aks+09}
{Atek}, H., {Kunth}, D., {Schaerer}, D., {Hayes}, M., {Deharveng}, J.~M.,  {{\"O}stlin}, G., \& {Mas-Hesse}, J.~M. 2009, \aap, 506, L1

\bibitem[{Baldwin} {et~al.}(1981){Baldwin}, {Phillips}, \&  {Terlevich}]{bpt81}
{Baldwin}, J.~A., {Phillips}, M.~M., \& {Terlevich}, R. 1981, \pasp, 93, 5

\bibitem[{Bottorff} {et~al.}(1998){Bottorff}, {Lamothe}, {Momjian}, {Verner},  {Vinkovi{\'c}}, \& {Ferland}]{blm+98}
{Bottorff}, M., {Lamothe}, J., {Momjian}, E., {Verner}, E., {Vinkovi{\'c}}, D.,  \& {Ferland}, G. 1998, \pasp, 110, 1040

\bibitem[{Bouwens} {et~al.}(2009){Bouwens}, {Illingworth}, {Franx}, {Chary},  {Meurer}, {Conselice}, {Ford}, {Giavalisco}, \& {van Dokkum}]{bif+09}
{Bouwens}, R.~J., {et al.} 2009, \apj, 705, 936

\bibitem[{Bouwens} {et~al.}(2010a){Bouwens}, {Illingworth},  {Oesch}, {Stiavelli}, {van Dokkum}, {Trenti}, {Magee}, {Labb{\'e}}, {Franx},  {Carollo}, \& {Gonzalez}]{bio+10b}
{Bouwens}, R.~J., {et al.} 2010a, \apjl, 709, L133

\bibitem[{Bouwens} {et~al.}(2010b){Bouwens}, {Illingworth},  {Oesch}, {Trenti}, {Stiavelli}, {Carollo}, {Franx}, {van Dokkum},  {Labb{\'e}}, \& {Magee}]{bio+10a}
{Bouwens}, R.~J., {et al.} 2010b, \apjl, 708, L69

\bibitem[{Brinchmann} {et~al.}(2008a){Brinchmann}, {Kunth}, \&  {Durret}]{bkd08}
{Brinchmann}, J., {Kunth}, D., \& {Durret}, F. 2008a, \aap, 485,  657

\bibitem[{Brinchmann} {et~al.}(2008b){Brinchmann}, {Pettini}, \&  {Charlot}]{bpc08}
{Brinchmann}, J., {Pettini}, M., \& {Charlot}, S. 2008b, \mnras,  385, 769

\bibitem[{Bruzual} \& {Charlot}(2003){Bruzual} \& {Charlot}]{bc03}
{Bruzual}, G. \& {Charlot}, S. 2003, \mnras, 344, 1000

\bibitem[{Bunker} {et~al.}(2009){Bunker}, {Wilkins}, {Ellis}, {Stark},  {Lorenzoni}, {Chiu}, {Lacy}, {Jarvis}, \& {Hickey}]{bwe+10}
{Bunker}, A., {et al.} 2009, ArXiv e-prints 0909.2255

\bibitem[{Cabanac} {et~al.}(2008){Cabanac}, {Valls-Gabaud}, \&  {Lidman}]{cvl08}
{Cabanac}, R.~A., {Valls-Gabaud}, D., \& {Lidman}, C. 2008, \mnras, 386, 2065

\bibitem[{Calzetti} {et~al.}(2000){Calzetti}, {Armus}, {Bohlin}, {Kinney},  {Koornneef}, \& {Storchi-Bergmann}]{cab+00}
{Calzetti}, D., {Armus}, L., {Bohlin}, R.~C., {Kinney}, A.~L., {Koornneef}, J.,  \& {Storchi-Bergmann}, T. 2000, \apj, 533, 682

\bibitem[{Campbell} {et~al.}(1986){Campbell}, {Terlevich}, \&  {Melnick}]{ctm86}
{Campbell}, A., {Terlevich}, R., \& {Melnick}, J. 1986, \mnras, 223, 811

\bibitem[{Carigi} {et~al.}(2005){Carigi}, {Peimbert}, {Esteban}, \&  {Garc{\'{\i}}a-Rojas}]{cpeg05}
{Carigi}, L., {Peimbert}, M., {Esteban}, C., \& {Garc{\'{\i}}a-Rojas}, J. 2005,  \apj, 623, 213

\bibitem[{Chabrier}(2003){Chabrier}]{c03}
{Chabrier}, G. 2003, \pasp, 115, 763

\bibitem[{Chandar} {et~al.}(2004){Chandar}, {Leitherer}, \&  {Tremonti}]{clt04}
{Chandar}, R., {Leitherer}, C., \& {Tremonti}, C.~A. 2004, \apj, 604, 153

\bibitem[{Chiappini} {et~al.}(2003){Chiappini}, {Romano}, \&  {Matteucci}]{crm03}
{Chiappini}, C., {Romano}, D., \& {Matteucci}, F. 2003, \mnras, 339, 63

\bibitem[{Conroy} {et~al.}(2008){Conroy}, {Shapley}, {Tinker}, {Santos}, \&  {Lemson}]{cst+08}
{Conroy}, C., {Shapley}, A.~E., {Tinker}, J.~L., {Santos}, M.~R., \& {Lemson},  G. 2008, \apj, 679, 1192

\bibitem[{Crowther} \& {Hadfield}(2006){Crowther} \& {Hadfield}]{ch06}
{Crowther}, P.~A. \& {Hadfield}, L.~J. 2006, \aap, 449, 711

\bibitem[{Dessauges-Zavadsky} {et~al.}(2010){Dessauges-Zavadsky}, {D'Odorico},  {Schaerer}, {Modigliani}, {Tapken}, \& {Vernet}]{dds+09}
{Dessauges-Zavadsky}, M., {D'Odorico}, S., {Schaerer}, D., {Modigliani}, A.,  {Tapken}, C., \& {Vernet}, J. 2010, \aap, 510, A26

\bibitem[{Dopita} {et~al.}(2006){Dopita}, {Fischera}, {Sutherland}, {Kewley},  {Tuffs}, {Popescu}, {van Breugel}, {Groves}, \& {Leitherer}]{dfs+06}
{Dopita}, M.~A., {et al.} 2006, \apj, 647, 244

\bibitem[{Eldridge} \& {Stanway}(2009){Eldridge} \& {Stanway}]{es09}
{Eldridge}, J.~J. \& {Stanway}, E.~R. 2009, \mnras, 400, 1019

\bibitem[{Erb} {et~al.}(2006a){Erb}, {Shapley}, {Pettini},  {Steidel}, {Reddy}, \& {Adelberger}]{esp+06}
{Erb}, D.~K., {Shapley}, A.~E., {Pettini}, M., {Steidel}, C.~C., {Reddy},  N.~A., \& {Adelberger}, K.~L. 2006a, \apj, 644, 813

\bibitem[{Erb} {et~al.}(2003){Erb}, {Shapley}, {Steidel}, {Pettini},  {Adelberger}, {Hunt}, {Moorwood}, \& {Cuby}]{ess+03}
{Erb}, D.~K., {Shapley}, A.~E., {Steidel}, C.~C., {Pettini}, M., {Adelberger},  K.~L., {Hunt}, M.~P., {Moorwood}, A.~F.~M., \& {Cuby}, J. 2003, \apj, 591,  101

\bibitem[{Erb} {et~al.}(2006b){Erb}, {Steidel}, {Shapley},  {Pettini}, {Reddy}, \& {Adelberger}]{ess+06stars}
{Erb}, D.~K., {Steidel}, C.~C., {Shapley}, A.~E., {Pettini}, M., {Reddy},  N.~A., \& {Adelberger}, K.~L. 2006b, \apj, 647, 128

\bibitem[{Erb} {et~al.}(2006c){Erb}, {Steidel}, {Shapley},  {Pettini}, {Reddy}, \& {Adelberger}]{ess+06mass}
---. 2006c, \apj, 646, 107

\bibitem[{Esteban} {et~al.}(2004){Esteban}, {Peimbert}, {Garc{\'{\i}}a-Rojas},  {Ruiz}, {Peimbert}, \& {Rodr{\'{\i}}guez}]{epg+04}
{Esteban}, C., {Peimbert}, M., {Garc{\'{\i}}a-Rojas}, J., {Ruiz}, M.~T.,  {Peimbert}, A., \& {Rodr{\'{\i}}guez}, M. 2004, \mnras, 355, 229

\bibitem[{Fazio} {et~al.}(2004){Fazio}, {Hora}, {Allen}, {Ashby}, {Barmby},  {Deutsch}, {Huang}, {Kleiner}, {Marengo}, {Megeath}, {Melnick}, {Pahre},  {Patten}, {Polizotti}, {Smith}, {Taylor}, {Wang}, {Willner}, {Hoffmann},  {Pipher}, {Forrest}, {McMurty}, {McCreight}, {McKelvey}, {McMurray}, {Koch},  {Moseley}, {Arendt}, {Mentzell}, {Marx}, {Losch}, {Mayman}, {Eichhorn},  {Krebs}, {Jhabvala}, {Gezari}, {Fixsen}, {Flores}, {Shakoorzadeh}, {Jungo},  {Hakun}, {Workman}, {Karpati}, {Kichak}, {Whitley}, {Mann}, {Tollestrup},  {Eisenhardt}, {Stern}, {Gorjian}, {Bhattacharya}, {Carey}, {Nelson},  {Glaccum}, {Lacy}, {Lowrance}, {Laine}, {Reach}, {Stauffer}, {Surace},  {Wilson}, {Wright}, {Hoffman}, {Domingo}, \& {Cohen}]{irac}
{Fazio}, G.~G., {et al.} 2004, \apjs, 154, 10

\bibitem[{Ferland} {et~al.}(1998){Ferland}, {Korista}, {Verner}, {Ferguson},  {Kingdon}, \& {Verner}]{cloudy}
{Ferland}, G.~J., {Korista}, K.~T., {Verner}, D.~A., {Ferguson}, J.~W.,  {Kingdon}, J.~B., \& {Verner}, E.~M. 1998, \pasp, 110, 761

\bibitem[{Ferland} \& {Osterbrock}(1985){Ferland} \& {Osterbrock}]{fo85}
{Ferland}, G.~J. \& {Osterbrock}, D.~E. 1985, \apj, 289, 105

\bibitem[{Finkelstein} {et~al.}(2009){Finkelstein}, {Papovich}, {Giavalisco},  {Reddy}, {Ferguson}, {Koekemoer}, \& {Dickinson}]{fpg+10}
{Finkelstein}, S.~L., {Papovich}, C., {Giavalisco}, M., {Reddy}, N.~A.,  {Ferguson}, H.~C., {Koekemoer}, A.~M., \& {Dickinson}, M. 2009, ArXiv  e-prints 0912.1338

\bibitem[{F{\"o}rster Schreiber} {et~al.}(2009){F{\"o}rster Schreiber},  {Genzel}, {Bouch{\'e}}, {Cresci}, {Davies}, {Buschkamp}, {Shapiro},  {Tacconi}, {Hicks}, {Genel}, {Shapley}, {Erb}, {Steidel}, {Lutz},  {Eisenhauer}, {Gillessen}, {Sternberg}, {Renzini}, {Cimatti}, {Daddi},  {Kurk}, {Lilly}, {Kong}, {Lehnert}, {Nesvadba}, {Verma}, {McCracken},  {Arimoto}, {Mignoli}, \& {Onodera}]{fgb+09}
{F{\"o}rster Schreiber}, N.~M., {et al.} 2009, \apj, 706, 1364

\bibitem[{F{\"o}rster Schreiber} {et~al.}(2006){F{\"o}rster Schreiber},  {Genzel}, {Lehnert}, {Bouch{\'e}}, {Verma}, {Erb}, {Shapley}, {Steidel},  {Davies}, {Lutz}, {Nesvadba}, {Tacconi}, {Eisenhauer}, {Abuter}, {Gilbert},  {Gillessen}, \& {Sternberg}]{fgl+06}
{F{\"o}rster Schreiber}, N.~M., {et al.} 2006, \apj, 645, 1062

\bibitem[{F{\"o}rster Schreiber} {et~al.}(2001){F{\"o}rster Schreiber},  {Genzel}, {Lutz}, {Kunze}, \& {Sternberg}]{fgl+01}
{F{\"o}rster Schreiber}, N.~M., {Genzel}, R., {Lutz}, D., {Kunze}, D., \&  {Sternberg}, A. 2001, \apj, 552, 544

\bibitem[{Fosbury} {et~al.}(2003){Fosbury}, {Villar-Mart{\'{\i}}n},  {Humphrey}, {Lombardi}, {Rosati}, {Stern}, {Hook}, {Holden}, {Stanford},  {Squires}, {Rauch}, \& {Sargent}]{fvh+03}
{Fosbury}, R.~A.~E., {et al.} 2003, \apj, 596,  797

\bibitem[{Garnett} {et~al.}(1991){Garnett}, {Kennicutt}, {Chu}, \&  {Skillman}]{gkc+91}
{Garnett}, D.~R., {Kennicutt}, Jr., R.~C., {Chu}, Y., \& {Skillman}, E.~D.  1991, \apj, 373, 458

\bibitem[{Garnett} {et~al.}(1999){Garnett}, {Shields}, {Peimbert},  {Torres-Peimbert}, {Skillman}, {Dufour}, {Terlevich}, \&  {Terlevich}]{gsp+99}
{Garnett}, D.~R., {Shields}, G.~A., {Peimbert}, M., {Torres-Peimbert}, S.,  {Skillman}, E.~D., {Dufour}, R.~J., {Terlevich}, E., \& {Terlevich}, R.~J.  1999, \apj, 513, 168

\bibitem[{Garnett} {et~al.}(1997){Garnett}, {Shields}, {Skillman}, {Sagan}, \&  {Dufour}]{gss97}
{Garnett}, D.~R., {Shields}, G.~A., {Skillman}, E.~D., {Sagan}, S.~P., \&  {Dufour}, R.~J. 1997, \apj, 489, 63

\bibitem[{Garnett} {et~al.}(1995){Garnett}, {Skillman}, {Dufour}, {Peimbert},  {Torres-Peimbert}, {Terlevich}, {Terlevich}, \& {Shields}]{gsd+95}
{Garnett}, D.~R., {Skillman}, E.~D., {Dufour}, R.~J., {Peimbert}, M.,  {Torres-Peimbert}, S., {Terlevich}, R., {Terlevich}, E., \& {Shields}, G.~A.  1995, \apj, 443, 64

\bibitem[{Genzel} {et~al.}(2008){Genzel}, {Burkert}, {Bouch{\'e}}, {Cresci},  {F{\"o}rster Schreiber}, {Shapley}, {Shapiro}, {Tacconi}, {Buschkamp},  {Cimatti}, {Daddi}, {Davies}, {Eisenhauer}, {Erb}, {Genel}, {Gerhard},  {Hicks}, {Lutz}, {Naab}, {Ott}, {Rabien}, {Renzini}, {Steidel}, {Sternberg},  \& {Lilly}]{gbb+08}
{Genzel}, R., {et al.} 2008, \apj, 687, 59

\bibitem[{Genzel} {et~al.}(2006){Genzel}, {Tacconi}, {Eisenhauer},  {F{\"o}rster Schreiber}, {Cimatti}, {Daddi}, {Bouch{\'e}}, {Davies},  {Lehnert}, {Lutz}, {Nesvadba}, {Verma}, {Abuter}, {Shapiro}, {Sternberg},  {Renzini}, {Kong}, {Arimoto}, \& {Mignoli}]{gte+06}
{Genzel}, R., {et al.} 2006, \nat, 442, 786

\bibitem[{Groves} {et~al.}(2006){Groves}, {Heckman}, \& {Kauffmann}]{ghk06}
{Groves}, B.~A., {Heckman}, T.~M., \& {Kauffmann}, G. 2006, \mnras, 371, 1559

\bibitem[{Hainline} {et~al.}(2009){Hainline}, {Shapley}, {Kornei}, {Pettini},  {Buckley-Geer}, {Allam}, \& {Tucker}]{hsk+09}
{Hainline}, K.~N., {Shapley}, A.~E., {Kornei}, K.~A., {Pettini}, M.,  {Buckley-Geer}, E., {Allam}, S.~S., \& {Tucker}, D.~L. 2009, \apj, 701, 52

\bibitem[{Henry} {et~al.}(2000){Henry}, {Edmunds}, \& {K{\"o}ppen}]{hek00}
{Henry}, R.~B.~C., {Edmunds}, M.~G., \& {K{\"o}ppen}, J. 2000, \apj, 541, 660

\bibitem[{Izotov} {et~al.}(2006){Izotov}, {Stasi{\'n}ska}, {Meynet}, {Guseva},  \& {Thuan}]{ism+06}
{Izotov}, Y.~I., {Stasi{\'n}ska}, G., {Meynet}, G., {Guseva}, N.~G., \&  {Thuan}, T.~X. 2006, \aap, 448, 955

\bibitem[{Jimenez} \& {Haiman}(2006){Jimenez} \& {Haiman}]{jh06}
{Jimenez}, R. \& {Haiman}, Z. 2006, \nat, 440, 501

\bibitem[{Kauffmann} {et~al.}(2003){Kauffmann}, {Heckman}, {Tremonti},  {Brinchmann}, {Charlot}, {White}, {Ridgway}, {Brinkmann}, {Fukugita}, {Hall},  {Ivezi{\'c}}, {Richards}, \& {Schneider}]{kht+03}
{Kauffmann}, G., {et al.} 2003, \mnras, 346, 1055

\bibitem[{Kennicutt}(1998a){Kennicutt}]{k98}
{Kennicutt}, R.~C. 1998a, \araa, 36, 189

\bibitem[{Kennicutt}(1998b){Kennicutt}]{k98schmidt}
---. 1998b, \apj, 498, 541

\bibitem[{Kewley} {et~al.}(2001){Kewley}, {Dopita}, {Sutherland}, {Heisler},  \& {Trevena}]{kds+01}
{Kewley}, L.~J., {Dopita}, M.~A., {Sutherland}, R.~S., {Heisler}, C.~A., \&  {Trevena}, J. 2001, \apj, 556, 121

\bibitem[{Kewley} \& {Ellison}(2008){Kewley} \& {Ellison}]{ke08}
{Kewley}, L.~J. \& {Ellison}, S.~L. 2008, \apj, 681, 1183

\bibitem[{Kobulnicky} {et~al.}(1999){Kobulnicky}, {Kennicutt}, \&  {Pizagno}]{kkp99}
{Kobulnicky}, H.~A., {Kennicutt}, R.~C., \& {Pizagno}, J.~L. 1999, \apj, 514,  544

\bibitem[{Kobulnicky} \& {Skillman}(1998){Kobulnicky} \& {Skillman}]{ks98}
{Kobulnicky}, H.~A. \& {Skillman}, E.~D. 1998, \apj, 497, 601

\bibitem[{Kornei} {et~al.}(2010){Kornei}, {Shapley}, {Erb}, {Steidel},  {Reddy}, {Pettini}, \& {Bogosavljevi{\'c}}]{kse+10}
{Kornei}, K.~A., {Shapley}, A.~E., {Erb}, D.~K., {Steidel}, C.~C., {Reddy},  N.~A., {Pettini}, M., \& {Bogosavljevi{\'c}}, M. 2010, \apj, 711, 693

\bibitem[{Labb{\'e}} {et~al.}(2010){Labb{\'e}}, {Gonz{\'a}lez}, {Bouwens},  {Illingworth}, {Franx}, {Trenti}, {Oesch}, {van Dokkum}, {Stiavelli},  {Carollo}, {Kriek}, \& {Magee}]{lgb+10}
{Labb{\'e}}, I., {et al.} 2010, \apjl, 716, L103

\bibitem[{Law} {et~al.}(2009){Law}, {Steidel}, {Erb}, {Larkin}, {Pettini},  {Shapley}, \& {Wright}]{lse+09}
{Law}, D.~R., {Steidel}, C.~C., {Erb}, D.~K., {Larkin}, J.~E., {Pettini}, M.,  {Shapley}, A.~E., \& {Wright}, S.~A. 2009, \apj, 697, 2057

\bibitem[{Leitherer} {et~al.}(2002){Leitherer}, {Calzetti}, \&  {Martins}]{lcm02}
{Leitherer}, C., {Calzetti}, D., \& {Martins}, L.~P. 2002, \apj, 574, 114

\bibitem[{Leitherer} {et~al.}(2010){Leitherer}, {Ortiz Ot{\'a}lvaro}, {Bresolin}, {Kudritzki},  {Lo Faro}, {Pauldrach}, {Pettini}, \& {Rix}]{lob+10}
{Leitherer}, C., {Ortiz Ot{\'a}lvaro}, P.~A., {Bresolin}, F., {Kudritzki}, R.-P., {Lo Faro}, B., {Pauldrach}, A.~W.~A., {Pettini}, M., \& {Rix}, S. 2010, \apjs, submitted

\bibitem[{Leitherer} {et~al.}(1995){Leitherer}, {Robert}, \&  {Heckman}]{lrh95}
{Leitherer}, C., {Robert}, C., \& {Heckman}, T.~M. 1995, \apjs, 99, 173

\bibitem[{Leitherer} {et~al.}(1999){Leitherer}, {Schaerer}, {Goldader},  {Delgado}, {Robert}, {Kune}, {de Mello}, {Devost}, \& {Heckman}]{sb99}
{Leitherer}, C., {et al.} 1999, \apjs, 123, 3

\bibitem[{Liu} {et~al.}(2008){Liu}, {Shapley}, {Coil}, {Brinchmann}, \&  {Ma}]{lsc+08}
{Liu}, X., {Shapley}, A.~E., {Coil}, A.~L., {Brinchmann}, J., \& {Ma}, C. 2008,  \apj, 678, 758

\bibitem[{Maiolino} {et~al.}(2008){Maiolino}, {Nagao}, {Grazian}, {Cocchia},  {Marconi}, {Mannucci}, {Cimatti}, {Pipino}, {Ballero}, {Calura}, {Chiappini},  {Fontana}, {Granato}, {Matteucci}, {Pastorini}, {Pentericci}, {Risaliti},  {Salvati}, \& {Silva}]{mng+08}
{Maiolino}, R., {et al.} 2008, \aap, 488, 463

\bibitem[{McGaugh}(1991){McGaugh}]{m91}
{McGaugh}, S.~S. 1991, \apj, 380, 140

\bibitem[{McLean} {et~al.}(1998){McLean}, {Becklin}, {Bendiksen}, {Brims},  {Canfield}, {Figer}, {Graham}, {Hare}, {Lacayanga}, {Larkin}, {Larson},  {Levenson}, {Magnone}, {Teplitz}, \& {Wong}]{mbb+98}
{McLean}, I.~S., {et al.} 1998, in Proc. SPIE Vol. 3354, p. 566-578, Infrared Astronomical  Instrumentation, Albert M. Fowler; Ed., Vol. 3354, 566--578

\bibitem[{McLure} {et~al.}(2006){McLure}, {Cirasuolo}, {Dunlop}, {Sekiguchi},  {Almaini}, {Foucaud}, {Simpson}, {Watson}, {Hirst}, {Page}, \&  {Smail}]{mcd+06}
{McLure}, R.~J., {et al.} 2006, \mnras, 372, 357

\bibitem[{McLure} {et~al.}(2010){McLure}, {Dunlop}, {Cirasuolo}, {Koekemoer},  {Sabbi}, {Stark}, {Targett}, \& {Ellis}]{mdc+10}
{McLure}, R.~J., {Dunlop}, J.~S., {Cirasuolo}, M., {Koekemoer}, A.~M., {Sabbi},  E., {Stark}, D.~P., {Targett}, T.~A., \& {Ellis}, R.~S. 2010, \mnras, 403,  960

\bibitem[{Morton}(2003){Morton}]{morton03}
{Morton}, D.~C. 2003, \apjs, 149, 205

\bibitem[{Oesch} {et~al.}(2010){Oesch}, {Bouwens}, {Illingworth}, {Carollo},  {Franx}, {Labb{\'e}}, {Magee}, {Stiavelli}, {Trenti}, \& {van  Dokkum}]{obi+10}
{Oesch}, P.~A., {et al.} 2010, \apjl, 709, L16

\bibitem[{Oke} {et~al.}(1995){Oke}, {Cohen}, {Carr}, {Cromer}, {Dingizian},  {Harris}, {Labrecque}, {Lucinio}, {Schaal}, {Epps}, \& {Miller}]{occ+95}
{Oke}, J.~B., {et al.} 1995, \pasp, 107, 375

\bibitem[{P{\'e}rez-Montero} \& {D{\'{\i}}az}(2003){P{\'e}rez-Montero} \& {D{\'{\i}}az}]{pd03}
{P{\'e}rez-Montero}, E. \& {D{\'{\i}}az}, A.~I. 2003, \mnras, 346, 105

\bibitem[{P{\'e}rez-Montero} {et~al.}(2010){P{\'e}rez-Montero},  {Garc{\'{\i}}a-Benito}, {H{\"a}gele}, \& {D{\'{\i}}az}]{pghd10}
{P{\'e}rez-Montero}, E., {Garc{\'{\i}}a-Benito}, R., {H{\"a}gele}, G.~F., \&  {D{\'{\i}}az}, {\'A}.~I. 2010, \mnras, 404, 2037

\bibitem[{Pettini} \& {Pagel}(2004){Pettini} \& {Pagel}]{pp04}
{Pettini}, M. \& {Pagel}, B.~E.~J. 2004, \mnras, 348, L59

\bibitem[{Pettini} {et~al.}(2002){Pettini}, {Rix}, {Steidel}, {Adelberger},  {Hunt}, \& {Shapley}]{prs+02}
{Pettini}, M., {Rix}, S.~A., {Steidel}, C.~C., {Adelberger}, K.~L., {Hunt},  M.~P., \& {Shapley}, A.~E. 2002, \apj, 569, 742

\bibitem[{Pettini} {et~al.}(2001){Pettini}, {Shapley}, {Steidel}, {Cuby},  {Dickinson}, {Moorwood}, {Adelberger}, \& {Giavalisco}]{pss+01}
{Pettini}, M., {Shapley}, A.~E., {Steidel}, C.~C., {Cuby}, J., {Dickinson}, M.,  {Moorwood}, A.~F.~M., {Adelberger}, K.~L., \& {Giavalisco}, M. 2001, \apj,  554, 981

\bibitem[{Pettini} {et~al.}(1997){Pettini}, {Smith}, {King}, \&  {Hunstead}]{pskh97}
{Pettini}, M., {Smith}, L.~J., {King}, D.~L., \& {Hunstead}, R.~W. 1997, \apj,  486, 665

\bibitem[{Pettini} {et~al.}(2000){Pettini}, {Steidel}, {Adelberger},  {Dickinson}, \& {Giavalisco}]{psa+00}
{Pettini}, M., {Steidel}, C.~C., {Adelberger}, K.~L., {Dickinson}, M., \&  {Giavalisco}, M. 2000, \apj, 528, 96

\bibitem[{Pettini} {et~al.}(2008){Pettini}, {Zych}, {Steidel}, \&  {Chaffee}]{pzsc08}
{Pettini}, M., {Zych}, B.~J., {Steidel}, C.~C., \& {Chaffee}, F.~H. 2008,  \mnras, 385, 2011

\bibitem[{Prochaska} \& {Wolfe}(1999){Prochaska} \& {Wolfe}]{pw99}
{Prochaska}, J.~X. \& {Wolfe}, A.~M. 1999, \apjs, 121, 369

\bibitem[{Quider} {et~al.}(2009){Quider}, {Pettini}, {Shapley}, \&  {Steidel}]{qpss09}
{Quider}, A.~M., {Pettini}, M., {Shapley}, A.~E., \& {Steidel}, C.~C. 2009,  \mnras, 398, 1263

\bibitem[{Quider} {et~al.}(2010){Quider}, {Shapley}, {Pettini}, {Steidel}, \&  {Stark}]{qsp+10}
{Quider}, A.~M., {Shapley}, A.~E., {Pettini}, M., {Steidel}, C.~C., \& {Stark},  D.~P. 2010, \mnras, 402, 1467

\bibitem[{Razoumov} \& {Sommer-Larsen}(2010){Razoumov} \& {Sommer-Larsen}]{rs10}
{Razoumov}, A.~O. \& {Sommer-Larsen}, J. 2010, \apj, 710, 1239

\bibitem[{Reddy} {et~al.}(2010){Reddy}, {Erb}, {Pettini}, {Steidel}, \&  {Shapley}]{rep+10}
{Reddy}, N.~A., {Erb}, D.~K., {Pettini}, M., {Steidel}, C.~C., \& {Shapley},  A.~E. 2010, \apj, 712, 1070

\bibitem[{Reddy} {et~al.}(2006a){Reddy}, {Steidel}, {Erb}, {Shapley}, \&  {Pettini}]{rse+06}
{Reddy}, N.~A., {Steidel}, C.~C., {Erb}, D.~K., {Shapley}, A.~E., \& {Pettini},  M. 2006, \apj, 653, 1004

\bibitem[{Reddy} {et~al.}(2006b){Reddy}, {Steidel}, {Fadda}, {Yan}, {Pettini},  {Shapley}, {Erb}, \& {Adelberger}]{rsf+06}
{Reddy}, N.~A., {Steidel}, C.~C., {Fadda}, D., {Yan}, L., {Pettini}, M.,  {Shapley}, A.~E., {Erb}, D.~K., \& {Adelberger}, K.~L. 2006, \apj, 644, 792

\bibitem[{Rix} {et~al.}(2004){Rix}, {Pettini}, {Leitherer}, {Bresolin},  {Kudritzki}, \& {Steidel}]{rpl+04}
{Rix}, S.~A., {Pettini}, M., {Leitherer}, C., {Bresolin}, F., {Kudritzki}, R.,  \& {Steidel}, C.~C. 2004, \apj, 615, 98

\bibitem[{Schaerer}(2003){Schaerer}]{s03}
{Schaerer}, D. 2003, \aap, 397, 527

\bibitem[{Schwartz} {et~al.}(2006){Schwartz}, {Martin}, {Chandar},  {Leitherer}, {Heckman}, \& {Oey}]{smc+06}
{Schwartz}, C.~M., {Martin}, C.~L., {Chandar}, R., {Leitherer}, C., {Heckman},  T.~M., \& {Oey}, M.~S. 2006, \apj, 646, 858

\bibitem[{Shapley} {et~al.}(2001){Shapley}, {Steidel}, {Adelberger},  {Dickinson}, {Giavalisco}, \& {Pettini}]{ssa+01}
{Shapley}, A.~E., {Steidel}, C.~C., {Adelberger}, K.~L., {Dickinson}, M.,  {Giavalisco}, M., \& {Pettini}, M. 2001, \apj, 562, 95

\bibitem[{Shapley} {et~al.}(2005){Shapley}, {Steidel}, {Erb}, {Reddy},  {Adelberger}, {Pettini}, {Barmby}, \& {Huang}]{sse+05}
{Shapley}, A.~E., {Steidel}, C.~C., {Erb}, D.~K., {Reddy}, N.~A., {Adelberger},  K.~L., {Pettini}, M., {Barmby}, P., \& {Huang}, J. 2005, \apj, 626, 698

\bibitem[{Shapley} {et~al.}(2003){Shapley}, {Steidel}, {Pettini}, \&  {Adelberger}]{ssp+03}
{Shapley}, A.~E., {Steidel}, C.~C., {Pettini}, M., \& {Adelberger}, K.~L. 2003,  \apj, 588, 65

\bibitem[{Siana} {et~al.}(2010){Siana}, {Teplitz}, {Ferguson}, {Brown},  {Giavalisco}, {Dickinson}, {Chary}, {de Mello}, {Conselice}, {Bridge},  {Gardner}, {Colbert}, \& {Scarlata}]{stf+10}
{Siana}, B., {et al.} 2010,  ArXiv e-prints 1001.3412

\bibitem[{Simcoe} {et~al.}(2006){Simcoe}, {Sargent}, {Rauch}, \&  {Becker}]{ssrb06}
{Simcoe}, R.~A., {Sargent}, W.~L.~W., {Rauch}, M., \& {Becker}, G. 2006, \apj,  637, 648

\bibitem[{Smith} {et~al.}(2002){Smith}, {Norris}, \& {Crowther}]{snc02}
{Smith}, L.~J., {Norris}, R.~P.~F., \& {Crowther}, P.~A. 2002, \mnras, 337,  1309

\bibitem[{Stark} {et~al.}(2009){Stark}, {Ellis}, {Bunker}, {Bundy}, {Targett},  {Benson}, \& {Lacy}]{seb+09}
{Stark}, D.~P., {Ellis}, R.~S., {Bunker}, A., {Bundy}, K., {Targett}, T.,  {Benson}, A., \& {Lacy}, M. 2009, \apj, 697, 1493

\bibitem[{Stark} {et~al.}(2010){Stark}, {Ellis}, {Chiu}, {Ouchi}, \&  {Bunker}]{sec+10}
{Stark}, D.~P., {Ellis}, R.~S., {Chiu}, K., {Ouchi}, M., \& {Bunker}, A. 2010,  ArXiv e-prints 1003.5244

\bibitem[{Stark} {et~al.}(2008){Stark}, {Swinbank}, {Ellis}, {Dye}, {Smail},  \& {Richard}]{sse+08}
{Stark}, D.~P., {Swinbank}, A.~M., {Ellis}, R.~S., {Dye}, S., {Smail}, I.~R.,  \& {Richard}, J. 2008, \nat, 455, 775

\bibitem[{Steidel} {et~al.}(2003){Steidel}, {Adelberger}, {Shapley},  {Pettini}, {Dickinson}, \& {Giavalisco}]{sas+03}
{Steidel}, C.~C., {Adelberger}, K.~L., {Shapley}, A.~E., {Pettini}, M.,  {Dickinson}, M., \& {Giavalisco}, M. 2003, \apj, 592, 728

\bibitem[{Steidel} {et~al.}(2010){Steidel}, {Erb}, {Shapley}, {Pettini},  {Reddy}, {Bogosavljevi{\'c}}, {Rudie}, \& {Rakic}]{ses+10}
{Steidel}, C.~C., {Erb}, D.~K., {Shapley}, A.~E., {Pettini}, M., {Reddy}, N.,  {Bogosavljevi{\'c}}, M., {Rudie}, G.~C., \& {Rakic}, O. 2010, \apj, 717, 289

\bibitem[{Steidel} {et~al.}(2002){Steidel}, {Hunt}, {Shapley}, {Adelberger},  {Pettini}, {Dickinson}, \& {Giavalisco}]{shs+02}
{Steidel}, C.~C., {Hunt}, M.~P., {Shapley}, A.~E., {Adelberger}, K.~L.,  {Pettini}, M., {Dickinson}, M., \& {Giavalisco}, M. 2002, \apj, 576, 653

\bibitem[{Steidel} {et~al.}(2004){Steidel}, {Shapley}, {Pettini},  {Adelberger}, {Erb}, {Reddy}, \& {Hunt}]{ssp+04}
{Steidel}, C.~C., {Shapley}, A.~E., {Pettini}, M., {Adelberger}, K.~L., {Erb},  D.~K., {Reddy}, N.~A., \& {Hunt}, M.~P. 2004, \apj, 604, 534

\bibitem[{Terlevich} {et~al.}(2004){Terlevich}, {Silich}, {Rosa-Gonz{\'a}lez},  \& {Terlevich}]{tsrt04}
{Terlevich}, R., {Silich}, S., {Rosa-Gonz{\'a}lez}, D., \& {Terlevich}, E.  2004, \mnras, 348, 1191

\bibitem[{Vanzella} {et~al.}(2009){Vanzella}, {Giavalisco}, {Dickinson},  {Cristiani}, {Nonino}, {Kuntschner}, {Popesso}, {Rosati}, {Renzini}, {Stern},  {Cesarsky}, {Ferguson}, \& {Fosbury}]{vgd+09}
{Vanzella}, E., {et al.} 2009, \apj,  695, 1163

\bibitem[{Veilleux} \& {Osterbrock}(1987){Veilleux} \& {Osterbrock}]{vo87}
{Veilleux}, S. \& {Osterbrock}, D.~E. 1987, \apjs, 63, 295

\bibitem[{Verhamme} {et~al.}(2008){Verhamme}, {Schaerer}, {Atek}, \&  {Tapken}]{vsat08}
{Verhamme}, A., {Schaerer}, D., {Atek}, H., \& {Tapken}, C. 2008, \aap, 491, 89

\bibitem[{Verhamme} {et~al.}(2006){Verhamme}, {Schaerer}, \&  {Maselli}]{vsm06}
{Verhamme}, A., {Schaerer}, D., \& {Maselli}, A. 2006, \aap, 460, 397

\bibitem[{Villar-Mart{\'{\i}}n} {et~al.}(2004){Villar-Mart{\'{\i}}n},  {Cervi{\~n}o}, \& {Gonz{\'a}lez Delgado}]{vcd04}
{Villar-Mart{\'{\i}}n}, M., {Cervi{\~n}o}, M., \& {Gonz{\'a}lez Delgado}, R.~M.  2004, \mnras, 355, 1132

\bibitem[{Westera} {et~al.}(2004){Westera}, {Cuisinier}, {Telles}, \&  {Kehrig}]{wctk04}
{Westera}, P., {Cuisinier}, F., {Telles}, E., \& {Kehrig}, C. 2004, \aap, 423,  133

\bibitem[{Wilson} {et~al.}(2003){Wilson}, {Eikenberry}, {Henderson},  {Hayward}, {Carson}, {Pirger}, {Barry}, {Brandl}, {Houck}, {Fitzgerald}, \&  {Stolberg}]{wirc}
{Wilson}, J.~C., {et al.} 2003, in Instrument Design  and Performance for Optical/Infrared Ground-based Telescopes. Edited by Iye,  Masanori; Moorwood, Alan F. M. Proceedings of the SPIE, Volume 4841, pp.  451-458 (2003)., 451--458

\bibitem[{Wise} \& {Cen}(2009){Wise} \& {Cen}]{wc09}
{Wise}, J.~H. \& {Cen}, R. 2009, \apj, 693, 984

\bibitem[{Yan} {et~al.}(2005){Yan}, {Dickinson}, {Stern}, {Eisenhardt},  {Chary}, {Giavalisco}, {Ferguson}, {Casertano}, {Conselice}, {Papovich},  {Reach}, {Grogin}, {Moustakas}, \& {Ouchi}]{yds+05}
{Yan}, H., {et al.} 2005, \apj, 634, 109

\bibitem[{Yan} {et~al.}(2009){Yan}, {Windhorst}, {Hathi}, {Cohen}, {Ryan},  {O'Connell}, \& {McCarthy}]{ywh+10}
{Yan}, H., {Windhorst}, R., {Hathi}, N., {Cohen}, S., {Ryan}, R., {O'Connell},  R., \& {McCarthy}, P. 2009, ArXiv e-prints 0910.0077

\bibitem[{Yuan} \& {Kewley}(2009){Yuan} \& {Kewley}]{yk09}
{Yuan}, T. \& {Kewley}, L.~J. 2009, \apjl, 699, L161

\end{thebibliography}
\end{document}